\documentclass[twocolumn]{jpsj3}
\usepackage{graphicx}
\usepackage{dcolumn}
\usepackage{bm}
\usepackage{ulem}
\usepackage{braket}
\usepackage{amsmath,amssymb}
\usepackage{color}
\usepackage{setspace}
\usepackage{multirow}

\title{
Unified description of electronic orderings and cross correlations\\ 
by complete multipole representation
}

\author{Satoru Hayami$^1$ and Hiroaki Kusunose$^{2,3}$}
\inst{
$^1$Graduate School of Science, Hokkaido University, Sapporo 060-0810, Japan \\
$^2$Department of Physics, Meiji University, Kawasaki 214-8571, Japan \\
$^3$Quantum Research Center for Chirality, Institute for Molecular Science, Okazaki 444-8585, Japan
}

\abst{
We overview recent developments of electronic orderings and associated cross correlations in condensed matter physics based on a complete set of multipole representations (electric, magnetic, electric toroidal, and magnetic toroidal multipoles) with distinct space-time inversion symmetries. 
By means of the symmetry-adapted complete basis set in any physical Hilbert space including atomic and cluster degrees of freedom, it provides a unified description of electronic orderings, deformations of band structures, cross correlations, and transport properties in solids. 
We review that a multipole representation is a powerful tool to attain a comprehensive understanding of various electronic properties and physical phenomena observed in materials.  
We also demonstrate how to identify active multipoles and expected physical phenomena in real materials with several examples. 
Our review would serve as a solid foundation for future studies toward observations and identifications of unknown electronic phases and their related physical phenomena. 
}

\begin{document}
\maketitle

\section{Introduction}

One of the fascinating aspects of condensed matter physics is its diversity. 
In spite of only about 110 elements in the world, a myriad of materials with various physical properties, such as magnetism, dielectricity, and superconductivity, appear depending on the types of constituent elements, their chemical composition, and crystal structures.
Such diversity has often been classified according to the symmetry breaking in the electron system. 
For example, the emergence of ferroelectricity is related to the spatial inversion symmetry breaking and that of ferromagnetism is related to the time-reversal symmetry breaking. 
Many studies have been devoted to exploring and controlling new physical properties induced by symmetry breaking, such as multiferroicity in the breakings of both spatial inversion and time-reversal symmetries, chirality in the breakings of both spatial inversion and mirror symmetries with keeping time-reversal symmetry, and ferroaxiality in the breaking of the vertical mirror symmetries. 
Depending on the symmetry breakings, various cross correlations and transport phenomena have been discovered, which should be the foundations for the developments and applications of modern science and technology. 

By taking a spirit of multipole expansion in ordinary electromagnetism to describe any \textit{anisotropy} of electromagnetic media, the concept of ``electronic multipole'' has been introduced as ``symmetry-adapted basis set" to describe any anisotropy of combined electronic degrees of freedom in solids, e.g., charge, spin, and orbital, in a unified manner. 
Such a concept has been mainly used to express unconventional electronic phases in rare-earth and actinide compounds with $f$ electrons in the early stage of research~\cite{kuramoto2008electronic,  Santini_RevModPhys.81.807,kuramoto2009multipole, Kusunose_JPSJ.77.064710, suzuki2018first}; 
higher-rank atomic-scale multipole orderings like electric quadrupole and magnetic octupole orderings have been established by intensive experimental and theoretical investigations.  
In this context, higher-rank atomic-scale multipole orderings have also been found in $d$-electron systems~\cite{Hirai_doi:10.7566/JPSJ.88.064712, Hirai_PhysRevResearch.2.022063, takayama2021spin}. 
Meanwhile, the conventional description within one type of atomic orbital within a single atom limits the spatial inversion parity of multipoles to even parity owing to the restriction of the relevant Hilbert space. 

Since the last decade, attempts to extend the concept of atomic-scale (electronic) multipoles have been extensively performed mainly from the theoretical side. 
One of the extensions is the application to a cluster consisting of multiple atomic sites, which is so-called a cluster multipole~\cite{Yanase_JPSJ.83.014703, Hayami_PhysRevB.90.024432, Hayami_PhysRevB.90.081115, hayami2016emergent, Suzuki_PhysRevB.95.094406}. 
By regarding an antiferroic alignment of atomic-scale multipoles as the ferroic alignment of higher-rank multipoles in the cluster unit from the symmetry viewpoint, one finds that unconventional ``odd-parity" multipoles with spatial inversion odd can be activated in the spanned Hilbert space. 
Similarly, any bond and local-current (imaginary electron hopping) orderings that originate from the bond degree of freedom in a cluster can be also expressed as the cluster multipole~\cite{Hayami_PhysRevLett.122.147602, Hayami_PhysRevB.102.144441}. 
Another extension is the application to multi atomic orbitals with different types, which is so-called a hybrid multipole~\cite{hayami2018microscopic}.  

In the development of these extensions, it was recognized that two types of ``toroidal-type" multipoles, electric toroidal (ET) and magnetic toroidal (MT) multipoles, are indispensable for a unified description, where ET and MT multipoles have the opposite spatial inversion parity to the conventional electric (E) and magnetic (M) multipoles, respectively~\cite{dubovik1975multipole, dubovik1986axial, dubovik1990toroid, kopaev2009toroidal, Spaldin_0953-8984-20-43-434203, talebi2018theory}. 
Recently, it was clarified that four types of multipoles consisting of E, M, MT, and ET multipoles give a complete set to describe arbitrary multi-site and multi-orbital degrees of freedom at the quantum-mechanical operator level~\cite{hayami2018microscopic, Hayami_PhysRevB.98.165110, kusunose2020complete, Kusunose_PhysRevB.107.195118}.  
In parallel with theoretical studies, prototype materials to exhibit unconventional multipole orderings have been discovered and/or recognized in experiments: e.g., Mn$_3$Sn~\cite{nakatsuji2015large, ikhlas2017large,kuroda2017evidence,higo2018large}, Cd$_2$Re$_2$O$_7$~\cite{yamaura2002low, hiroi2018pyrochlore}, NiTiO$_3$~\cite{hayashida2020visualization, Hayashida_PhysRevMaterials.5.124409}, and UNi$_4$B~\cite{Mentink1994, Oyamada2007, saito2018evidence}.

\begin{figure*}[tb!]
\centering
\includegraphics[width=1.0 \hsize]{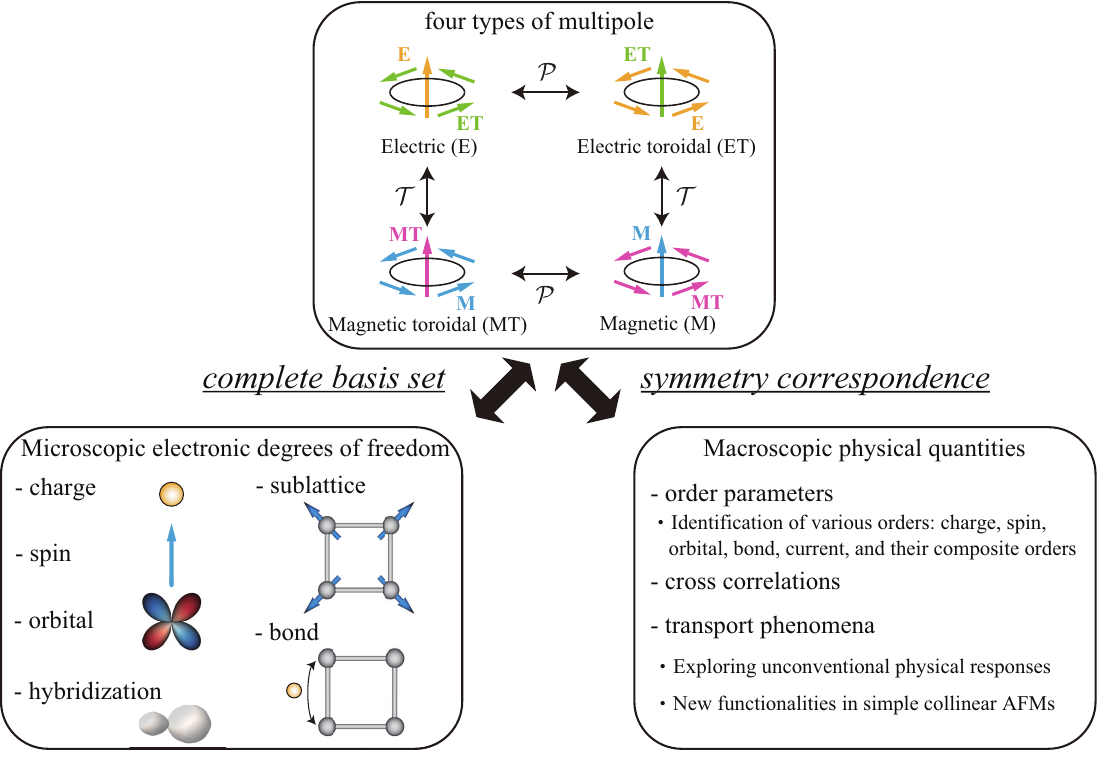} 
\caption{
\label{fig: concept}
Concept of electronic multipoles.
E, M, MT, and ET stand for electric, magnetic, magnetic toroidal, and electric toroidal multipoles, respectively. 
$\mathcal{P}$ and $\mathcal{T}$ represent spatial inversion and time-reversal operations, respectively.
}
\end{figure*}

The state-of-the-art concept of electronic multipoles is summarized in Fig.~\ref{fig: concept}. 
Since four types of multipoles (E, M, MT, and ET) constitute a symmetry-adapted complete basis set in any Hilbert space spanned by atomic and cluster degrees of freedom, they play a role in communicating microscopic electronic degrees of freedom activated in relevant Hilbert space and macroscopic physical quantities~\cite{Yatsushiro_PhysRevB.104.054412}. 
The advantages of using the multipole representation are as follows: 
\begin{itemize}
\item Systematic identification and classification of electronic order parameters 
\item Predictability of overlooked physical phenomena under antiferromagnetic (AFM), charge, orbital, and other electronic orderings
\item Exploration of cross correlations and linear, nonlinear, and nonreciprocal transports
\item Intuitive understanding of the interplay between orderings and phenomena via a coupling of electronic multipoles 
\end{itemize}
These advantages would provide a unified understanding of diverse physical phenomena beyond the symmetry argument and would bring about bottom-up engineering of desired functionalities based on microscopic electronic degrees of freedom. 

In this paper, we review recent developments in the research of multipole representations and its application to materials. 
In Sect.~\ref{sec: Four types of multipoles}, we introduce the quantum-mechanical operator expressions of four types of multipoles with distinct spatial inversion and time-reversal parities. 
We briefly explain how four types of multipole constitute a symmetry-adapted basis set. 
In Sect.~\ref{sec: Electronic band structure}, we introduce multipole representation in momentum space and discuss the relation to the spin splitting and modulations in the energy dispersions. 
We also discuss the microscopic origins to induce band modulations with or without the relativistic spin-orbit coupling in terms of microscopic multipole couplings. 
Then, we discuss cross correlations triggered by the multipole orderings in Sect.~\ref{sec: Cross-correlation phenomena}. 
In Sect.~\ref{sec: Classification of multipoles}, we show the classification of multipoles under 32 point groups and 122 magnetic point groups. 
Based on the multipole classification, we discuss prototype and candidate materials to exhibit unconventional electronic orderings and physical phenomena in Sect.~\ref{sec: Prototype materials}. 
Finally, Sect.~\ref{sec: Summary and perspectives} is devoted to the summary and perspectives.

\section{Four types of multipoles}
\label{sec: Four types of multipoles}

\begin{figure*}[htb!]
\centering
\includegraphics[width=1.0 \hsize]{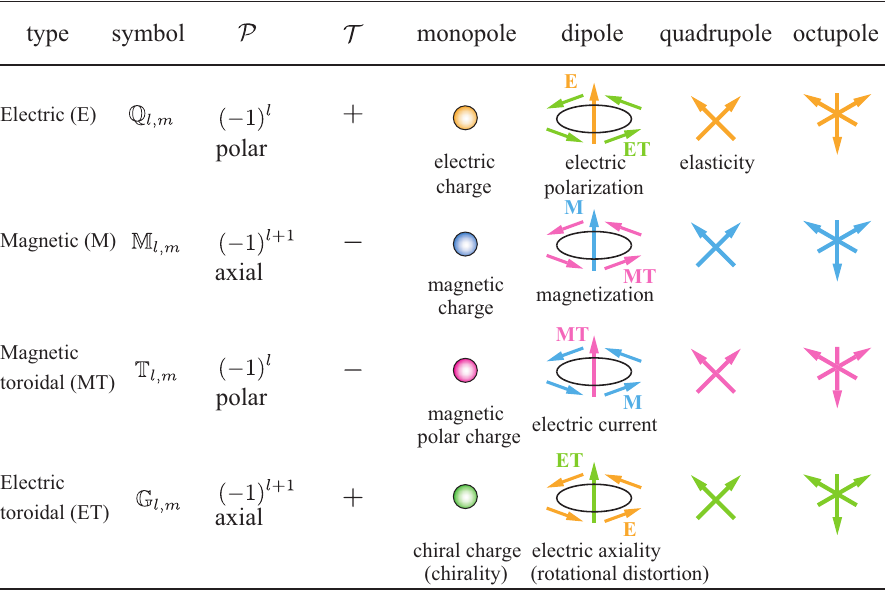} 
\caption{
\label{fig: four_mp}
Four types of multipoles. 
$\mathcal{P}$ and $\mathcal{T}$ represent the spatial inversion and time-reversal parities, respectively.
}
\end{figure*}

The concept of electronic multipoles has been used to describe the spatial anisotropy of physical quantities. 
It was originally introduced in classical electromagnetism in order to characterize the spatial anisotropy of the electric charge and current distributions~\cite{Schwartz_PhysRev.97.380, BlattWeisskopf201111, Jackson3rd1999}; 
E multipoles, which correspond to the polar tensor with time-reversal even, appear in the expansion of the scalar potential, while M multipoles, which correspond to the axial tensor with time-reversal odd, appear in the expansion of the vector potential. 
Accordingly, corresponding electric and magnetic fields are related to E and M multipoles, respectively. 
In other words, arbitrary electric and magnetic fields can be expressed as a linear combination of E and M multipoles, respectively.

In analogy with this argument, electronic multipole \textit{bases} were introduced in condensed matter physics in order to characterize the spatial anisotropy of atomic-scale charge, current, and spin distributions; the quantum-mechanical operator expressions for E and M multipoles as defined in classical electromagnetism were firstly derived~\cite{Santini_RevModPhys.81.807,kuramoto2009multipole, Kusunose_JPSJ.77.064710}. 
On the other hand, there is a crucial difference from the multipole in classical electromagnetism. 
Two additional multipoles distinct from E and M multipoles appear in quantum physics, since the electron systems can naturally have the internal degrees of freedom corresponding to the polar tensor with time-reversal odd and axial tensor with time-reversal even, which are different from E and M multipoles: The former corresponds to MT multipoles and the latter corresponds to ET multipoles~\cite{dubovik1975multipole, dubovik1986axial, dubovik1990toroid, kopaev2009toroidal, Spaldin_0953-8984-20-43-434203, talebi2018theory, hayami2018microscopic}. 
In the end, four types of multipoles are defined in condensed matter physics according to the spatial inversion and time-reversal parities, as shown in Fig.~\ref{fig: four_mp}~\cite{kusunose2022generalization}: E multipole $\mathbb{Q}_{l,m}$, M multipole $\mathbb{M}_{l,m}$, MT multipole $\mathbb{T}_{l,m}$, and ET multipole $\mathbb{G}_{l,m}$, where $l$ and $m$ stand for the rank of multipole and its $z$ component with $-l \leq m \leq l$. 
From the viewpoint of the space-time inversion symmetry, the time-reversal counterpart of the E (M) multipole is the MT (ET) multipole, and the spatial inversion counterpart of the E (M) multipole is the ET (MT) multipole. 
The latter means that the dipole component of MT (ET) multipole is represented by the vortex-like alignment of M (E) multipole~\cite{kopaev2009toroidal, Spaldin_0953-8984-20-43-434203, naumov2004unusual, Prosandeev_PhysRevLett.96.237601, Gao_PhysRevB.97.134423, Gao_PhysRevB.98.060402}. 

As detailed in subsequent sections, four types of multipole bases describe any internal electronic degrees of freedom in solids owing to their completeness in the Hilbert space. 
In other words, any physical quantities and external fields can be mapped into any of four multipoles at the microscopic level.  
The well-known examples are the correspondence between E multipoles and physical quantities; the electric charge characterized by the (polar) scalar with time-reversal even is proportional to the E monopole basis, the electric polarization characterized by the polar vector with time-reversal even is proportional to the E dipole basis, and the elasticity characterized by the 2nd-rank polar tensor with time-reversal even is proportional to the E quadrupole basis, and so on. 
In a similar manner, the correspondence between M multipoles and physical quantities is also well-known; magnetic charge or magnetic flux characterized by the pseudo (axial) scalar with time-reversal odd and spin or orbital angular momentum characterized by the axial vector with time-reversal odd are proportional to the M monopole and M dipole bases, respectively. 
Similar to the E and M multipoles, MT and ET multipoles have related physical quantities. 
In the MT multipoles, the electric current characterized by the polar tensor with time-reversal odd is proportional to the MT dipole basis. 
In the ET multipoles, the chirality characterized by the pseudo (axial) scalar with time-reversal even is proportional to the ET monopole basis~\cite{kishine2022definition}, and electric axiality (rotational distortion) characterized by the axial vector with time-reversal even is proportional to the ET dipole basis~\cite{dubovik1986axial, jin2020observation, Hlinka_PhysRevLett.116.177602, Hlinka_PhysRevLett.113.165502, Hayami_doi:10.7566/JPSJ.91.113702}. 

Furthermore, exotic electronic orderings are systematically classified into any of the multipole orderings based on the above multipole bases. 
For example, the nematic ordering is categorized into the E quadrupole ordering~\cite{chuang2010nematic, Goto_JPSJ.80.073702, Yoshizawa_JPSJ.81.024604} and the anapole ordering is categorized into the MT dipole ordering~\cite{jeong2017time, Matsuda_PhysRevX.11.011021}.  
Since electronic orderings belonging to the same category of multipole basis exhibit the common physical phenomena at the qualitative level as discussed in Sect.~\ref{sec: Cross-correlation phenomena}, it is useful to systematically classify the electronic orderings in terms of the multipole bases. 

In the following subsections, we introduce the operator expressions for four types of multipole bases in Sect.~\ref{sec: Operator definition}. 
Then, we show when and how each multipole is activated in the Hilbert space: the conventional multipole within one type of atomic orbital in a single atom in Sect.~\ref{sec: Conventional multipole}, the hybrid multipoles with multi types of atomic orbital in Sect.~\ref{sec: Hybrid multipole}, and the cluster multipoles with multi sites in Sect.~\ref{sec: Cluster multipole}. 
We also explain the concept of symmetry-adapted multipole basis in Sect.~\ref{sec: Symmetry-adapted multipole basis}.

\subsection{Definition of multipole operators}
\label{sec: Operator definition}

\subsubsection{Spinless space}

\begin{table}[tb!]
\begin{center}
\caption{
Electric (E) multipole operators up to rank 3~\cite{hayami2018microscopic}. 
The superscript ${\rm (orb)}$ is omitted for simplicity. 
}
\label{tab_mp_Q}
\begingroup
\renewcommand{\arraystretch}{1.1}
\scalebox{0.92}{
 \begin{tabular}{ccccc}
 \multicolumn{4}{l}{\fbox{even-parity}} \\
 \hline \hline
rank & type & symbol & definition \\ \hline
$0$ & E  & $\mathbb{Q}_0$ & $1$ \\ 
   \hline
$2$ & E  & $\mathbb{Q}_{u}$, $\mathbb{Q}_{v}$ & $\displaystyle \frac{1}{2}(3z^2-r^2)$, $\displaystyle \frac{\sqrt{3}}{2}(x^2-y^2)$ \\
    &    & $\mathbb{Q}_{yz}$, $\mathbb{Q}_{zx}$, $\mathbb{Q}_{xy}$ & $\sqrt{3}yz$, $\sqrt{3}zx$, $\sqrt{3}xy$ \\
\hline\hline \\
 \multicolumn{4}{l}{\fbox{odd-parity}} \\
  \hline \hline
rank & type & symbol & definition \\ \hline
$1$ & E  &  $\mathbb{Q}_x$, $\mathbb{Q}_y$, $\mathbb{Q}_z$ & $x$, $y$, $z$ \\
\hline
3 & E  & $\mathbb{Q}_{xyz}$ & $\sqrt{15}xyz$ \\
  &    & $\mathbb{Q}_{x}^{\alpha}$, $\mathbb{Q}_{y}^{\alpha}$, $\mathbb{Q}_{z}^{\alpha}$ & $\displaystyle \frac{1}{2}x(5x^{2}-3r^{2})$, (cyclic) \\
  &    & $\mathbb{Q}_{x}^{\beta}$, $\mathbb{Q}_{y}^{\beta}$, $\mathbb{Q}_{z}^{\beta}$ & $\displaystyle \frac{\sqrt{15}}{2}x(y^{2}-z^{2})$, (cyclic) \\
\hline\hline
\end{tabular}
}
\endgroup
\end{center}
\end{table}

The operator expressions of four types of multipoles in spinless Hilbert space are defined by~\cite{hayami2018microscopic, kusunose2020complete}
\begin{align}
\label{eq: Qlm}
\mathbb{Q}_{l,m}^{\rm (orb)}=&O_{l,m},
\\
\label{eq: Mlm}
\mathbb{M}_{l,m}^{\rm (orb)}=&\frac{1}{2}\left[(\bm{\nabla}O_{l,m})\cdot\hat{\bm{m}_l}+\hat{\bm{m}_l}\cdot(\bm{\nabla}O_{l,m})\right],
\\
\label{eq: Tlm}
\mathbb{T}_{l,m}^{\rm (orb)}=&\frac{1}{2}\left[(\bm{\nabla}O_{l,m})\cdot \hat{\bm{t}}-\hat{\bm{t}}\cdot(\bm{\nabla}O_{l,m})\right],
\\
\label{eq: Glm}
\mathbb{G}_{l,m}^{\rm (orb)}=&\frac{1}{2}\left[(\nabla_\alpha \nabla_\beta O_{l,m})\hat{g}_l^{\alpha\beta}-(\hat{g}_l^{\alpha\beta})^{\dagger}(\nabla_\alpha \nabla_\beta O_{l,m})\right], \nonumber \\
=&\frac{1}{2}\frac{4i}{(l+1)^{2}(l+2)}\left[(\bm{\nabla}O_{l,m})\cdot\hat{\bm{l}}\,\hat{\bm{l}}^{2}-\hat{\bm{l}}^{2}\,\hat{\bm{l}}\cdot(\bm{\nabla}O_{l,m})\right],
\end{align}
with 
\begin{align}
\hat{\bm{m}}_l&=\frac{2\hat{\bm{l}}}{l+1},  \\
\hat{\bm{t}}_l &= \frac{2}{(l+1)(l+2)}(\bm{r}\times \hat{\bm{l}}), \\
\hat{g}_l^{\alpha\beta}&=\hat{t}^\alpha_l\hat{m}^\beta_l, 
\end{align}
where the black-board font is used to represent the orthogonal basis. 
$\hat{\bm{l}}=-i(\bm{r}\times\bm{\nabla})$ is the dimensionless orbital angular momentum operator, and the prefactor $1/2$ is due to symmetrization of the operators.
$O_{l,m}$ is proportional to the spherical harmonics of the orbital angular momentum (rank of multipole), $l=0$ (monopole), $1$ (dipole), $2$ (quadrupole), $3$ (octupole), $4$ (hexadecapole), $\cdots$ and its $z$-component, $m=-l,-l+1,\cdots,l$, which is represented by 
\begin{align}
O_{l,m}(\bm{r})=\sqrt{\frac{4\pi}{2l+1}}r^{l}Y_{l,m}(\hat{\bm{r}}),
\quad
\hat{\bm{r}}=\frac{\bm{r}}{r}. 
\end{align}
The explicit expressions of $\mathbb{Q}_{l,m}$ up to rank 3 are shown in Table~\ref{tab_mp_Q}, where the symbol is defiend for the monopole as $\mathbb{Q}_0$, dipole as $\mathbb{Q}_{1,m}=(\mathbb{Q}_x, \mathbb{Q}_y, \mathbb{Q}_z)$, quadrupole as $\mathbb{Q}_{2,m}=(\mathbb{Q}_u, \mathbb{Q}_v, \mathbb{Q}_{yz}, \mathbb{Q}_{zx}, \mathbb{Q}_{xy})$, and octupole as $\mathbb{Q}_{3,m}=(\mathbb{Q}_{xyz}, \mathbb{Q}^{\alpha}_x, \mathbb{Q}^{\alpha}_y, \mathbb{Q}^{\alpha}_z, \mathbb{Q}^{\beta}_x, \mathbb{Q}^{\beta}_y, \mathbb{Q}^{\beta}_z)$. 
Since $\mathbb{Q}_{l,m}$ is a polar tensor, the even-rank E multipole has the spatial parity $+1$, while the odd-rank one has $-1$. 

The expressions for the other three multipoles are easily derived by using Eqs.~(\ref{eq: Mlm})--(\ref{eq: Glm}). 
For example, the expressions of the M quadrupole are given by 
\begin{align}
\mathbb{M}^{\rm (orb)}_{u}&=3zm_{2}^{z}-\bm{r}\cdot \bm{m}_2, \\
\mathbb{M}^{\rm (orb)}_v&=\sqrt{3} (xm_{2}^{x}-ym_{2}^{y}), \\
\mathbb{M}^{\rm (orb)}_{yz}&=\sqrt{3}(ym_{2}^{z}+zm_{2}^{y}), \\
\mathbb{M}^{\rm (orb)}_{zx}&=\sqrt{3}(zm_{2}^{x}+xm_{2}^{z}), \\
\mathbb{M}^{\rm (orb)}_{xy}&=\sqrt{3}(xm_{2}^{y}+ym_{2}^{x}). 
\end{align}
The detailed expressions for other multipoles are shown in Refs.~\citen{hayami2018microscopic, Hayami_PhysRevB.98.165110, kusunose2020complete}. 
Owing to the expressions in Eqs.~(\ref{eq: Mlm})--(\ref{eq: Glm}), $\mathbb{M}^{\rm (orb)}_{l,m}$ and $\mathbb{T}^{\rm (orb)}_{l,m}$ become nonzero for $l \geq 1$, while $\mathbb{G}^{\rm (orb)}_{l,m}$ becomes nonzero for $l \geq 2$; the M monopole $\mathbb{M}^{\rm (orb)}_0$, MT monopole $\mathbb{T}^{\rm (orb)}_0$, ET monopole $\mathbb{G}^{\rm (orb)}_0$, and ET dipole $\mathbb{G}_{1,m}^{\rm (orb)}=(\mathbb{G}^{\rm (orb)}_x, \mathbb{G}^{\rm (orb)}_y, \mathbb{G}^{\rm (orb)}_z)$ are identically zero in the spinless space.  
The spatial parity of the MT multipole is the same as that of the E multipole, while the spatial parity of the M and ET multipoles is opposite to that of the E multipole; the M and ET multipoles have the spatial parity of $(-1)^{l+1}$. 

Since the operator expressions are well-defined, the expectation value of arbitrary multipole bases can be evaluated once the basis atomic wave function is given. 
For example, the matrix elements of the E quadrupole $\mathbb{Q}^{\rm (orb)}_{2,m}=(\mathbb{Q}^{\rm (orb)}_u, \mathbb{Q}^{\rm (orb)}_v, \mathbb{Q}^{\rm (orb)}_{yz}, \mathbb{Q}^{\rm (orb)}_{zx}, \mathbb{Q}^{\rm (orb)}_{xy})$ for the three $p$-wave functions $\phi^p=\{ \phi_x, \phi_y, \phi_z\}=\sqrt{3/4\pi}\{ x/r, y/r, z/r \}$ are given by 
\begin{align}
&
\mathbb{Q}^{\rm (orb)}_{u}=\frac{1}{5}
\left(
\begin{array}{ccc}
 -1 & 0 & 0 \\
 0 & -1 & 0 \\
 0 & 0 & 2 \\
\end{array}
\right), 
\nonumber \\&
\mathbb{Q}^{\rm (orb)}_{v}=\frac{\sqrt{3}}{5}
\left(
\begin{array}{ccc}
 1 & 0 & 0 \\
 0 & -1 & 0 \\
 0 & 0 & 0 \\
\end{array}
\right),
\nonumber \\&
\mathbb{Q}^{\rm (orb)}_{yz}=\frac{\sqrt{3}}{5}
\left(
\begin{array}{ccc}
 0 & 0 & 0 \\
 0 & 0 & 1 \\
 0 & 1 & 0 \\
\end{array}
\right),
\nonumber \\&
\mathbb{Q}^{\rm (orb)}_{zx}=\frac{\sqrt{3}}{5}
\left(
\begin{array}{ccc}
 0 & 0 & 1 \\
 0 & 0 & 0 \\
 1 & 0 & 0 \\
\end{array}
\right),
\nonumber \\&
\mathbb{Q}^{\rm (orb)}_{xy}=\frac{\sqrt{3}}{5}
\left(
\begin{array}{ccc}
 0 & 1 & 0 \\
 1 & 0 & 0 \\
 0 & 0 & 0 \\
\end{array}
\right).
\end{align}
All the E quadrupoles are activated in this basis wave function, and their expectation values with respect to a linear combination of $\phi^{p}$ (denoted as $\phi$) are evaluated by $\braket{\phi|\mathbb{Q}^{\rm (orb)}_{2,m}|\phi}$. 
The systematic derivation of the multipole matrices based on the Wigner-Eckart theorem is shown in Ref.~\citen{kusunose2020complete}. 
It is noted that the multipole matrices are orthonormal with each other by multiplying an appropriate normalization constant; ${\rm Tr}[\mathbb{X}^{\rm (orb)}_{l,m} \mathbb{X'}^{\rm (orb)}_{l',m'}]=\delta_{XX'}\delta_{l l'}\delta_{mm'}$ for $\mathbb{X}^{\rm (orb)}, \mathbb{X'}^{\rm (orb)}=\mathbb{Q}^{\rm (orb)}, \mathbb{M}^{\rm (orb)}, \mathbb{T}^{\rm (orb)}, \mathbb{G}^{\rm (orb)}$.

\subsubsection{Spinful space}

The multipole operator in spinful space is obtained by using the addition rule of the orbital angular momenta for the spinless multipole operator $\mathbb{X}^{\rm (orb)}_{l,m}$ and identity and Pauli matrices $\sigma_{sn}$ in spin space~\cite{kusunose2020complete}, which is given by 
\begin{align}
\label{eq: mp_spinful}
\mathbb{X}_{l,m}(s, k)&\equiv
i^{s+k}
\sum_n \braket{l+k,m-n;sn |lm}
\mathbb{X}_{l+k,m-n}^{\rm (orb)}\sigma_{sn}, 
\end{align}
where $\braket{l_1, m_1; l_2, m_2 |lm}$ is the Clebsch-Gordan (CG) coefficient, $\sigma_{00}=\sigma_0$, $\sigma_{10}=\sigma_z$, and $\sigma_{1\pm1}= \mp (\sigma_x \pm i \sigma_y)/\sqrt{2}$. 
The spinful multipole basis describes any physical quantities in spinful space; it also satisfies the orthonormal relation as ${\rm Tr}[\mathbb{X}_{l,m} (s,k) \mathbb{X'}_{l',m'}(s', k')] =\delta_{XX'}\delta_{l l'}\delta_{mm'}\delta_{s, s'} \delta_{k,k'}$. 

In the spinful space, the operator expressions for the M monopole $\mathbb{M}_0$, ET monopole $\mathbb{G}_0$, and ET dipole $\mathbb{G}_{1,m}=(\mathbb{G}_x, \mathbb{G}_y, \mathbb{G}_z)$, that identically vanish in the \textit{spinless} space, are defined as follows: 
\begin{align}
\label{eq: M0}
\mathbb{M}_0 (1,1)&= \frac{1}{\sqrt{3}}(\bm{r}\cdot \bm{\sigma}), \\
\label{eq: G0}
\mathbb{G}_0 (1,1)&= \frac{1}{\sqrt{3}}(\bm{t}_{1}\cdot \bm{\sigma}),  \\
\label{eq: G1}
\mathbb{G}_\mu (1,0)&= \frac{1}{\sqrt{2}}(\bm{\sigma}\times \bm{l})_\mu, 
\end{align} 
for $\mu=x,y,z$. 
Thus, the spin degree of freedom is essential to activate these multipoles within the atomic wave function. 
Moreover, the anisotropic M dipole is also defined in the spinful space as
\begin{align}
\label{eq: Ma}
\mathbb{M}_{\mu}(1,1)=\frac{1}{\sqrt{10}}\sum_{\nu}(3r_{\mu}r_{\nu}-\bm{r}^{2}\delta_{\mu\nu})\sigma_{\nu},
\end{align}
which can be observed in x-ray magnetic circular dichroism (XMCD) measurements~\cite{stohr1995x, stohr1999exploring, yamasaki2020augmented, Sasabe_PhysRevLett.131.216501}, and plays an important role in the anomalous Hall effect as discussed later.
On the other hand, the MT monopole $\mathbb{T}_0$ is not defined even in the spinful space at least within the atomic wave function. 
The systematic derivation of the multipole matrices in spinful space is also found in Ref.~\citen{kusunose2020complete}.

\subsection{Conventional multipole}
\label{sec: Conventional multipole}

\begin{figure*}[htb!]
\centering
\includegraphics[width=1.0 \hsize]{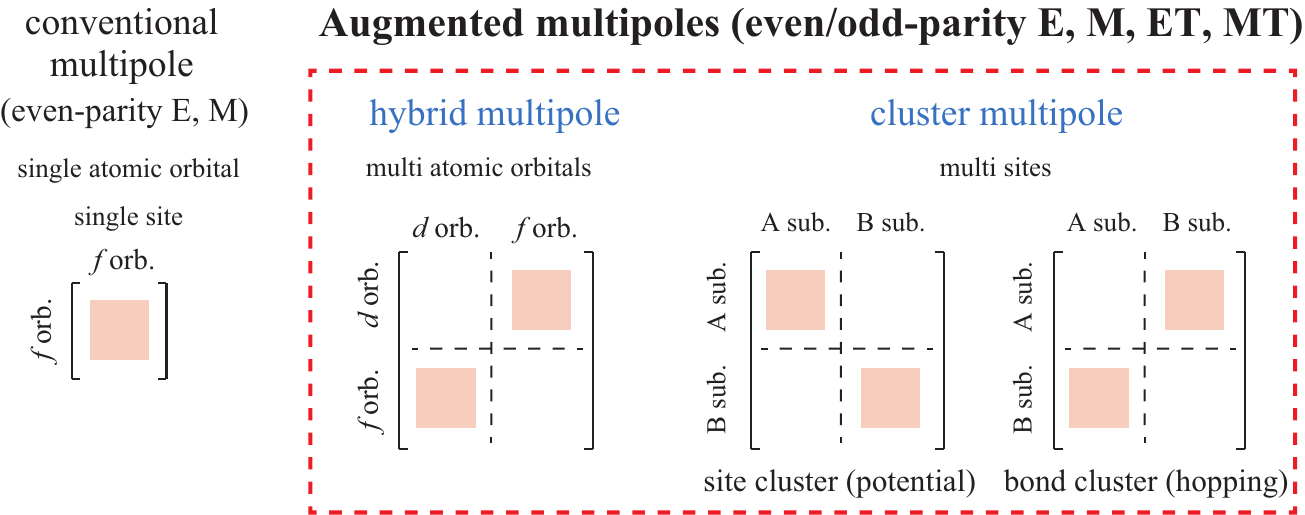} 
\caption{
\label{fig: augmented_mp}
Relevant Hilbert spaces of conventional, hybrid, and cluster multipoles. 
The conventional multipoles are activated within one type of atomic orbital in a single atom, while the hybrid and cluster multipoles are activated in multi atomic orbitals and multi sites, respectively. 
All of the four types of multipoles can be active in the latter space. 
The hybrid and cluster multipoles together are referred to as augmented multipoles.
}
\end{figure*}

\begin{table*}[t!]
\begin{center}
\caption{
Active multipoles in the atomic $s$, $p$, $d$, and $f$ wave functions~\cite{hayami2018microscopic, Hayami_PhysRevB.98.165110}. 
$\mathcal{P}$ represents the spatial inversion operation. 
E, M, MT, and ET represent electric, magnetic, magnetic toroidal, and electric toroidal multipoles, respectively. 
\# represents the number of independent multipoles. 
}
\label{tab:active_mp}
\begingroup
\renewcommand{\arraystretch}{1.2}
\scalebox{0.9}{
 \begin{tabular}{ccccccccccccc}
 \multicolumn{6}{l}{\fbox{Conventional multipole}} \\
 \multicolumn{6}{l}{\underline{$\cdot$ spinless space}} \\
$L$ & orbital & \# & $\mathcal{P}$ & $l=0$ & 1 & 2 & 3 & 4 & 5 & 6     \\
\hline\hline
$0$& $s$-$s$  &1 & $+$   &E& --& --&--&-- &--&--\\ 
$1$& $p$-$p$ &9  &        &E&M&E&--&--&--&--\\ 
$2$& $d$-$d$  &25  &     &E&M&E&M&E&--&--      \\ 
$3$&  $f$-$f$ &49   &      &E&M&E&M&E&M&E \\ 
\hline\hline
 \multicolumn{6}{l}{\underline{$\cdot$ spinful space}} \\
$J$ & orbital & \# & $\mathcal{P}$ &$l=0$ & 1 & 2 & 3 & 4 & 5 & 6 & 7     \\
\hline\hline
$\frac{1}{2}$ & $s$-$s$, $p$-$p$& 4  & $+$  &E& M& --&--&-- &--&--&--\\ 
$\frac{3}{2}$ &  $p$-$p$, $d$-$d$&  16 &       &E&M&E&M&--&--&--&--\\ 
$\frac{5}{2}$ & $d$-$d$, $f$-$f$&   36  &    &E&M&E&M&E&M&-- &--     \\ 
$\frac{7}{2}$ & $f$-$f$&   64  &    &E&M&E&M&E&M&E &M     \\  
\hline\hline
\\
\multicolumn{6}{l}{\fbox{Hybrid multipole}} \\
  \multicolumn{6}{l}{\underline{$\cdot$ spinless space}} \\
 $L_{1}$-$L_{2}$ & orbital & \# & $\mathcal{P}$ & $l=0$ & 1 & 2 & 3 & 4 & 5 & 6     \\
\hline\hline
0-2 & $s$-$d$ &10 & $+$  &--&--&E/MT&--&--&--&-- \\ 
1-3 & $p$-$f$ &   42  &         &--&--&E/MT&M/ET&E/MT&--&--\\ 
\hline
0-1 & $s$-$p$ &6  & -   &--&E/MT&--&--&--&--&-- \\ 
0-3 & $s$-$f$ &14  &         & --&--&--&E/MT&--&--&-- \\ 
1-2 & $p$-$d$ &30  &       &--&E/MT&M/ET&E/MT &--&--&--\\ 
2-3 & $d$-$f$ &70   &        & --&E/MT&M/ET&E/MT&M/ET&E/MT&--      \\ \hline\hline
 \multicolumn{6}{l}{\underline{$\cdot$ spinful space}} \\
 $J_{1}$-$J_{2}$ & orbital & \# & $\mathcal{P}$ &$l=0$ & 1 & 2 & 3 & 4 & 5 & 6 & 7     \\
 \hline\hline
$\frac{1}{2}$-$\frac{3}{2}$& $s$-$d$, $p$-$p$& 16 & $+$  &--&M/ET&E/MT&--&--&--&--&--\\ 
$\frac{1}{2}$-$\frac{5}{2}$ & $s$-$d$, $p$-$f$ & 24  &      &--&--&E/MT&M/ET &--&--&--&--\\ 
$\frac{1}{2}$-$\frac{7}{2}$ & $p$-$f$ & 32  &      &--&--&--&M/ET &E/MT&--&--&--\\ 
$\frac{3}{2}$-$\frac{5}{2}$ &$p$-$f$, $d$-$d$& 48   &        &--&M/ET&E/MT&M/ET&E/MT&--&--&--\\ 
$\frac{3}{2}$-$\frac{7}{2}$ & $p$-$f$& 64   &       & --&--&E/MT&M/ET&E/MT&M/ET&--&--   \\
$\frac{5}{2}$-$\frac{7}{2}$ & $f$-$f$&  96    &     & --&M/ET&E/MT&M/ET&E/MT&M/ET&E/MT&--   \\ 
\hline
$\frac{1}{2}$-$\frac{1}{2}$& $s$-$p$ & 8 & -   &M/ET& E/MT& --&--&-- &--&--&--\\ 
$\frac{3}{2}$-$\frac{3}{2}$& $p$-$d$&  32   &      &M/ET&E/MT&M/ET&E/MT&--&--&--&--\\ 
$\frac{5}{2}$-$\frac{5}{2}$& $d$-$f$&  72   &     &M/ET&E/MT&M/ET&E/MT&M/ET&E/MT&--&-- \\ 
$\frac{1}{2}$-$\frac{3}{2}$& $s$-$p$, $p$-$d$& 16 &  &--&E/MT&M/ET&--&--&--&--&-- \\ 
$\frac{1}{2}$-$\frac{5}{2}$&  $s$-$f$, $p$-$d$&  24 &  &--&--&M/ET&E/MT &--&--&--&--\\ 
$\frac{1}{2}$-$\frac{7}{2}$&  $s$-$f$& 32  &    &--&--&--&E/MT &M/ET&--&--&--\\ 
$\frac{3}{2}$-$\frac{5}{2}$& $p$-$d$, $d$-$f$& 48  &   &--&E/MT&M/ET&E/MT&M/ET&--&--&--\\ 
$\frac{3}{2}$-$\frac{7}{2}$&  $d$-$f$& 64  &   & --&--&M/ET&E/MT&M/ET&E/MT&--&--  \\
$\frac{5}{2}$-$\frac{7}{2}$&  $d$-$f$&  96 &  &--&E/MT&M/ET&E/MT&M/ET&E/MT&M/ET&--\\  
\hline\hline
\end{tabular}
}
\endgroup
\end{center}
\end{table*}

By using the expressions in Eqs.~(\ref{eq: Qlm})--(\ref{eq: Glm}) in spinless space and Eq.~(\ref{eq: mp_spinful}) in spinful space, one can identify active multipoles in a given Hilbert space. 
The necessary condition to activate the multipoles is understood from the addition rule of the angular momentum, since the matrix element of the multipole operator is proportional to the overlap integral of the spherical harmonics. 
For example, the matrix element $\bra{L_1, M_1}\mathbb{X}^{\rm (orb)}_{l,m} \ket{L_2, M_2} \propto \int d\hat{\bm{r}} Y^*_{L_1, M_1}(\hat{\bm{r}})Y_{l,m} (\hat{\bm{r}}) Y_{L_2, M_2}(\hat{\bm{r}})$ can remain when the rank-$l$ multipoles satisfy $|L_1-L_2| \leq l \leq L_1 + L_2$, where $\ket{L, M}$ is a state with the orbital angular momentum $L$ and its component $M$. 

First, let us consider the active multipoles in the conventional situation, where the system consists of one type of atomic orbital in a single atom, i.e., $L_1=L_2$. 
In this case, only the even-parity multipoles, $\mathcal{P}\mathbb{X}_{l,m}=\mathbb{X}_{l,m}$, become active, where $\mathcal{P}$ is the spatial inversion operation. 
Besides, the rank-$2l$ (rank-$2l+1$) multipole should be a time-reversal-even polar (time-reversal-odd axial) one owing to the addition rule regarding the wave function with the polar property. 
Thus, the active multipoles within one type of atomic orbital in a single atom are even-rank E multipoles and odd-rank M multipoles; the odd-parity multipoles and toroidal-type multipoles are not activated in this case even when a system is noncentrosymmetric, as schematically shown in the leftmost panel of Fig.~\ref{fig: augmented_mp}. 
We show the active multipoles in spinless and spinful spaces in the upper panel of Table~\ref{tab:active_mp}. 
The $N$-dimensional matrix in the spinless space is spanned by the $N(N+1)/2$ symmetric real matrices corresponding to the E multipole, and the $N(N-1)/2$ antisymmetric pure imaginary matrices corresponding to the M multipole when the basis function is real. 
Similarly, the 2$N$-dimensional matrix in spinful space is spanned by the $N(2N-1)$ E multipoles and $N(2N+1)$ M multipoles.

\subsection{Hybrid multipole}
\label{sec: Hybrid multipole}

\begin{figure*}[htb!]
\centering
\includegraphics[width=0.95 \hsize]{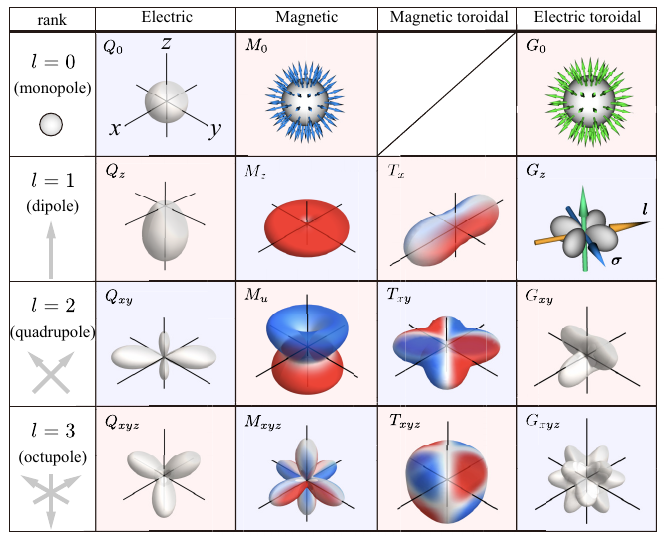} 
\caption{
\label{fig: table_mp_hybrid}
Examples of hybrid multipoles up to rank 3. 
The shape and color map in each cell stand for the electric charge density and the $z$ component of the orbital angular momentum density, respectively. 
The blue, green, and orange arrows represent the spin, ET dipole, and orbital magnetic moments, respectively. 
}
\end{figure*}

The hybrid multipole describes the electronic degrees of freedom in hybrid atomic orbital systems. 
In contrast to the conventional multipole in the previous section, it can describe two types of toroidal multipole degrees of freedom. 
The concept of the hybrid multipole is useful for systematic discussion of electronic degrees of freedom and order parameters that arise from atomic-scale hybridization among atomic orbitals, as schematically shown in Fig.~\ref{fig: augmented_mp}, for instance, the $f$ electron systems hybridized with conduction electrons~\cite{watanabe2019charge, iwasa2019neutron}, excitonic insulators with different orbital characters of valence/conduction bands~\cite{Tsubouchi_PhysRevB.66.052418, Hasan_RevModPhys.82.3045, Kunes_PhysRevB.90.235112, Kaneko_PhysRevB.94.125127, yamaguchi2017multipole}, and cluster systems including quantum dots and organic molecules~\cite{Hanson_RevModPhys.79.1217, Greene_RevModPhys.89.035006}. 
More recently, it has been argued as an order parameter for the hidden order in URu$_2$Si$_2$, where the staggered alignment of the odd-parity electric dotriacontapole or electric toroidal monopole can be an order-parameter candidate~\cite{Kambe_PhysRevB.97.235142, kambe2020symmetry, hayami2023chiral}. 
Although the staggered alignment of these odd-parity multipoles in the case of URu$_2$Si$_2$ does not lead to the apparent symmetry lowering within the x-ray diffraction measurement, it can provide an alternation experimental direction of detecting the hidden order parameter that has never been performed~\cite{hayami2023chiral}.

The hybrid multipoles are activated between different orbital angular momenta for integer angular momentum in the spineless space, while those are activated between different orbital angular momenta or between different $J$ multiplets with the same orbital angular momentum for half-integer angular momentum in the spinful space~\cite{hayami2018microscopic}. 
As shown in the lower panel of Table~\ref{tab:active_mp}, a pair of the odd-rank E and MT multipoles and a pair of even-rank M and ET multipoles appear under the odd-parity hybridization ($\mathcal{P}=-1$), such as $s$-$p$, $p$-$d$, and $d$-$f$ hybridizations in the spinless space. 
For example, the matrix elements of E and MT dipoles in the $s$-$p$ hybridized system with the atomic basis function $(\phi^s, \phi^p)$ are given by 
\begin{align}
&
\label{eq:spQ}
\mathbb{Q}_{x}=\frac{1}{\sqrt{3}}
\left(
\begin{array}{ccc}
1 & 0 & 0  \\
\end{array}
\right),
\ \
\mathbb{Q}_{y}=\frac{1}{\sqrt{3}}
\left(
\begin{array}{ccc}
 0 & 1 & 0 \\ 
 \end{array}
\right),
\nonumber \\ &
\mathbb{Q}_{z}=\frac{1}{\sqrt{3}}
\left(
\begin{array}{ccc}
 0 & 0 & 1 \\ 
\end{array}
\right),
\\&
\label{eq:spMT}
\mathbb{T}_{x}=\frac{1}{3\sqrt{3}}
\left(
\begin{array}{ccc}
 i & 0 & 0 \\
\end{array}
\right),
\ \ 
\mathbb{T}_{y}=\frac{1}{3\sqrt{3}}
\left(
\begin{array}{ccc}
 0 & i & 0 \\ 
\end{array}
\right),
\nonumber \\ &
\mathbb{T}_{z}=\frac{1}{3\sqrt{3}}
\left(
\begin{array}{ccc}
 0 & 0 & i \\ 
\end{array}
\right),
\end{align}
where only the off-diagonal elements $\braket{\phi^{s}|X_{l,m}|\phi^{p}}$ are shown.
The real $s$-$p$ hybridization is proportional to the E dipole basis, which is related to the $sp$-hybrid orbital as discussed in the context of molecules. 
On the other hand, its magnetic counterpart characterized by the pure imaginary $s$-$p$ hybridization is proportional to the MT dipole basis. 
It can be the origin of atomic-scale MT dipole ordering and linear magnetoelectric effect, which are usually discussed in the presence of the vortex spin cluster~\cite{yatsushiro2019atomic}. 
The M monopole in Eq.~(\ref{eq: M0}) and ET monopole in Eq.~(\ref{eq: G0}) are also activated by taking into account the spin-dependent hybridization. 

Similar to the odd-parity hybridization, a pair of the even-rank E and MT multipoles and a pair of odd-rank M and ET multipoles appear under the even-parity hybridization ($\mathcal{P}=+1$), such as $s$-$d$ and $p$-$f$ hybridizations in the spinless space. 
The ET dipole in Eq.~(\ref{eq: G1}) is classified in this category. 
We show the schematic pictures of the hybrid multipoles up to rank 3 in Fig.~\ref{fig: table_mp_hybrid}.

\subsection{Cluster multipole}
\label{sec: Cluster multipole}

The cluster multipole describes the electronic degrees of freedom over multi sites, which is divided into the site-cluster multipole describing the onsite electronic degrees of freedom in Sect.~\ref{sec: Site-cluster multipole} and the bond-cluster multipole describing the offsite ones in Sect.~\ref{sec: Bond-cluster multipole}, as schematically shown in Fig.~\ref{fig: augmented_mp}. 

\subsubsection{Site-cluster multipole}
\label{sec: Site-cluster multipole}
The site-cluster multipole was introduced to represent the spatial distribution of electronic degrees of freedom such as charge, spin, and orbital, over multi sites in a cluster as a ferroic multipole~\cite{Arima_doi:10.7566/JPSJ.82.013705, Yanase_JPSJ.83.014703,Hayami_PhysRevB.90.024432, Hayami_PhysRevB.90.081115, hayami2016emergent, Suzuki_PhysRevB.95.094406, yatsushiro2020odd, Yatsushiro_doi:10.7566/JPSJ.91.104701}. 
Thanks to Neumann's principle ``the symmetry of macroscopic properties exhibited by crystals cannot be lower than the symmetry of crystal point groups'', the introduction of the symmetry-adapted cluster multipoles makes the connection between complicated charge/spin/orbital orderings and physical phenomena transparent. 
Based on this concept, various fascinating phenomena have been uncovered, such as the linear magnetoelectric effect under the cluster M quadrupole ordering in Cr$_2$O$_3$~\cite{dzyaloshinskii1960magneto, astrov1960magnetoelectric, astrov1961magnetoelectric, FolenPhysRevLett.6.607, Shitade_PhysRevB.98.020407}, Co$_4$Nb$_2$O$_9$~\cite{fischer1972new,Khanh_PhysRevB.93.075117,Khanh_PhysRevB.96.094434,Yanagi2017,Yanagi_PhysRevB.97.020404}, Ba(TiO)Cu$_4$(PO$_4$)$_4$~\cite{kimura2016magnetodielectric, Kato_PhysRevLett.118.107601, Kimura_PhysRevB.97.134418}, and KOsO$_4$~\cite{Hayami_PhysRevB.97.024414,Yamaura_PhysRevB.99.155113}, and the linear magnetoelectric effect under the cluster MT dipole ordering in UNi$_4$B~\cite{saito2018evidence,Hayami_PhysRevB.90.024432,Hayami_1742-6596-592-1-012101}, nonreciprocal magnon excitations in $\alpha$-Cu$_2$V$_2$O$_7$~\cite{Gitgeatpong_PhysRevB.92.024423,Gitgeatpong_PhysRevB.95.245119,Gitgeatpong_PhysRevLett.119.047201,Hayami_doi:10.7566/JPSJ.85.053705}, the magnetopiezo effect under the cluster M quadrupole (hexadecapole) ordering in Ba$T_2$As$_2$($T=$Mn, Fe), Eu$T$Bi$_2$($T={\rm Mn}, {\rm Zn}$)~\cite{Watanabe_PhysRevB.96.064432,shiomi2019observation}, anomalous Hall/Nernst effect and magneto-optical Kerr effect under the cluster M octupole ordering in Mn$_3$Sn~\cite{nakatsuji2015large,Suzuki_PhysRevB.95.094406,ikhlas2017large,kuroda2017evidence,higo2018large}, and the antisymmetric spin splitting under the cluster E octupole ordering in Sr$_3$Ru$_2$O$_7$~\cite{hitomi2014electric,hitomi2016electric}. 
Moreover, such a concept can be applied to quasicrystal Au$_{72}$Al$_{14}$Tb$_{14}$~\cite{Sato_PhysRevB.100.054417}, superconductivity~\cite{Sumita_PhysRevB.93.224507, Ishizuka_PhysRevB.103.094504}, and 
the coexistence of spin (or cluster M octupole) and cluster E quadrupole orderings~\cite{Ishitobi_PhysRevB.104.L241110, Ishitobi_PhysRevB.107.104413, Hattori_PhysRevB.107.205126, hayami2023multiple}

\begin{figure*}[htb!]
\centering
\includegraphics[width=1.0 \hsize]{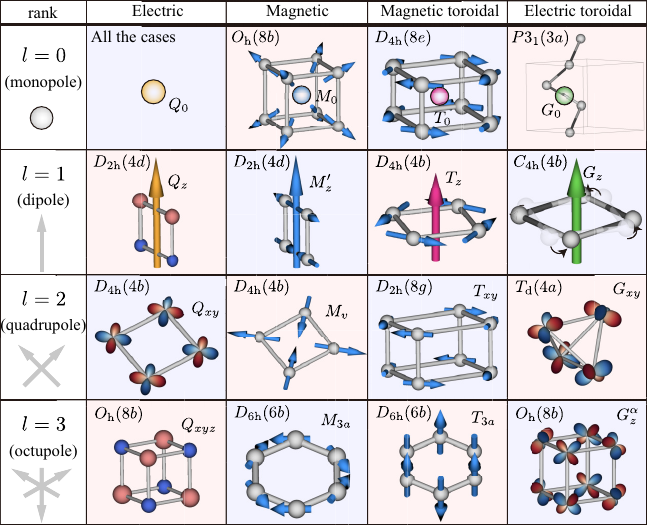} 
\caption{
\label{fig: table_mp}
Examples of the site-cluster electric, magnetic, magnetic toroidal, and electric toroidal multipoles up to the rank $l=3$. 
The red (blue) cells stand for odd-parity (even-parity) multipoles. 
We also show the corresponding space/point group and site symmetry (Wyckoff position) in each cell. 
The spheres and arrows represent the monopole and dipole, respectively.
}
\end{figure*}

Let us briefly explain the method to generate the symmetry-adapted site-cluster multipole basis set based on the virtual-cluster method, which was introduced in Refs.~\citen{Suzuki_PhysRevB.99.174407, Kusunose_PhysRevB.107.195118}. 
In this method, the general point of the crystallographic point group excluding the translational symmetry of the system is considered. 
Since the number of general points is equal to the number of symmetry operations under the point group, $N_g$, symmetry operations of a point group can be associated with $N_g$ vectors, which constitute a virtual cluster. 
The advantages of using this method are that (1) the choice of origin is no longer arbitrary and (2) this method can be applied to nonsymmorphic space groups that include screw and glide operations, since the general point is equidistant from the origin. 
Then, a symmetry-adapted multipole basis is generated as similar to that of the atomic multipoles described in Sect.~\ref{sec: Operator definition} using the general points. 
After generating the multipole basis in the virtual cluster, the components of the basis are mapped onto the site positions of the target system, $\bm{R}_1, \bm{R}_2, \cdots $. 

As an example, let us consider the symmetry-adapted multipole basis set for a four-site cluster with  $\bm{R}_1=(1,1,0)$, $\bm{R}_2=(-1,-1,0)$, $\bm{R}_3=(1,-1,0)$, and $\bm{R}_4=(-1,1,0)$ under the space group $P4/mmm$ (point group $D_{\rm 4h}$). 
Since the general point is given by 
\begin{align}
&\bm{r}_1=(x,y,z), \bm{r}_2=(-x,-y,z), \bm{r}_3=(-y,x,z),  \cr
&\bm{r}_4=(y,-x,z), \bm{r}_5=(-x,y,-z), \bm{r}_6=(x,-y,-z), \cr 
&\bm{r}_7=(y,x,-z), \bm{r}_8=(-y,-x,-z), \bm{r}_9=(-x,-y,-z), \cr 
&\bm{r}_{10}=(x,y,-z), \bm{r}_{11}=(y,-x,-z), \bm{r}_{12}=(-y,x,-z), \cr
&\bm{r}_{13}=(x,-y,z), \bm{r}_{14}=(-x,y,z), \bm{r}_{15}=(-y,-x,z), \cr 
&\bm{r}_{16}=(y,x,z), 
\end{align}
one can generate the symmetry-adapted multipole basis set for this cluster. 
Then, a sixteen-dimensional orthonormal basis is generated as follows: 
\begin{align}
\label{eq: QQ}
\overline{\mathbb{Q}}_{\alpha} \equiv [
\bar{q}_\alpha(\bm{r}_1), \bar{q}_\alpha(\bm{r}_2), \cdots, \bar{q}_\alpha(\bm{r}_{16})],  \quad \bar{q}_\alpha(\bm{r}_m) = O_{\alpha}(\bm{r}_m), 
\end{align}
where $\bar{q}_\alpha(\bm{r}_m)$ in the right side in Eq.~(\ref{eq: QQ}) represents the weight of ``charge" at site $\bm{r}_m$, which is derived by evaluating the spherical harmonics $O_{\alpha}$ with $\bm{r}_m$ under $D_{\rm 4h}$.

Next, mapping onto the four-site cluster should be done. 
Compared to $\bm{R}_1$--$\bm{R}_4$ with $\bm{r}_1$--$\bm{r}_{16}$, one finds the following relation as 
\begin{align}
\bm{R}_1=(1,1,0) &\leftrightarrow \bm{r}_1, \bm{r}_7, \bm{r}_{10}, \bm{r}_{16} \equiv \bm{r}^{(1)}_1, \bm{r}^{(1)}_2, \bm{r}^{(1)}_3, \bm{r}^{(1)}_4, \notag\\
\bm{R}_2=(-1, -1,0) &\leftrightarrow \bm{r}_2, \bm{r}_8, \bm{r}_9,  \bm{r}_{15} \equiv \bm{r}^{(2)}_1, \bm{r}^{(2)}_2, \bm{r}^{(2)}_3, \bm{r}^{(2)}_4, \notag\\
\bm{R}_3=(1,-1,0) &\leftrightarrow  \bm{r}_4, \bm{r}_6, \bm{r}_{11}, \bm{r}_{13} \equiv  \bm{r}^{(3)}_1, \bm{r}^{(3)}_2, \bm{r}^{(3)}_3, \bm{r}^{(3)}_4,\notag\\
\bm{R}_4=(-1,1,0) &\leftrightarrow \bm{r}_3, \bm{r}_5, \bm{r}_{12}, \bm{r}_{14} \equiv \bm{r}^{(4)}_1, \bm{r}^{(4)}_2, \bm{r}^{(4)}_3, \bm{r}^{(4)}_4. 
\end{align}
Then, four independent site-cluster multipoles are obtained by using a four-dimensional orthonormal basis as 
\begin{align}
\label{eq: sitemp_ex}
\mathbb{Q}^{\rm (s)}_{\alpha}=\left[
q_{\alpha}(1),
q_{\alpha}(2),
q_{\alpha}(3),
q_{\alpha}(4)
\right],
\quad
q_{\alpha}(i)=\sum_{k=1}^{4} \bar{q}_{\alpha}(\bm{r}_{k}^{(i)})
\end{align}
where the superscript $(\rm s)$ stands for the site-cluster multipole. 
By taking the direct product with the atomic multipole basis set in Sect.~\ref{sec: Conventional multipole} and \ref{sec: Hybrid multipole}, one can express any spatial distributions of charge, spin, and orbital in electrons as a site-cluster multipole. 
We show cluster E quadrupole and M quadrupole that are constructed from four-site orbital and spin configurations, respectively, in Fig.~\ref{fig: table_mp}, where other examples inducing a variety of site-cluster multipoles in different clusters are also shown. 

The above systematic generation of symmetry-adapted site-cluster multipole is useful to predict candidate magnetic structures by performing a high-throughput calculation for materials~\cite{Huebsch_PhysRevX.11.011031, huebsch2022magnetic}.  
In addition, by using the cluster degree of freedom, one can easily recognize the occurrence of unconventional multipole orderings in collinear and/or staggered alignments, such as the MT dipole~\cite{Hayami_doi:10.7566/JPSJ.84.064717, hayami2016emergent}, the MT quadrupole~\cite{hayami2022spinconductivity}, MT octupole~\cite{Hayami_PhysRevB.90.081115}, ET dipole~\cite{hayami2023cluster}, and ET quadrupole~\cite{yatsushiro2020odd, Yatsushiro_doi:10.7566/JPSJ.91.104701}. 
Especially, it is noteworthy that the MT monopole, which has never appeared in the atomic degree of freedom, also arises in the AFM ordering~\cite{Hayami_PhysRevB.108.L140409}.  
The concept was extended to a finite-$q$ magnetic ordering~\cite{Yanagi_PhysRevB.107.014407}. 

Furthermore, one can engineer various functionalities via the antiferroic alignment of multipoles. 
One of the typical examples is the anomalous Hall effect, which is usually expected to occur in the ferromagnetic ordering, i.e., in the presence of the uniform M dipole~\cite{Karplus_PhysRev.95.1154,smit1958spontaneous, Maranzana_PhysRev.160.421, Berger_PhysRevB.2.4559,nozieres1973simple, Nagaosa_RevModPhys.82.1539}. 
Meanwhile, recent studies have clarified that the anomalous Hall effect can be expected even in AFMs without a net magnetization once the symmetry under the AFM ordering is the same as that under the ferromagnetic one. 
It should be noted that the spin-orbit coupling is necessary for the anomalous Hall effect.
This aspect has a great advantage to efficient spintronics devices without the leakage of the magnetic field as ordinary ferromagnetism does~\cite{Baltz_RevModPhys.90.015005}.
Such situations have been found in various materials, such as La$M$O$_3$ ($M=$ Cr, Mn, and Fe)~\cite{Solovyev_PhysRevB.55.8060}, Mn$_3$Ir~\cite{Chen_PhysRevLett.112.017205, Chen_PhysRevB.101.104418}, Mn$_3$Sn~\cite{nakatsuji2015large,Suzuki_PhysRevB.95.094406,ikhlas2017large,kuroda2017evidence,higo2018large}, antiperovskite AFM Mn$_3$$A$N ($A=$ Ga, Sn, and Ni)~\cite{Gurung_PhysRevMaterials.3.044409, Zhou_PhysRevB.99.104428, Boldrin_PhysRevMaterials.3.094409, Huyen_PhysRevB.100.094426, you2021cluster}, NdMnP~\cite{kotegawa2023large}, the pyrochlore oxides~\cite{Tomizawa_PhysRevB.80.100401, kim2020strain}, the bilayer MnPSe$_3$~\cite{Sivadas_PhysRevLett.117.267203}, $\kappa$-type organic conductors~\cite{Naka_PhysRevB.102.075112}, and other materials/situations~\cite{vsmejkal2020crystal,yamasaki2020augmented, Chen_PhysRevB.106.024421}. 

Based on the symmetry-adapted multipole basis, the emergence of the anomalous Hall effect is understood from the appearance of the anisotropic M dipole in a unified way~\cite{Hayami_PhysRevB.103.L180407}. 
As similar to the atomic multipole $\mathbb{M}_{\mu}(1,1)$ in Eq.~(\ref{eq: Ma}), this cluster anisotropic M dipole is derived from the contraction of the rank-2 E quadrupole $(\mathbb{Q}^{\rm (s)}_u, \mathbb{Q}^{\rm (s)}_v, \mathbb{Q}^{\rm (s)}_{yz}, \mathbb{Q}^{\rm (s)}_{zx}, \mathbb{Q}^{\rm (s)}_{xy})$ and spin in Eq.~(\ref{eq: mp_spinful}), whose expression is given by 
\begin{align}
\label{eq:AMDx}
\mathbb{M}'_x&=\left[ \left(-\frac{1}{\sqrt{3}}\mathbb{Q}_u^{\rm (s)} + \mathbb{Q}_v^{\rm (s)}\right)\sigma_x+\mathbb{Q}_{xy}^{\rm (s)} \sigma_y + \mathbb{Q}_{zx}^{\rm (s)} \sigma_z\right], \\
\label{eq:AMDy}
\mathbb{M}'_y &=\left[\mathbb{Q}_{xy}^{\rm (s)}\sigma_x -\left(\frac{1}{\sqrt{3}}\mathbb{Q}_u^{\rm (s)} + \mathbb{Q}_v^{\rm (s)} \right)\sigma_y+\mathbb{Q}_{yz}^{\rm (s)} \sigma_z \right], \\
\label{eq:AMDz}
\mathbb{M}'_z &=\left[\mathbb{Q}_{zx}^{\rm (s)} \sigma_x + \mathbb{Q}_{yz}^{\rm (s)}\sigma_y+\frac{2}{\sqrt{3}}\mathbb{Q}_u^{\rm (s)} \sigma_z\right],
\end{align}
where the numerical coefficient and the superscript of multipoles are omitted for simplicity. 
This anisotropic M dipole is independent of the higher-rank M multipoles and MT multipoles as well as the M dipoles in the system as discussed in Sect.~\ref{sec: Operator definition} and is activated by the AFM structure, as shown in Fig.~\ref{fig: table_mp}. 
Indeed, the AFM structure observed in Mn$_{3}$Sn~\cite{nakatsuji2015large} is expressed mostly by the cluster anisotropic M dipole basis.

In contrast to the conventional M dipole with the uniform distribution, the anisotropic M dipole exhibits symmetric quadrupole distributions; $\mathbb{M}'_\mu$ for $\mu=x,y,z$ does not carry any magnetic moment due to the anisotropic spatial distribution of spins.
Meanwhile, the angle dependence of $\mathbb{M}'_\mu$ is the same as that of the ordinary M dipoles, $\sigma_\mu$ and $l_\mu$. 
The same anisotropy of $\mathbb{M}'_\mu$ results in common symmetry structure in physical responses such as the anomalous Hall/Nernst effect and magneto-optical Kerr effect. 
Moreover, the anisotropic M dipole is naturally applicable to the phenomenological linear-response tensor, in which the M dipole corresponds to the antisymmetric component of the conductivity tensor for instance, as will be discussed in Sect.~\ref{sec: Cross-correlation phenomena}~\cite{Hayami_PhysRevB.98.165110}. 
In recent experiments for Mn$_3$Sn, the anisotropic M dipole was observed by means of the XMCD measurement, which was often called $\bm{T}$ vector~\cite{yamasaki2020augmented, kimata2021x, Sasabe_PhysRevLett.126.157402}.

\subsubsection{Bond-cluster multipole}
\label{sec: Bond-cluster multipole}

\begin{figure}[t!]
\centering
\includegraphics[width=1.0 \hsize]{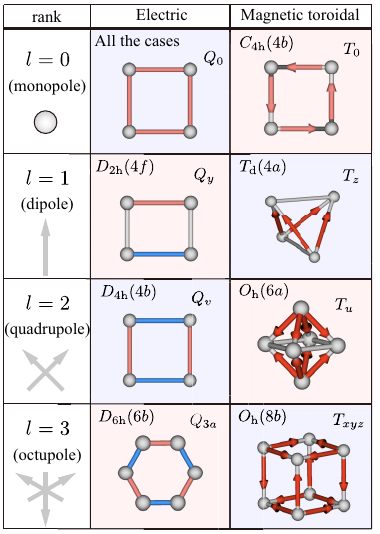} 
\caption{
\label{fig: table_mp_bond}
Examples of the bond-cluster electric and magnetic toroidal multipoles up to the rank $l=3$. 
The red (blue) cells stand for odd-parity (even-parity) multipoles. 
We also show the corresponding space/point group and site symmetry (Wyckoff position) in each cell. 
The red and blue colors in the objects represent the positive and negative weights of the electric monopole on each bond, while the red arrows represent the direction of the MT dipole (pure imaginary hopping). 
}
\end{figure}

The bond-cluster multipole describes the electron hopping and its spatial distribution, which corresponds to the off-diagonal matrix element in multi-site (sublattice) space. 
It is useful to understand various bond or loop-current orderings in a systematic manner~\cite{Hayami_PhysRevLett.122.147602, Hayami_PhysRevB.102.144441}. 
For example, anisotropic density waves including the staggered flux state~\cite{affleck1988large, Nayak_PhysRevB.62.4880, Chakravarty_PhysRevB.63.094503, Allais_PhysRevB.90.155114, Ikeda_PhysRevLett.81.3723, Fujimoto_PhysRevLett.106.196407}, the topological Mott insulating state~\cite{Raghu_PhysRevLett.100.156401,kurita2011topological}, and the loop current state~\cite{Varma_PhysRevB.55.14554, Simon_PhysRevLett.89.247003,Varma_PhysRevB.73.155113,li2010hidden,varma2010high,Zhehao_PhysRevB.97.174511} are systematically described by the bond-cluster multipoles. 
In addition, various bond orderings suggested in the kagome metals, such as $A$V$_3$Sb$_5$ ($A= $K, Rb, and Cs)~\cite{Kiesel_PhysRevLett.110.126405, Wang_PhysRevB.87.115135, Wu_PhysRevLett.127.177001, Denner_PhysRevLett.127.217601, Park_PhysRevB.104.035142, tazai2022mechanism}, are also classified in terms of the bond-cluster multipoles. 

The method for generating a symmetry-adapted bond-cluster multipole basis is obtained by a similar procedure to that for the site-cluster multipole in Sect.~\ref{sec: Site-cluster multipole}, although there are two differences between them~\cite{Kusunose_PhysRevB.107.195118}: 
One is to use the bond-centered position vector instead of the site-centered position vector like $\bm{c}_1=(\bm{R}_1+\bm{R}_2)/2$, and the other is to take into account the orientation of the bond like $\bm{b}_1=\bm{R}_1-\bm{R}_2$.
As for the latter, it is necessary to add a negative sign when taking a sum as in Eq.~(\ref{eq: sitemp_ex}) when the bond direction is reversed for a symmetry operation. 
Even-parity bond cluster multipole with respect to the bond orientation, such as the real hopping corresponding to a time-reversal even scalar quantity, are represented by a spatial distribution of the E monopoles $\mathbb{Q}^{\rm (b)}_\alpha$ [superscript (b) denotes bond-cluster multipoles], while odd-parity bond cluster multipole with respect to the bond orientation, such as the imaginary hopping corresponding to a time-reversal odd vector quantity, are represented by a spatial alignment of the MT dipoles $\mathbb{T}^{\rm (b)}_\mu$ ($\mu=x,y,z$). 
It is noted that no M and ET multipoles appear in the bond-cluster multipoles, since arbitrary bond modulations are expressed by a linear combination of polar quantities. 
We show examples of the bond-cluster multipoles up to rank 3 in Fig.~\ref{fig: table_mp_bond}. 
Similar to the site-cluster multipoles, the concept of the bond-cluster multipoles can be applied to not only an isolated cluster such as a molecule but also a periodic lattice by considering a phase factor arising from the translational symmetry; see Ref.~\citen{Kusunose_PhysRevB.107.195118} for a detailed procedure. 

\subsection{Hamiltonian represented by multipole basis}
\label{sec: Symmetry-adapted multipole basis}

\begin{table}[t]
\caption{\label{tbl:samb_ham}
Correspondence between physical quantities and symmetry-adapted multipole basis~\cite{Kusunose_PhysRevB.107.195118}.
The upper, middle, and lower panels represent one-body, two-body, and hopping terms, respectively.
The site (bond) dependence in the upper (middle) panel is expressed by the site-cluster $\mathbb{Q}_{l,m}^{\rm (s)}$ (bond-cluster $\mathbb{Q}_{l,m}^{\rm (b)}$, $\mathbb{T}_{l,m}^{\rm (b)}$) symmetry-adapted multipole bases.
DM int. is the Dzyaloshinskii-Moriya interaction.
The repeated indices are implicitly summed in the expression.
Reprinted table with permission from Ref.~\citen{Kusunose_PhysRevB.107.195118}, Copyright (2023) by the American Physical Society.
}
\begin{spacing}{1.5}
\scalebox{0.92}{
\begin{tabular}{ccc} \hline \hline
Type & Expression & Correspondence \\ \hline
Electric potential & $\phi q$ & $q\to\mathbb{Q}_{0}^{\rm (a)}$ \\
Crystal field & $\phi_{lm}Q_{l,m}$ & $Q_{l,m}\to\mathbb{Q}_{l,m}^{\rm (a)}$ \\
Zeeman term & $-h^{a}m^{a}$ & $m^{a}\to\mathbb{M}_{1m}^{\rm (a)}$ \\
Spin-orbit int. & $\zeta l^{a}\sigma^{a}$ & $l^{a},\sigma^{a}\to \mathbb{M}_{1m}^{\rm (a)}$ \\ \hline
Density-density int. & $V_{ij}n_{i}n_{j}$ & $n_{i}n_{j}\to\mathbb{Q}_{0}^{\rm (a)}$ \\
Elastic energy & $\epsilon_{ij}^{ab}u_{i}^{a}u_{j}^{b}$ & $u_{i}^{a}u_{j}^{b}\to\mathbb{Q}_{0}^{\rm (a)},\mathbb{Q}_{2m}^{\rm (a)}$ \\
Exchange int. & $J_{ij}^{ab}S_{i}^{a}S_{j}^{b}$ & $S_{i}^{a}S_{j}^{b}\to\mathbb{Q}_{0}^{\rm (a)},\mathbb{Q}_{2m}^{\rm (a)}$ \\
DM int. & $D_{ij}^{c}\epsilon_{abc}S_{i}^{a}S_{j}^{b}$ & $\epsilon_{abc}S_{i}^{a}S_{j}^{b}\to\mathbb{G}_{l,m}^{\rm (a)}$ \\ \hline
Real hopping & $t_{ij}c_{i}^{\dagger}c_{j}+{\rm h.c.}$ & $c_{i}^{\dagger}c_{j}+{\rm h.c.}\to\mathbb{Q}_{l,m}^{\rm (b)}$ \\
Imaginary hopping & $i t_{ij}c_{i}^{\dagger}c_{j}+{\rm h.c.}$ & $i c_{i}^{\dagger}c_{j}+{\rm h.c.}\to\mathbb{T}_{l,m}^{\rm (b)}$ \\
\hline\hline
\end{tabular}
}
\end{spacing}
\end{table}

To summarize the previous sections, the symmetry-adapted multipole basis is constructed by atomic multipoles described by conventional multipoles and hybrid multipoles, $\mathbb{X}_{\alpha}$, and cluster multipoles described by site-cluster multipoles and bond-cluster multipoles, $\mathbb{Y}_{\alpha}$, which is given by~\cite{Kusunose_PhysRevB.107.195118} 
\begin{align}
\label{eq: Z_XY}
\hat{\mathbb{Z}}_{\alpha}=\sum_{\alpha_1\alpha_2}C^{\alpha_1\alpha_2}_{\alpha}(X, Y| Z) \mathbb{X}_{\alpha_1} \otimes \mathbb{Y}_{\alpha_2}, 
\end{align}
where $\mathbb{Z}, \mathbb{X}, \mathbb{Y}$ represent any of $(\mathbb{Q}, \mathbb{M}, \mathbb{T}, \mathbb{G})$ and its subscripts $\alpha, \alpha_1, \alpha_2$ represent a set of the internal degrees of freedom in crystals to uniquely identify the basis, which consists of the irreducible representation, its component, rank $l$, and the multiplicity to distinguish independent harmonics belonging to the same irreducible representation.
$C^{\alpha_1\alpha_2}_{\alpha}(X, Y| Z)$ represents the Clebsch-Gordan coefficient that arises from the reduction to the irreducible representation $\alpha$ from the direct product of $\alpha_1$ and $\alpha_2$. 
The symmetry-adapted multipole basis $\hat{\mathbb{Z}}_{\alpha}$ satisfies the orthonormal conditions and completeness: 
\begin{align}
& {\rm Tr} \left(\hat{\mathbb{Z}}_{\alpha_1}^\dagger \hat{\mathbb{Z}}_{\alpha_2}\right)=\delta_{\alpha_1 \alpha_2}, \\
&
\sum_{\alpha} \braket{a|\hat{\mathbb{Z}}^{}_{\alpha}|a'}\braket{b|\hat{\mathbb{Z}}_{\alpha}^{\dagger}|b'} =\delta_{ab}\delta_{a'b'}. 
\end{align}
In this way, any physical quantities can be expanded by the symmetry-adapted multipole basis.

For example, the Hamiltonian of the system is expressed as a linear combination of symmetry-adapted multipole bases that belong to the totally symmetric irreducible representation of the crystallographic point group under consideration as follows~\cite{Kusunose_PhysRevB.107.195118}. 
\begin{align}
\label{eq: Hamexpand}
H=\sum_{\alpha} z_{\alpha} \hat{\mathbb{Z}}_{\alpha}, 
\end{align}
where $z_\alpha$ is a coefficient, which includes the information about the model parameters, such as the crystalline electric field, relativistic spin-orbit coupling, Coulomb interaction, and electron hoppings. 
We show the correspondence between physical quantities and symmetry-adapted multipole basis in Table~\ref{tbl:samb_ham}. 

The momentum-space representation in Eq.~(\ref{eq: Hamexpand}) under the periodic system is given by 
\begin{align}
\label{eq: Hamexpand_k}
&H(\bm{k})=\sum_{\alpha} z_{\alpha} \hat{\mathbb{Z}}_{\alpha}(\bm{k}),   \\
&\hat{\mathbb{Z}}_{\alpha}(\bm{k})=\sum_{\alpha_1\alpha_2} C^{\alpha_1\alpha_2}_{\alpha}(X, Y| Z) \left[\mathbb{X}_{\alpha_1} \otimes \mathbb{Y}_{\alpha_2}(\bm{k})\right], 
\label{eq: zkbs}
\end{align}
where $\mathbb{Y}_{\alpha_2}(\bm{k})$ is obtained by taking into account the phase factor arising from the translational symmetry of crystals for the cluster multipoles in Sect.~\ref{sec: Cluster multipole}; see Ref.~\citen{Kusunose_PhysRevB.107.195118} for the derivation.

\section{Electronic band structure}
\label{sec: Electronic band structure}

Multipole degrees of freedom are also closely related to the electronic band structure in crystals; there is a correspondence between arbitrary band deformations/spin splittings and multipoles. 
We discuss the momentum representation of the multipoles in Sect.~\ref{sec: Momentum representation}. 
Then, we show several microscopic mechanisms of band deformations and spin splittings based on the multipole description in Sect.~\ref{sec: Microscopic origin of band deformation and spin splitting}. 

\subsection{Momentum representation}
\label{sec: Momentum representation}

\begin{figure}[t!]
\begin{center}
\includegraphics[width=1.0 \hsize]{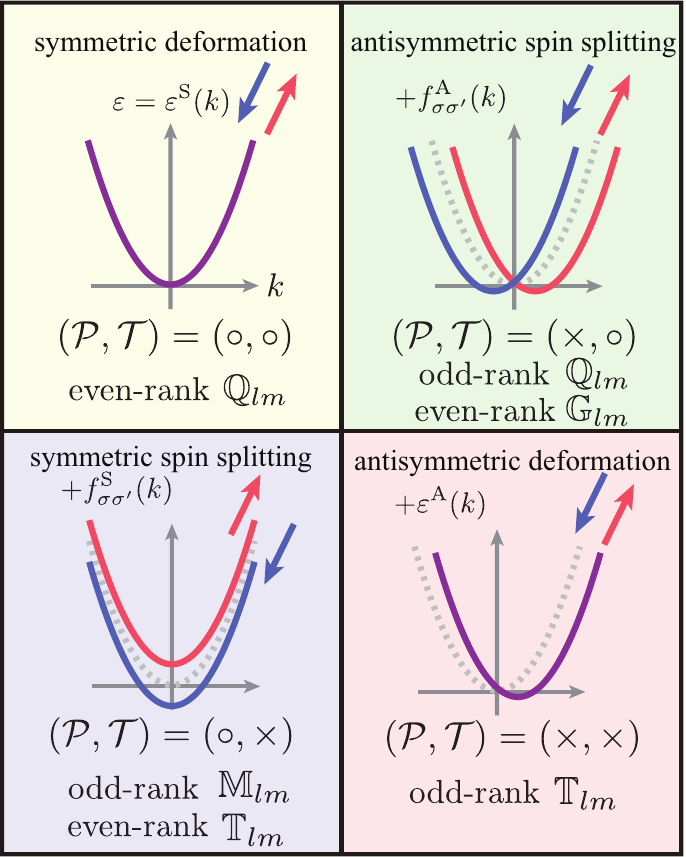}
\caption{
\label{Fig: band_4type}
Deformations and spin splittings in the band structure in the presence ($\circ$) or absence ($\times$) of spatial inversion symmetry $\mathcal{P}$ and time-reversal symmetry $\mathcal{T}$.
The blue (red) arrows stand for the up(down)-spin moments. 
The dashed curves represent the energy bands in the presence of $\mathcal{P}$ and $\mathcal{T}$. 
The relevant multipoles are also shown at the bottom of each panel. 
}
\end{center}
\end{figure}

According to the presence/absence of $\mathcal{P}$ and $\mathcal{T}$ symmetries, the band structures are classified into four types: symmetric deformation with $(\mathcal{P}, \mathcal{T})=(+1,+1)$, symmetric spin splitting with $(\mathcal{P}, \mathcal{T})=(+1,-1)$, antisymmetric spin splitting with $(\mathcal{P}, \mathcal{T})=(-1, +1)$, and antisymmetric band deformation with $(\mathcal{P}, \mathcal{T})=(-1,-1)$, as shown in Fig.~\ref{Fig: band_4type}. 
Since the concept of multipoles also describes the anisotropy in momentum space as well as real space, four types of multipoles can cover these deformations and spin splittings in a unified way.

\begin{table}[htb!]
\caption{
Multipoles in momentum space up to rank 2 in the limit of $\bm{k} \to \bm{0}$~\cite{Hayami_PhysRevB.98.165110}. 
$k^2=k_x^2+k_y^2+k_z^2$. 
}
\label{tab_k-multipole}
\centering
\begingroup
\scalebox{1.0}{
\renewcommand{\arraystretch}{1.4}
\scalebox{0.8}{
 \begin{tabular}{cccc}
   \multicolumn{4}{l}{\fbox{symmetric band deformation: even-rank $\mathbb{Q}_{l,m}(\bm{k})$}} \\
 \hline 
rank & type & symbol & definition \\ \hline
$0$ & E & $\mathbb{Q}_0$ & $\sigma_{0}$ \\ \hline
$2$ & E & $\mathbb{Q}_{u}$ & $\displaystyle \frac{1}{2}(3k_{z}^2-k^2)\sigma_{0}$ \\
&  & $\mathbb{Q}_{v}$ & $\displaystyle \frac{\sqrt{3}}{2}(k_{x}^2-k_{y}^2)\sigma_{0}$ \\
    &      & $\mathbb{Q}_{yz}$ & $\sqrt{3}k_{y}k_{z}\sigma_{0}$ \\
    &      & $\mathbb{Q}_{zx}$ & $\sqrt{3}k_{z}k_{x}\sigma_{0}$ \\
    &      & $\mathbb{Q}_{xy}$ & $\sqrt{3}k_{x}k_{y}\sigma_{0}$ \\
 \hline \\
     \multicolumn{4}{l}{\fbox{symmetric spin splitting: odd-rank $\mathbb{M}_{l,m}(\bm{k})$ and even-rank $\mathbb{T}_{l,m}(\bm{k})$}} \\ \hline
rank & type  & symbol & definition \\ \hline
$1$ & M  & $\mathbb{M}_{x}$,$\mathbb{M}_{y}$,$\mathbb{M}_{z}$ & $ \sigma_{x}$, $ \sigma_{y}$, $ \sigma_{z}$  \\ 
\hline
 $2$    & MT  & $\mathbb{T}_{u}$ & $3 k_x k_z \sigma_y-3 k_y k_z \sigma_x$ \\
  & & $\mathbb{T}_{v}$ & $ \sqrt{3}(2 k_x k_y \sigma_z-k_y k_z \sigma_x-k_x k_z \sigma_y)$ \\
    &      & $\mathbb{T}_{yz}$ & $\sqrt{3}[(k_z^2-k_y^2) \sigma_x+k_x k_y \sigma_y-k_z k_x \sigma_z]$\\    
    &      & $\mathbb{T}_{zx}$ & $\sqrt{3}[(k_x^2-k_z^2) \sigma_y+k_y k_z \sigma_z-k_x k_y \sigma_x]$ \\    
    &      & $\mathbb{T}_{xy}$ & $\sqrt{3}[(k_y^2-k_x^2) \sigma_z+k_z k_x \sigma_x-k_y k_z \sigma_y]$ \\    
\hline \\
   \multicolumn{4}{l}{\fbox{antisymmetric band deformation: odd-rank $\mathbb{T}_{l,m}(\bm{k})$}} \\
   rank & type & symbol & definition \\ \hline
1   & MT & $\mathbb{T}_{x}$, $\mathbb{T}_{y}$, $\mathbb{T}_{z}$ & $k_{x}\sigma_{0}$, $k_{y}\sigma_{0}$, $k_{z}\sigma_{0}$ \\ \hline
\hline \\
\multicolumn{4}{l}{\fbox{antisymmetric spin splitting: odd-rank $\mathbb{Q}_{l,m}(\bm{k})$ and even-rank $\mathbb{G}_{l,m}(\bm{k})$}} \\
rank & type  & symbol & definition \\ \hline
$0$    & ET  & $\mathbb{G}_0$ & $\bm{k}\cdot\bm{\sigma}$ \\ \hline
$1$ & E   & $\mathbb{Q}_{x}$, $\mathbb{Q}_{y}$, $\mathbb{Q}_{z}$ & $(\bm{k} \times \bm{\sigma})_{x}$, $(\bm{k} \times \bm{\sigma})_{y}$, $(\bm{k} \times \bm{\sigma})_{z}$ \\ \hline
$2$    & ET  & $\mathbb{G}_{u}$ & $3k_{z}\sigma_{z}-\bm{k}\cdot\bm{\sigma}$ \\
   &  & $\mathbb{G}_{v}$ & $\sqrt{3}(k_{x}\sigma_{x}-k_{y}\sigma_{y})$ \\
    &     & $\mathbb{G}_{yz}$ & $\sqrt{3}(k_{y}\sigma_{z}+k_{z}\sigma_{y})$ \\ 
    &     & $\mathbb{G}_{zx}$ & $\sqrt{3}(k_{z}\sigma_{x}+k_{x}\sigma_{z})$ \\ 
    &     & $\mathbb{G}_{xy}$ & $\sqrt{3}(k_{x}\sigma_{y}+k_{y}\sigma_{x})$ \\ \hline
\end{tabular}
}
}
\endgroup
\end{table}

We present the momentum-space multipoles in the limit of $\bm{k} \to \bm{0}$ for simplicity~\cite{Hayami_PhysRevB.98.165110}. 
In the single-band system, the Hamiltonian can be generally expressed as 
\begin{align}
\label{eq:Hamkspace2}
\mathcal{H}=&\sum_{\bm{k}\sigma\sigma'} \biggl[ 
\varepsilon^{\rm S}(\bm{k})\delta_{\sigma\sigma'}+\varepsilon^{\rm A}(\bm{k})\delta_{\sigma\sigma'}
\cr&\quad\quad\quad\quad\quad
+f^{\rm S}_{\sigma\sigma'}(\bm{k})+f_{\sigma\sigma'}^{\rm A}(\bm{k}) 
\biggr] c^{\dagger}_{\bm{k}\sigma}c^{}_{\bm{k}\sigma'}, 
\end{align}
where $\varepsilon$($f_{\sigma\sigma'}$) is the charge(spin) sector and the superscript ${\rm S}$(${\rm A}$) represents symmetric(antisymmetric) contribution with respect to $\bm{k}$.
The coefficients $[\varepsilon^{\rm S}(\bm{k}), \varepsilon^{\rm A}(\bm{k}), f^{\rm S}_{\sigma \sigma'}(\bm{k}), f^{\rm A}_{\sigma \sigma'}(\bm{k})]$ are expanded by momentum-space multipoles as 
\begin{align}
\label{eq:dispersion_sym_spinless}
\varepsilon^{\rm S}(\bm{k})
&= \sum_{lm}^{{\rm even}} Q^{\rm ext}_{l,m} \mathbb{Q}_{l,m} (\bm{k}), \\
\label{eq:dispersion_asym_spinless}
\varepsilon^{\rm A}(\bm{k})
&= \sum_{lm}^{{\rm odd}} T^{\rm ext}_{l,m} \mathbb{T}_{l,m} (\bm{k}), \\
\label{eq:dispersion_sym_spinful}
f^{\rm S}_{\sigma \sigma'}(\bm{k})
&= \sum_{lm}^{{\rm odd}} M^{\rm ext}_{l,m} \mathbb{M}_{l,m}^{\sigma\sigma'} (\bm{k}) + \sum_{lm}^{{\rm even}}  T^{\rm ext}_{l,m} \mathbb{T}_{l,m}^{\sigma\sigma'} (\bm{k}), \\
\label{eq:dispersion_asym_spinful}
f^{\rm A}_{\sigma \sigma'}(\bm{k})
&= \sum_{lm}^{{\rm even}} G^{\rm ext}_{l,m} \mathbb{G}_{l,m}^{\sigma\sigma'} (\bm{k}) + \sum_{lm}^{{\rm odd}} Q^{\rm ext}_{l,m} \mathbb{Q}_{l,m}^{\sigma\sigma'} (\bm{k}), 
\end{align}
where \begin{align}
&
\mathbb{Q}_{l,m}(\bm{k})\equiv\begin{cases}
\sigma_{0}O_{l,m}(\bm{k}) & (l=0,2,4,6,\cdots) \\
(\bm{k}\times\bm{\sigma})\cdot\bm{\nabla}_{\bm{k}}O_{l,m}(\bm{k}) & (l=1,3,5,\cdots)
\end{cases}
\label{eq:qmom}
\\&
\mathbb{M}_{l,m}(\bm{k})\equiv\begin{cases}
0 & (l=0,2,4,6,\cdots) \\ 
\bm{\sigma}\cdot\bm{\nabla}_{\bm{k}}O_{l,m}(\bm{k}) & (l=1,3,5,\cdots)
\end{cases}
\label{eq:mmom}
\\&
\mathbb{T}_{l,m}(\bm{k})\equiv\begin{cases}
0 & (l=0) \\
(\bm{k}\times\bm{\sigma})\cdot\bm{\nabla}_{\bm{k}}O_{l,m}(\bm{k}) & (l=2,4,6,\cdots) \\
\sigma_{0}O_{l,m}(\bm{k}) & (l=1,3,5,\cdots)
\end{cases}
\\&
\mathbb{G}_{l,m}(\bm{k})\equiv\begin{cases}
\bm{k}\cdot\bm{\sigma} & (l=0) \\
\bm{\sigma}\cdot\bm{\nabla}_{\bm{k}}O_{l,m}(\bm{k}) & (l=2,4,6,\cdots) \\
0 & (l=1,3,5,\cdots) 
\end{cases}
\label{eq:etmom}
\end{align}
The expressions of $[\mathbb{Q}_{l,m}(\bm{k}), \mathbb{M}_{l,m}(\bm{k}), \mathbb{T}_{l,m}(\bm{k}), \mathbb{G}_{l,m}(\bm{k})]$ up to rank 2 are shown in Table~\ref{tab_k-multipole}. 
The spin-independent symmetric (antisymmetric) dispersion is described by the even-rank E (odd-rank MT) multipoles. In contrast, the spin-dependent symmetric (antisymmetric) dispersion is described by the odd-rank M and even-rank MT (odd-rank E and even-rank ET) multipoles.

\begin{figure*}[htb!]
\begin{center}
\includegraphics[width=1.0 \hsize]{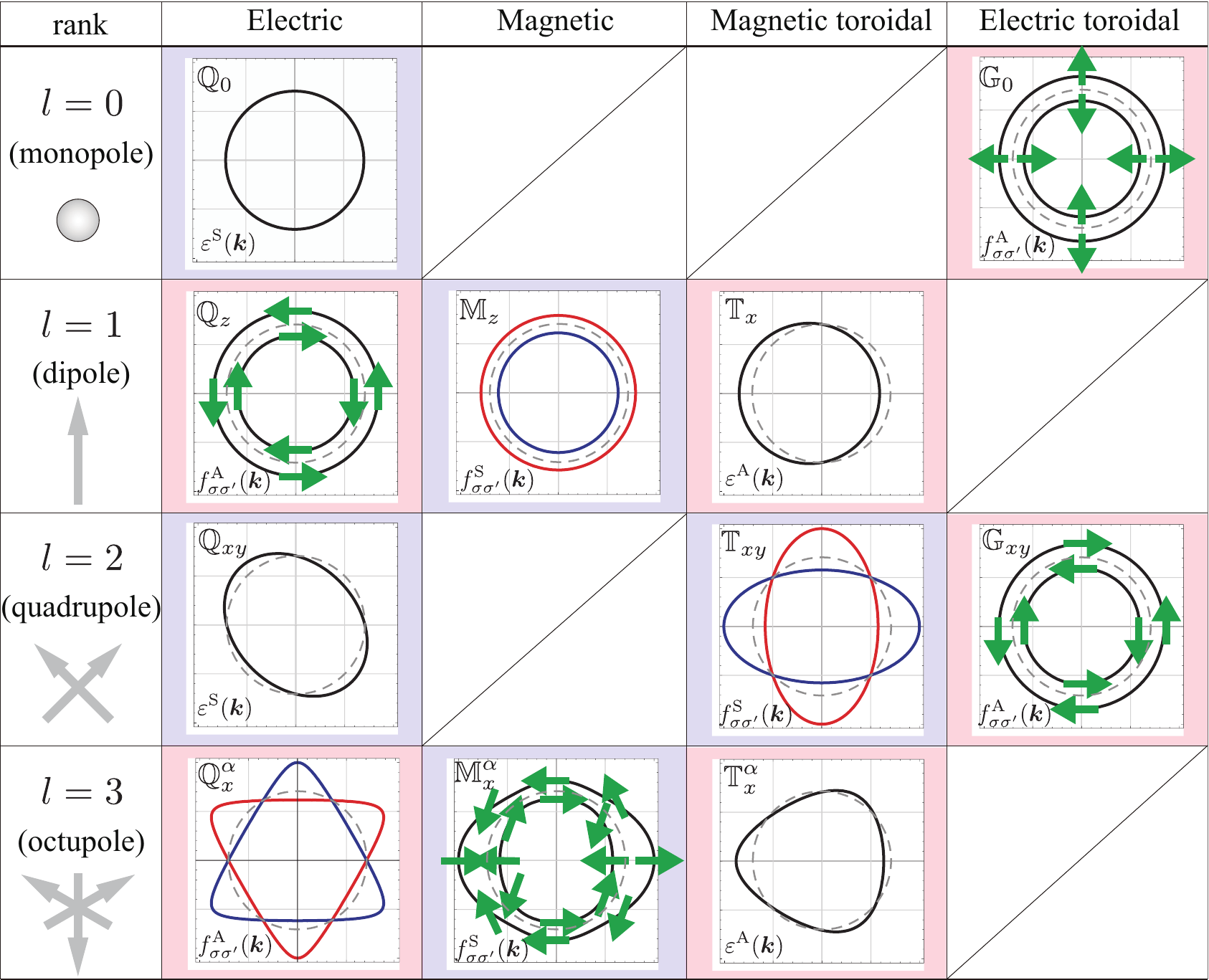}
\caption{
\label{fig: table_mp_band}
Examples of the typical electronic band modulations in the $k_x$-$k_y$ plane at $k_z=0$ in the presence of the E, M, MT, and ET multipoles up to the rank $l=3$ in the single-band picture~\cite{Hayami_PhysRevB.98.165110}. 
The blue (red) cells stand for the band modulations by the even-parity (odd-parity) multipoles. 
The green arrows represent the in-plane spin moments at $\bm{k}$, while the red (blue) curves show the up(down)-spin moments along the out-of-plane direction. 
The dashed circles in each cell represent the energy contours in the absence of the multipole, which corresponds to the energy contour for $\mathbb{Q}_0(\bm{k})$.
This figure is taken and modified from Ref.~\citen{Hayami_PhysRevB.98.165110}.
}
\end{center}
\end{figure*}

The coefficients $[\varepsilon^{\rm S}(\bm{k}), \varepsilon^{\rm A}(\bm{k}), f^{\rm S}_{\sigma \sigma'}(\bm{k}), f^{\rm A}_{\sigma \sigma'}(\bm{k})]$ become nonzero once the corresponding multipoles shown in Eqs.~(\ref{eq:dispersion_sym_spinless})--(\ref{eq:etmom}), i.e., ($Q^{\rm ext}_{l,m}$, $M^{\rm ext}_{l,m}$, $T^{\rm ext}_{l,m}$, $G^{\rm ext}_{l,m}$), are activated irrespective of intrinsic and extrinsic factors; electronic bands exhibit deformations and spin splittings according to the multipoles, as exemplified in Fig.~\ref{fig: table_mp_band}~\cite{Hayami_PhysRevB.98.165110}. 
$\varepsilon^{\rm S}(\bm{k})$ in Eq.~(\ref{eq:dispersion_sym_spinless}) describes the quadrupole-type symmetric deformation under E quadrupole (or nemtaic) ordering, as discussed in iron-based superconductors~\cite{Fang_PhysRevB.77.224509, Fernandes_PhysRevLett.105.157003, Kruger_PhysRevB.79.054504,Lv_PhysRevB.80.224506, Lee_PhysRevLett.103.267001, Onari_PhysRevLett.109.137001} and Sr$_3$Ru$_2$O$_7$~\cite{Raghu_PhysRevB.79.214402, Lee_PhysRevB.80.104438,Tsuchiizu_PhysRevLett.111.057003}. 
$f^{\rm S}_{\sigma \sigma'}(\bm{k})$ in Eq.~(\ref{eq:dispersion_sym_spinful}) is caused by the ferromagnetic ordering and external magnetic field accompanying the M dipole, and AFM ordering accompanying the MT quadrupole. 
For example, the band dispersions in the presence of the MT quadrupole $T_{xy}^{\rm ext}$, as found in AFMs Gd$_2$Sn$_2$O$_7$~\cite{wills2006magnetic} and Er$_2$Ru$_2$O$_7$~\cite{taira2003magnetic}, are given by 
\begin{align}
\varepsilon_\sigma (\bm{k})= 
\displaystyle \frac{\hbar^2 \bm{k}^2}{2m} + \sigma T^{\rm ext}_{xy}  \sqrt{(k_x^2-k_y^2)^2+ (k_x^2+k_y^2)k_z^2}, 
\end{align}
where $T^{\rm ext}_{xy}$ is an effective molecular field for inducing $\mathbb{T}_{xy}(\bm{k})$. 
This indicates the quadrupole-type spin splitting, as shown in Fig.~\ref{fig: table_mp_band}~\cite{hayami2022spinconductivity}.   
$f^{\rm A}_{\sigma \sigma'}(\bm{k})$ in Eq.~(\ref{eq:dispersion_asym_spinful}) is caused by the inversion symmetry breaking without time-reversal symmetry breaking. 
For example, the Rashba-type spin splitting with the form of $k_y \sigma_x -k_x \sigma_y$ corresponds to the E dipole basis $\mathbb{Q}_{z}(\bm{k})$, chiral(hedgehog)-type spin splitting $\bm{k} \cdot \bm{\sigma}$ corresponds to the ET monopole basis $\mathbb{G}_0(\bm{k})$, and the Dresselhaus-type spin splitting $k_x (k_y^2-k_z^2)\sigma_x+k_y (k_z^2-k_x^2)\sigma_y+k_z (k_x^2-k_y^2)\sigma_z$ corresponds to electric octupole basis $\mathbb{Q}_{xyz}(\bm{k})$, as schematically shown in Fig.~\ref{fig: table_mp_band}. 
The antisymmetric spin-orbit coupling (ASOC), which originates from the relativistic spin-orbit coupling in noncentrosymmetric crystals, is categorized in this category; we show its functional forms for the periodic crystal and $\bm{k} \to \bm{0}$ limit under all the noncentrosymmetric point groups in Table~\ref{table: asoc}, where the relevant odd-rank E and even-rank ET multipoles are also shown. 
$\varepsilon^{\rm A}(\bm{k})$ in Eq.~(\ref{eq:dispersion_asym_spinless}) is caused by the breakings of both the spatial inversion and time-reversal symmetries. 
For example, the MT dipole $T_x$ order activates $T_{x}^{\rm ext}$ which leads to $\mathbb{T}_{x}(\bm{k})$, i.e., the $\bm{k}$-linear dispersions with the spin degeneracy, namely, the antisymmetric band deformation, as shown in Fig.~\ref{fig: table_mp_band}.

\begin{table*}
\caption{Functional forms of the ASOC for the periodic crystal and $\bm{k}\to \bm{0}$ limit under noncentrosymmetric point groups (PGs). 
For $D_{\rm 2d}$, $D_{\rm 3h}$, $C_{\rm 3v}$, and $D_{3}$, the setting of $\bar{4}2m$, $\bar{6}m2$, $3m1$, and $312$ is adopted, respectively. 
The principle axis for the monoclinic crystal is taken along the $y$ axis. 
We use the abbreviations like $s_{x}=\sin k_x$ and $c_{y/2}=\cos k_y/2$ and so on for notational simplicity. 
$f_x=\frac{2}{\sqrt{3}}\left(2 c_{x/2}+c_{\sqrt{3}y/2}\right)s_{x/2}$, $f_y=2 c_{x/2}s_{\sqrt{3}y/2}$, $f'_x=\frac{4}{\sqrt{6}}\left(-c_x+ c_{x/2}\cos_{\sqrt{3}y/2}\right)s_z$, $f'_y=-2\sqrt{2}s_{x/2} s_{\sqrt{3}y/2} s_z$, $f_1=\frac{4}{\sqrt{6}}\left(-c_{3x/2}+ c_{\sqrt{3}y/2}\right)s_{\sqrt{3}y/2}$, $f_2=\frac{4}{\sqrt{6}}\left(-c_{x/2}+c_{\sqrt{3}y/2}\right) s_{x/2}$. 
}\label{table: asoc}
\footnotesize
\centering{
\begin{tabular}{c|c|c|c|c|c} \hline\hline
PG & irrep. & ASOC for periodic crystal and $\bm{k} \to \bm{0} $ limit & active PG & momentum multipole & note \\ \hline
\multirow{4}{*}{$O_{\rm h}$} & ${\rm A}_{1u}$ & $s_x \sigma_x+s_y\sigma_y+s_z \sigma_z \underset{(\bm{k}\to \bm{0})}{\to} k_x \sigma_x + k_y \sigma_y + k_z \sigma_z$ & $O$, $T$ & $\mathbb{G}_0$ & hedgehog \\ 
\cline{2-6}
& ${\rm A}_{2u}$ & $(-c_y + c_z)s_x \sigma_x+(-c_z + c_x)s_y\sigma_y+(-c_x + c_y)s_z\sigma_z$ & $T_{\rm d}$, $T$ & $\mathbb{Q}_{xyz}$ & Dresselhaus \\ 
& & $\underset{(\bm{k}\to \bm{0})}{\to} (k_y^2-k_z^2)\sigma_x+(k_z^2-k_x^2)\sigma_y+(k_z^2-k_x^2)\sigma_y$ & & \\
\hline\hline
\multirow{6}{*}{$D_{\rm 4h}$} & ${\rm A}_{1u}$ & $c_1(s_x \sigma_x+s_y \sigma_y)+c_2s_z \sigma_z \underset{(\bm{k}\to \bm{0})}{\to} c_1(k_x \sigma_x+k_y \sigma_y)+c_2k_z \sigma_z$ & $D_{4}$, $C_{4}$ & $\mathbb{G}_0, \mathbb{G}_u$ & hedgehog \\ \cline{2-6}
& ${\rm A}_{2u}$ & $s_y \sigma_x-s_x \sigma_y \underset{(\bm{k}\to \bm{0})}{\to} k_y \sigma_x - k_x \sigma_y$ & $C_{\rm 4v}$, $C_{4}$ & $\mathbb{Q}_z$ & Rashba \\ \cline{2-6}
& ${\rm B}_{1u}$ & $s_x \sigma_x-s_y \sigma_y \underset{(\bm{k}\to \bm{0})}{\to} k_x \sigma_x - k_y \sigma_y$ & $D_{\rm 2d}$, $S_{4}$ & $\mathbb{G}_v$ & $v$ type \\ \cline{2-6}
& ${\rm B}_{2u}$ & $s_y\sigma_x+s_x \sigma_y \underset{(\bm{k}\to \bm{0})}{\to} k_y \sigma_x + k_x \sigma_y$ & $S_{4}$ & $\mathbb{G}_{xy}$ & $xy$ type \\ \hline\hline
\multirow{4}{*}{$D_{\rm 2h}$} & ${\rm A}_{u}$ & $c_1 s_x \sigma_x+c_2 s_y \sigma_y+c_3 s_z \sigma_z \underset{(\bm{k}\to \bm{0})}{\to} c_1 k_x \sigma_x+c_2 k_y \sigma_y+c_3 k_z \sigma_z$ & $D_{2}$ & $\mathbb{G}_0, \mathbb{G}_u, \mathbb{G}_v$ \\ \cline{2-6}
& ${\rm B}_{1u}$ & $c_1 s_y \sigma_x + c_2 s_x \sigma_y + c_3 s_x s_y s_z \sigma_z $ & $C_{\rm 2v}$ & $\mathbb{Q}_z, \mathbb{G}_{xy}, \mathbb{Q}^{\beta}_z$ \\
& & $\underset{(\bm{k}\to \bm{0})}{\to} c_1 k_y \sigma_x + c_2 k_x \sigma_y + c_3 k_x k_y k_z \sigma_z$ & & \\
 \hline\hline
\multirow{5}{*}{$C_{\rm 2h}$} & ${\rm A}_{u}$ & $(c_1 s_x + c_2 s_z)\sigma_x + c_3 s_y \sigma_y + (c_4 s_x  + c_5 s_z )\sigma_z$ & $C_{2}$ & $\mathbb{G}_0, \mathbb{Q}_z, \mathbb{G}_u, \mathbb{G}_v, \mathbb{G}_{zx}$ &  \\
& & $\underset{(\bm{k}\to \bm{0})}{\to} (c_1 k_x + c_2 k_z)\sigma_x + c_3 k_y \sigma_y + (c_4 k_x  + c_5 k_z )\sigma_z$ & & \\
\cline{2-6}
& ${\rm B}_{u}$ & $c_1 s_y \sigma_x + (c_2 s_x + c_3 s_z) \sigma_y + c_4 s_y \sigma_z$ & $C_{\rm s}$  & $\mathbb{Q}_x, \mathbb{Q}_z, \mathbb{G}_{yz}, \mathbb{G}_{xy}$ &  \\
& & $\underset{(\bm{k}\to \bm{0})}{\to} c_1 k_y \sigma_x + (c_2 k_x + c_3 k_z) \sigma_y + c_4 k_y \sigma_z$ & & \\
\hline\hline
\multirow{4.5}{*}{$C_{\rm i}$} & ${\rm A}_{u}$ & $(c_1 s_x + c_2 s_y + c_3 s_z)\sigma_x+(c_4 s_x + c_5 s_y + c_6 s_z)\sigma_y $ & $C_{1}$ & $\mathbb{G}_0, \mathbb{Q}_x, \mathbb{Q}_y, \mathbb{Q}_z$ & \\
& & $+(c_7s_x + c_8 s_y + c_9 s_z)\sigma_z$ & & $\mathbb{G}_u, \mathbb{G}_v, \mathbb{G}_{yz}, \mathbb{G}_{zx}, \mathbb{G}_{xy}$ & \\
& & $\underset{(\bm{k}\to \bm{0})}{\to} (c_1 k_x + c_2 k_y + c_3 k_z)\sigma_x+(c_4 k_x + c_5 k_y + c_6 k_z)\sigma_y$ & & \\
& & $+(c_7 k_x + c_8 k_y + c_9 k_z)\sigma_z$ & & \\
\hline\hline
\multirow{8}{*}{$D_{\rm 6h}$} & ${\rm A}_{1u}$ & $c_1 (f_x \sigma_x+f_y\sigma_y ) + c_2 s_z \sigma_z \underset{(\bm{k}\to \bm{0})}{\to} c_1 (k_x \sigma_x+k_y\sigma_y ) + c_2 k_z \sigma_z$ & $D_{6}$, $C_{6}$ & $\mathbb{G}_0, \mathbb{G}_u$ & hedgehog \\
\cline{2-6}
& ${\rm A}_{2u}$ & $f_y \sigma_x-f_x \sigma_y \underset{(\bm{k}\to \bm{0})}{\to} k_y \sigma_x-k_x \sigma_y$ & $C_{\rm 6v}$, $C_{6}$ & $\mathbb{Q}_z$ & Rashba \\ \cline{2-6}
& ${\rm B}_{1u}$ & $c_1 (f'_x\sigma_x+f'_{y}\sigma_y )+c_2 f_2 \sigma_z $ & $D_{\rm 3h}$, $C_{\rm 3h}$ & $\mathbb{Q}_{3b}, \mathbb{G}_{4b}$ &  \\
& & $\underset{(\bm{k}\to \bm{0})}{\to} c_1 \left[(k_{x}^{2}-k_{y}^{2})k_{z}\sigma_x-k_{x}k_{y}k_{z}\sigma_y \right]+c_2 (k_{x}^{2}-3k_{y}^{2})k_{x} \sigma_z$ & & \\
\cline{2-6}
& ${\rm B}_{2u}$ & $c_1 (f_{y}'\sigma_x-f_{x}'\sigma_y)+ c_2 f_{1} \sigma_z $ & $C_{\rm 3h}$ & $\mathbb{Q}_{3a}, \mathbb{G}_{4a}$ &  \\
& & $\underset{(\bm{k}\to \bm{0})}{\to} c_1 \left[ -k_{x}k_{y}k_{z}\sigma_x-(k_{x}^{2}-k_{y}^{2})k_{z}\sigma_y \right]+ c_2 (3k_{x}^{2}-k_{y}^{2})k_{y} \sigma_z$ & & \\
 \hline\hline
\multirow{3}{*}{$D_{\rm 3d}$} & ${\rm A}_{1u}$ & $c_1 (f_x \sigma_x + f_y \sigma_y )+c_2 s_z \sigma_z \underset{(\bm{k}\to \bm{0})}{\to} c_1 (k_x \sigma_x + k_y \sigma_y )+c_2 k_z \sigma_z$ & $D_{3}$, $C_{3}$ & $\mathbb{G}_0, \mathbb{G}_u$ & hedgehog \\
\cline{2-6}
& ${\rm A}_{2u}$ & $f_y \sigma_x - f_x \sigma_y \underset{(\bm{k}\to \bm{0})}{\to} k_y \sigma_x - k_x \sigma_y$ & $C_{\rm 3v}$, $C_{3}$ & $\mathbb{Q}_z$ & Rashba \\
\hline\hline
\end{tabular}
}
\end{table*}

Beyond the single-band argument, further intriguing band dispersions can be expected. 
For example, multi-orbital degrees of freedom enable us to describe the momentum representation of M quadrupole, which does not appear in the single-band system~\cite{Hayami_PhysRevB.104.045117}. 
Specifically, the $xy$ component of the M quadrupole is represented by 
\begin{align}
\label{eq:Mxy_k1}
\mathbb{M}^{\rm (I)}_{xy}(\bm{k})&=\sqrt{3}( \mathcal{G}_x k_y+  \mathcal{G}_y k_x), \\
\label{eq:Mxy_k2}
\mathbb{M}^{\rm (II)}_{xy}(\bm{k})&=\sqrt{3}\left[\sqrt{2}(\mathcal{Q}_{yy}-\mathcal{Q}_{xx}) k_z + \mathcal{Q}_{zx} k_x - \mathcal{Q}_{yz}k_y \right],
\end{align}
where $\mathcal{Q}_{\mu\nu}=(l_{\mu} \sigma_{\nu}+l_{\nu} \sigma_{\mu})/\sqrt{2}$ for $\mu,\nu=x,y,z$ is the rank-2 polar tensor and $\bm{\mathcal{G}}=\bm{l}\times \bm{\sigma}$ is the rank-1 axial tensor, which are constructed from orbital $\bm{l}$ and spin $\bm{\sigma}$ angular momenta.  
Thus, the M quadrupole gives rise to the antisymmetric spin-orbital polarization with respect to $\bm{k}$ in the multi-orbital band structure, which is referred to as ``spin-orbital-momentum locking"~\cite{Hayami_PhysRevB.104.045117}. 
This unconventional momentum locking becomes the microscopic origin of the current-induced distortion (magneto-piezo effect) in metals~\cite{shiomi2019observation}.  
It is noted that all of the above momentum multipoles are generated in a unified way by using $\hat{\mathbb{Z}}_{\alpha}(\bm{k})$ in Eq.~(\ref{eq: zkbs}).

\subsection{Microscopic origin of band deformation and spin splitting}
\label{sec: Microscopic origin of band deformation and spin splitting}

\subsubsection{With relativistic spin-orbit coupling}

\begin{figure}[t!]
\begin{center}
\includegraphics[width=1.0 \hsize]{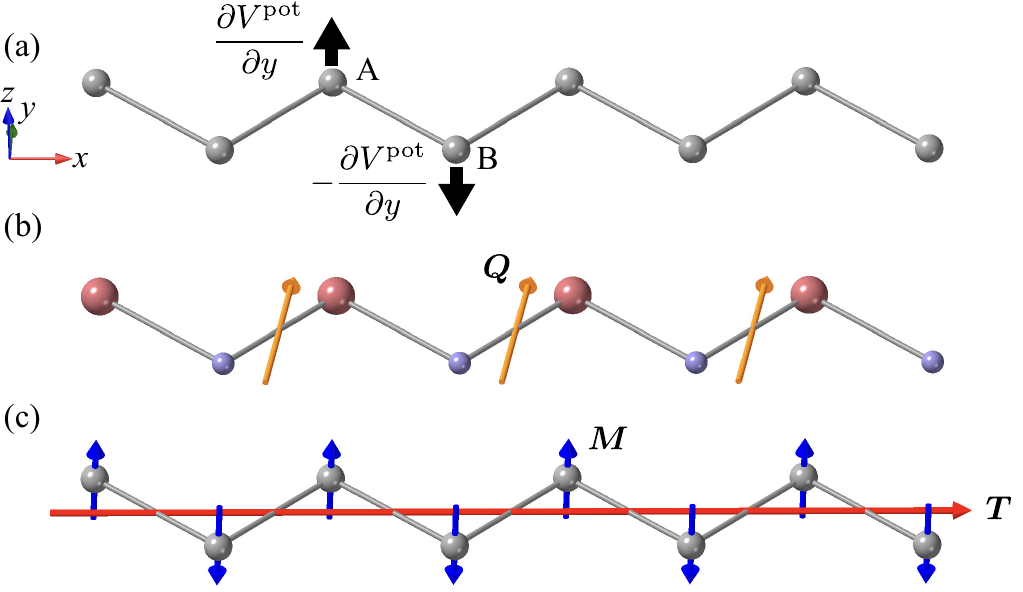}
\caption{
\label{Fig: zigzag}
(a) Zigzag chain consisting of two sublattices A and B~\cite{Hayami_PhysRevB.108.094106}. The black arrows represent the direction of sublattice-dependent potential gradient $\pm \partial V^{\rm pot}/\partial y$ corresponding to the odd-parity crystalline electric field. 
(b) Staggered alignment of charge ordering leading to a uniform alignment of the E dipole $\bm{Q}$ in the $y$ direction and (c) staggered alignment of the M dipole $\bm{M}$ along the $z$ direction leading to a uniform alignment of the MT dipole $\bm{T}$ in the $x$ direction.
Reprinted figures with permission in (a) and (c) from Ref.~\citen{Hayami_PhysRevB.108.094106}, Copyright (2023) by the American Physical Society.
}
\end{center}
\end{figure}

The relativistic spin-orbit coupling plays an important role in inducing the modulations and spin splittings in the band structure. 
Let us discuss a fundamental example of a one-dimensional zigzag chain with two sublattices A and B along the $x$ direction in Fig.~\ref{Fig: zigzag}, where the antisymmetric spin splitting and antisymmetric modulation are caused by simple spontaneous staggered electronic orderings~\cite{hayami2016emergent}. 
Owing to the lack of the inversion center at each sublattice site, the potential gradient (local electric field) occurs in the $y$ direction, $\pm \partial V^{\rm pot}/\partial y$, with an opposite sign between the sublattices A and B. 
In such a situation, the ASOC in the form of $\bm{g}_s (\bm{k}) \cdot \bm{\sigma}$ appears in the Hamiltonian, where $\bm{g}_s (\bm{k})$ is a so-called sublattice-dependent $g$-vector as $\bm{g}_s (\bm{k}) =   (0, 0,  p(s) \alpha k_x )$ for the $s=$ A and B sublattices [$p(s)=+1(-1)$ for the A (B) sublattice]~\cite{Yanase_JPSJ.83.014703, Hayami_doi:10.7566/JPSJ.84.064717}; $\alpha$ represents the magnitude of the ASOC. 
The microscopic origin of the ASOC is the cooperative effect between the relativistic atomic spin-orbit coupling and hoppings/hybridizations between orbitals with different parity~\cite{Bauer_Sigrist201201}.

Then, spontaneous symmetry breaking due to electron correlations can give rise to uniform odd-parity multipoles, such as the E dipole and the MT dipole. 
For example, when a charge ordering occurs as $n_{{\rm A}} \neq n_{{\rm B}}$ [$n_{{\rm A}({\rm B})}$ is the electron density in the A (B) sublattice], a uniform component of the E dipole is induced in the $y$ direction, $Q_y \propto (n_{{\rm A}} - n_{{\rm B}})|\partial V^{\rm pot}/\partial y|$, as shown in Fig.~\ref{Fig: zigzag}(b); the antisymmetric spin-split band is expected as similar to the polar systems. 
Meanwhile, when a staggered collinear magnetic ordering in the $z$ direction occurs as given by $m_{{\rm A}}^z =- m_{{\rm B}}^z$ [$m_{{\rm A}(\rm B)}^z$ is the magnetic moment in the $z$ direction in the A (B) sublattice], a uniform component of the MT dipole is induced in the $x$ direction $T_x \propto (m_{{\rm A}}^{z}-m_{{\rm B}}^{z}) |\partial V^{\rm pot}/\partial y|$, as shown in Fig.~\ref{Fig: zigzag}(c); the antisymmetric band modulation is expected. 

Such a symmetry argument can be seen by a simple single-orbital Hamiltonian, whose matrix for the basis $\phi=\{\phi_{{\rm A}\uparrow},\phi_{{\rm A}\downarrow},\phi_{{\rm B}\uparrow},\phi_{{\rm B}\downarrow} \}$ is given by 
\begin{align}
H_{k} 
&= 
\left( 
\begin{array}{cccc}
A_k   + h & 0  & B_k & 0 \\
0   & A_{-k}+z h & 0   & B_k \\
B_k  & 0 & A_{-k} -h & 0 \\
0  & B_k & 0  & A_k - z h
\end{array} \right),   
\label{eq:Txext_cluster}
\end{align}
where $z=+1 (-1)$ for the staggered charge (spin) ordering, and $A_k=2t_2 \cos 2 k_x  + 2\alpha \sin 2 k_x$ and $B_k=2 t_1 \cos k_x$; $t_1$ represents the hopping between A and B sublattices, $t_2$ represents one between the same sublattice, and $\alpha$ represents the magnitude of the ASOC; the distance between the nearest-neighbor A sublattice is set as 2. 

By diagonalizing $H_{k}$, the eigenvalues for up and down spins are obtained as 
\begin{align}
\label{eq: zigzag}
\varepsilon_{\uparrow} (k_x)&=2 t_2 \cos 2 k_x\pm \sqrt{C_k^2+4 \alpha h \sin 2 k_x}, \nonumber \\
\varepsilon_{\downarrow} (k_x)&=2 t_2 \cos 2 k_x\pm \sqrt{C_k^2-4 z\alpha h \sin 2 k_x},
\end{align}
where $C_k^2=2 t_1^2+2 \alpha^2+ h^2+2 t_1^2 \cos 2 k_x-2 \alpha^2 \cos 4 k_x$. 
By taking the limit of $k_x \to 0$ for the second term for simplicity, Eq.~(\ref{eq: zigzag}) turns into 
\begin{align}
\varepsilon_{\uparrow} (k_x) &\sim \pm \sqrt{h^2 + 4 t_1^2} \pm \frac{4  \alpha h}{\sqrt{h^2+4 t_1^2}}k_x, \nonumber \\
\varepsilon_{\downarrow} (k_x) &\sim \pm \sqrt{h^2 + 4 t_1^2}  \mp z\frac{4  \alpha h}{\sqrt{h^2+4 t_1^2}}k_x. 
\end{align}
These results clearly indicate that the staggered charge (spin) ordering leads to spin-dependent (spin-independent) antisymmetric band modulation: The former corresponds to the antisymmetric spin splitting, while the latter corresponds to the antisymmetric band deformation. 
In addition, one finds that the ASOC, $\alpha$, is necessary to cause such antisymmetric band modulations.

\subsubsection{Without relativistic spin-orbit coupling}

Although it has been recognized that the large spin-orbit coupling is important to induce the spin splitting as discussed above, recent studies have clarified that momentum-dependent ``symmetric" spin splitting emerges in AFMs with a collinear spin texture even when the effect of the spin-orbit coupling is negligibly small. 
The pioneering examples are the oxides MnO$_2$~\cite{noda2016momentum}, Ln$M$O$_3$ ($M=$Cr, Mn, Fe)~\cite{okugawa2018weakly}, and RuO$_2$~\cite{Ahn_PhysRevB.99.184432}, the organic conductor $\kappa$-(BETD-TTF)$_2$Cu[N(CN)$_2$]Cl~\cite{naka2019spin,hayami2019momentum,Hayami2020b,Naka_PhysRevB.102.075112}, the fluoride MnF$_2$~\cite{Yuan_PhysRevB.102.014422}, and MnTe~\cite{Lovesey_PhysRevB.108.174437, Mazin_PhysRevB.107.L100418, Gonzalez_PhysRevLett.130.036702, aoyama2023piezomagnetic, osumi2023observation, hariki2023x}. 
Subsequently, some authors often refer to this magnetism as ``altermagnetism"~\cite{Smejkal_PhysRevX.12.031042}. 
In addition, the emergence of the ``antisymmetric" spin splitting without the spin-orbit coupling has also been elucidated in noncollinear/noncoplanar AFMs, such as Ba$_{3}$MnNb$_{2}$O$_{9}$~\cite{Hayami_PhysRevB.101.220403, Hayami_PhysRevB.105.024413, Hayami_PhysRevB.108.094416}. 
To this date, the conditions and mechanisms of these unconventional spin splittings have been understood from the viewpoints of symmetry~\cite{hayami2019momentum, Hayami_PhysRevB.102.144441,egorov2021colossal, Yuan_PhysRevB.103.224410, Yuan_PhysRevMaterials.5.014409} and multipole~\cite{hayami2019momentum, Hayami_PhysRevB.101.220403, Hayami_PhysRevB.102.144441}. 
In particular, the use of the symmetry-adapted multipole basis allows us to systematically investigate the conditions for the spin splitting in terms of the electronic internal degrees of freedom and model parameters (e.g., which hopping and electron-electron interactions are important) irrespective of lattice structures and electron wave functions. 
We briefly introduce the mechanisms of both symmetric and antisymmetric spin splittings based on the symmetry-adapted multipole basis~\cite{hayami2019momentum, Hayami_PhysRevB.101.220403, Hayami_PhysRevB.102.144441}. 

\begin{figure}[t!]
\begin{center}
\includegraphics[width=1.0 \hsize]{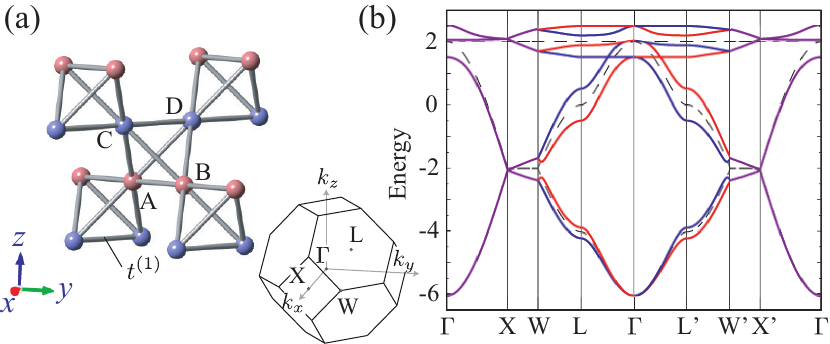}
\caption{
\label{Fig: spin_split_sym}
(a) Collinear AFM patterns in the pyrochlore structure~\cite{hayami2019momentum}. 
The red and blue spheres represent the opposite spin alignment, in which the spin axis is chosen arbitrarily due to a lack of relativistic spin-orbit coupling. 
(b) The corresponding band structures with the nearest-neighbor hopping $t^{(1)}=-1$ and the magnitude of the AFM mean-field $m_{xy}=0.5$.
The red (blue) lines show the up-(down-)spin bands. 
The dashed lines show spin-degenerate bands with $m_{xy}=0$. 
In the first Brillouin zone in (b), the prime symbols are related with $(k_x, k_y, k_z) \to (-k_x, k_y, -k_z)$.
Reprinted figure with permission from Ref.~\citen{hayami2019momentum}, Copyright (2019) by the Physical Society of Japan.
}
\end{center}
\end{figure}

First, let us discuss the symmetric spin splitting under the collinear AFM ordering~\cite{hayami2019momentum}. 
Considering a single-orbital tight-binding model for the four-sublattice pyrochlore structure, the Hamiltonian is given by 
\begin{align}
\mathcal{H}_{0}&=\sum_{\bm{k}\sigma}\sum_{ij}c_{\bm{k}i\sigma}^{\dagger}H_{t}^{ij}c_{\bm{k}j\sigma}^{},
\cr \quad&
H_{t}=2 t^{(1)}
\begin{pmatrix}
0 & c_{xy}^{+} & c_{zx}^{+} & c_{yz}^{+} \\
c_{xy}^{+} & 0 & c_{yz}^{-} & c_{zx}^{-} \\
c_{zx}^{+} & c_{yz}^{-} & 0 & c_{xy}^{-} \\
c_{yz}^{+} & c_{zx}^{-} & c_{xy}^{-} & 0
\end{pmatrix}
\quad
\begin{matrix} {\rm A} \\ {\rm B} \\ {\rm C} \\ {\rm D} \end{matrix}
\label{eq:HamtPyrochlore}
\end{align}
where $c^{\dagger}_{\bm{k}i\sigma}$ ($c_{\bm{k}i\sigma}^{}$) is the creation (annihilation) operator for wave vector $\bm{k}$, sublattice $i=$A-D, and spin $\sigma=\uparrow, \downarrow$, and $c_{\mu\nu}^{\pm}=\cos[(k_{\mu}\pm k_{\nu})a/4]$ for $\mu,\nu=x,y,z$.
Besides, the mean-field Hamiltonian to induce the collinear AFM ordering in Fig.~\ref{Fig: spin_split_sym}(a) is introduced as 
\begin{eqnarray}
H_m=  m_{xy}\rho_0\tau_{z} \sigma,   
\label{eq:HamMF}
\end{eqnarray}
where the product of two Pauli matrices $\rho_\mu$ and $\tau_\nu$ represent the four sublattice degrees of freedom, i.e., A-B and C-D, or (AB)-(CD) space, respectively. 
Owing to the collinear spin structure without the spin-orbit coupling, the spin axis can be taken arbitrarily without loss of generality.

\begin{table}[t!]
\caption{
Multipoles classified by $O_{\rm h}$ ($T_{\rm d}$) symmetry of the pyrochlore structure (tetrahedron unit)~\cite{hayami2019momentum}.
The superscript represents the time-reversal parity. 
$c_{\mu \nu}=\cos (k_\mu a/4) \cos (k_\nu a/4)$, $s_{\mu \nu}=\sin (k_\mu a/4) \sin (k_\nu a/4)$ for $\mu,\nu=x,y,z$, and $c_{r}=c_{yz}+c_{zx}+c_{xy}$.
This table is taken and modified from Ref.~\citen{hayami2019momentum}. 
}
\label{tab_multipoles1}
\centering
\scalebox{0.9}{
\begin{tabular}{ccccc} \hline\hline
irrep. & type & $\mathbb{Q}^{({\rm s})}_{l,m}$ & $\mathbb{Q}^{({\rm b})}_{l,m}$ & $\mathbb{Q}_{l,m}(\bm{k})$ \\ \hline
${\rm A}_{1g}^{+}$ (${\rm A}_{1}^{+}$) & $Q_{0}$ & $1$ & $\rho_{x}+\tau_{x}+\rho_{x}\tau_{x}$ & $\frac{2}{3}c_{r}$ \\ \hline
${\rm E}_g^{+}$ (${\rm E}^{+}$) & $Q_{u}$ & & $\tau_{x}-2\rho_{x}+\rho_{x}\tau_{x}$ & $(\frac{1}{3}c_{r}-c_{xy})$ \\
& $Q_{v}$ & & $\tau_{x}-\rho_{x}\tau_{x}$ & $(c_{zx}-c_{yz})$ \\ \hline
${\rm T}_{2g}^{+}$ (${\rm T}_{2}^{+}$) & $Q_{yz}$ & $\rho_{z}\tau_{z}$ & $-\rho_{y}\tau_{y}$ & $-2s_{yz}$ \\
& $Q_{zx}$ & $\rho_{z}$ & $\rho_{z}\tau_{x}$ & $-2s_{zx}$ \\
& $Q_{xy}$ & $\tau_{z}$ & $\rho_{x}\tau_{z}$ & $-2s_{xy}$ \\ \hline\hline
\end{tabular}
}
\end{table}

As shown in Fig.~\ref{Fig: spin_split_sym}(b), the model under the collinear AFM ordering for $m_{xy}\ne 0$ exhibits the symmetric spin splitting in the band structure; its functional form is represented by $k_x k_y \sigma$.  
The origin of the spin splitting becomes transparent if one expresses the Hamiltonians in Eqs.~(\ref{eq:HamtPyrochlore}) and (\ref{eq:HamMF}) in terms of symmetry-adapted multipole basis; 
$H_{t}$ and $H_{m}$ are rewritten as
\begin{align}
&\frac{H_{t}}{t^{(1)}}=
\mathbb{Q}_{0}^{({\rm b})}\mathbb{Q}_{0}(\bm{k})+\left[\mathbb{Q}_{u}^{({\rm b})}\mathbb{Q}_{u}(\bm{k})+\mathbb{Q}_{v}^{({\rm b})}\mathbb{Q}_{v}(\bm{k})\right]
\cr&\quad \quad
+\left[\mathbb{Q}_{yz}^{({\rm b})}\mathbb{Q}_{yz}(\bm{k})+\mathbb{Q}_{zx}^{({\rm b})}\mathbb{Q}_{zx}(\bm{k})+\mathbb{Q}_{xy}^{({\rm b})}\mathbb{Q}_{xy}(\bm{k})\right],
\cr&
H_{m}= m_{xy}\sigma\mathbb{Q}_{xy}^{\rm (s)}, 
\label{eq:hammmul}
\end{align}
where the multipoles are defined as shown in Table~\ref{tab_multipoles1}.  
It is noted that we express the mean-field Hamiltonian as the product of spin and electric multipoles rather than higher-rank magnetic-type multipoles, since it is convenient to use the representation decoupling spin and orbital degrees of freedom in the absence of the spin-orbit coupling, which is related to the concept of spin group, in the following analysis.

Since the same irreducible representations are coupled with each other, $m_{xy}$ term induces $\mathbb{Q}^{\rm (b)}_{xy}$ and $\mathbb{Q}_{xy}(\bm{k})$ as well, which leads to the spin splitting. 
In order to derive such an effective coupling, the following quantity at wave vector $\bm{k}$ in the magnetic unit cell is introduced~\cite{Hayami_PhysRevB.101.220403}, 
\begin{align}
\label{eq:expec_spin}
\mathrm{Tr}[e^{-\beta H_{\bm{k}}} \sigma_\mu]
= \sum_{s} \frac{(-\beta)^{s}}{s!}
g_s^{\mu}(\bm{k}), 
\end{align} 
where $\mu=0,x,y,z$, $\mathcal{H}=\sum_{\bm{k}}H_{\bm{k}}$ and $\beta$ is the inverse temperature.
By means of a sort of high-temperature expansion, the $s$th-order expansion coefficient of the $\mu$-component, $g_s^{\mu}(\bm{k})$, gives the corresponding effective multipole coupling as $g_{s}^{\mu}(\bm{k})\sigma_{\mu}/2$.
In the above pyrochlore case, the effective Hamiltonian in the lowest order is given by 
\begin{align}
&
H_{\rm eff}(\bm{k})=\frac{1}{2}g_{3}^{z}(\bm{k})\sigma,
\cr&\quad
g_3^{\sigma}(\bm{k})={\rm Tr}[(H_t+H_m)^3 \sigma]
\cr&\hspace{1.2cm}
= 48 m_{xy}[t^{(1)}]^{2} \mathbb{Q}_{xy}(\bm{k}) [\mathbb{Q}_{0}(\bm{k})-2 \mathbb{Q}_{u}(\bm{k})]
\cr&\hspace{1.2cm}
\underset{(\bm{k}\to \bm{0})}{\to} \propto m_{xy}[t^{(1)}]^{2} k_x k_y. 
\end{align}
Thus, the spin splitting is characterized by $\mathbb{Q}_{xy}(\bm{k}) \sigma$ in the ${\rm T}_{2g}$ irreducible representation.

In terms of $g_s^{\mu}(\bm{k})$, the conditions for the symmetric spin splitting are summarized as follows~\cite{Hayami_PhysRevB.102.144441}: 
\begin{enumerate}
\item Bond E multipoles or even number of bond MT multipoles are involved. 
\item Odd number of cluster E multipoles are involved.  
\item Trace of the sublattice degree of freedom (product of cluster multipoles) remains finite. 
\end{enumerate}
According to the rank of the relevant E multipoles, the momentum dependence of symmetric spin-split band structures is different; the E quadrupole induces the $d$-wave spin splitting, the E hexadecapole induces the $g$-wave spin splitting, the E tetrahexacontapole induces the $i$-wave spin splitting, and so on.

\begin{figure}[htb!]
\begin{center}
\includegraphics[width=1.0 \hsize]{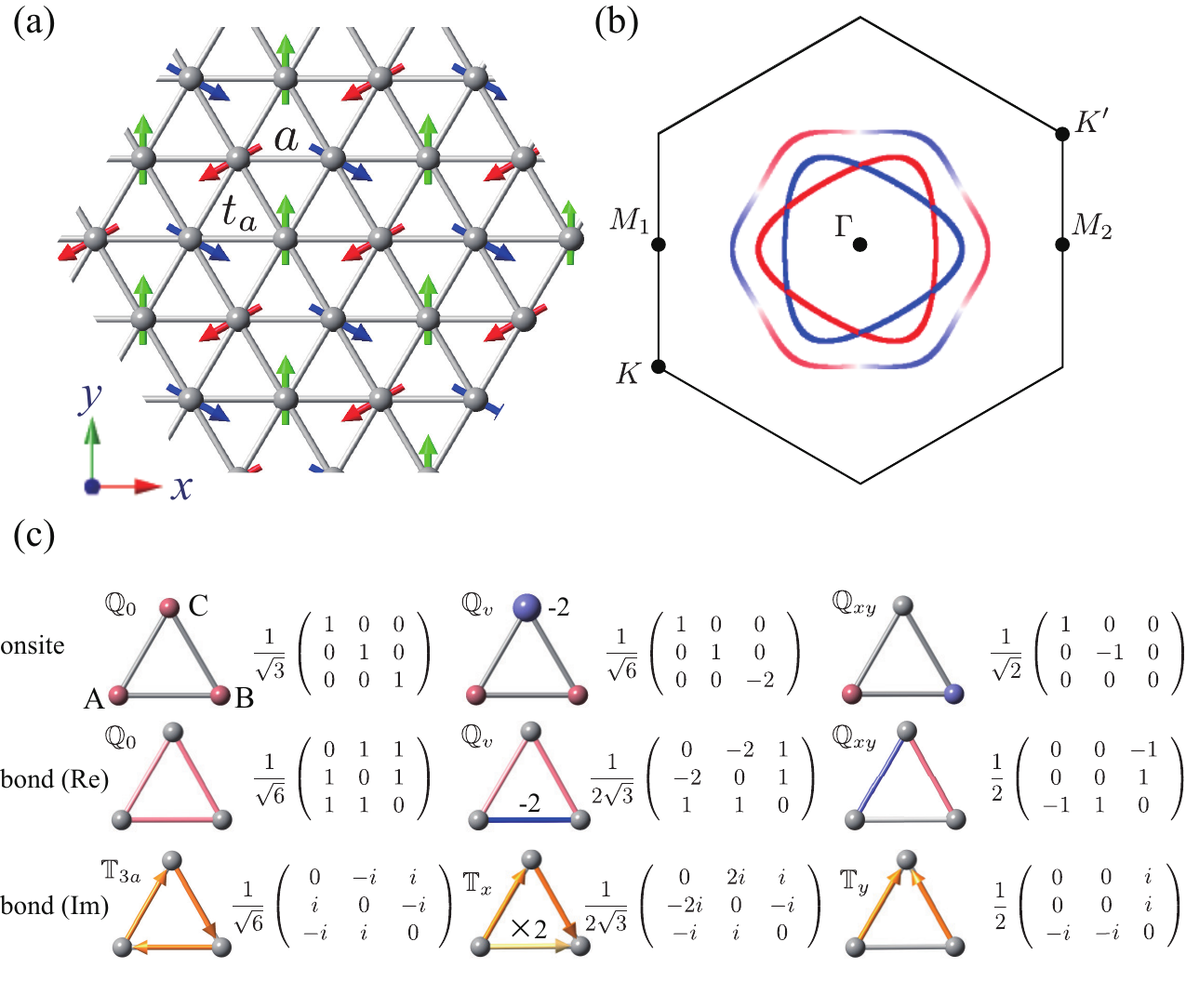}
\caption{
\label{Fig: spin_split_asym}
(a) The 120$^{\circ}$ AFM ordering on a triangular lattice~\cite{Hayami_PhysRevB.101.220403, Hayami_PhysRevB.102.144441}.
(b) The corresponding isoenergy surfaces where the contour shows the $z$-spin component. The model parameters are given by $t_a=1$, $m=0.5$, and $\mu=-2.5$. 
(c) Cluster and bond multipoles in a triangle cluster. 
The correspondence between multipoles and matrix elements is shown. 
The red (blue) circles represent the positive (negative) onsite potential, and the red (blue) lines and orange arrows on each bond represent the positive (negative) real and imaginary hoppings, respectively. 
The gray lines represent no hoppings. 
This figure is taken and modified from Ref.~\citen{Hayami_PhysRevB.101.220403}.
}
\end{center}
\end{figure}

Next, let us show the antisymmetric spin splitting under noncollinear magnetic orderings. 
We show an example of noncollinear 120$^{\circ}$ AFM ordering on a triangular lattice with the lattice constant $a$, as shown in Fig.~\ref{Fig: spin_split_asym}(a); the spin moments lie on the $xy$ plane. 
Each site consists of a single orbital, and the hopping between the nearest-neighbor sites is considered. 
The energy contour in the band structure is shown in Fig.~\ref{Fig: spin_split_asym}(b), where the results clearly show the antisymmetric spin splitting along the M$_1$-$\Gamma$-M$_2$ line in terms of the $z$-spin component, while there is no spin splitting along the K-$\Gamma$-K$'$ line; the functional form of the antisymmetric spin splitting is characterized by $k_x\left(k_x^2-3 k_y^2\right)$. 

The microscopic origin of the antisymmetric spin splitting is also understood from the symmetry-adapted multipole basis. 
The matrices of the hopping and mean-field Hamiltonians in the three-sublattice triangular system are given by 
\begin{align}
&H_{t}=t_{a}\left[
\mathbb{Q}^{\rm (b)}_{0}\mathbb{Q}_{0}(\bm{k})+\mathbb{T}^{\rm (b)}_{3a}\mathbb{T}_{3a}(\bm{k})\right],
\cr&
H_{m}=-m (\sigma_x\mathbb{Q}^{\rm (s)}_{xy}  +\sigma_y\mathbb{Q}^{\rm (s)}_{v} ), 
\label{eq:ham_TL}
\end{align}
where $\mathbb{Q}^{\rm (s, b)}$ and $\mathbb{T}^{\rm (b)}$ are the $3 \times 3$ matrices defined in a triangle cluster in Fig.~\ref{Fig: spin_split_asym}(c), and the form factor is given by 
\begin{align}
\mathbb{Q}_{0}(\bm{k})&=\sqrt{6}(\cos k_{x}a+2\cos\tilde{k}_{x}a\cos\tilde{k}_{y}a), \nonumber \\
\mathbb{T}_{3a}(\bm{k})&=-\sqrt{6}(\sin k_xa-2\sin\tilde{k}_{x}a\cos\tilde{k}_{y}a), 
\end{align}
with $\tilde{k}_{x}=k_{x}/2$ and $\tilde{k}_{y}=\sqrt{3}k_{y}/2$. 
The appearance of the MT octupole degrees of freedom, $\mathbb{T}^{\rm (b)}_{3a}$, in the hopping Hamiltonian $H_{t}$ is owing to the lack of local inversion symmetry under the three-sublattice ordering, which plays an important role in the emergent antisymmetric spin splitting as a result of the coupling with $H_m$. 
The mean-field Hamiltonian $H_m$ consists of two spin components to express the $120^{\circ}$ noncollinear magnetic order with the amplitude $m$.

As similar to the symmetric spin splitting, $g_s^{\mu}(\bm{k})$ gives information about the microscopic coupling between multipole degrees of freedom. 
The lowest-order contribution in $g^{z}_s(\bm{k}) $ is given by 
\begin{align}
\label{eq:spinsplit_asym_TL}
g^{z}_5(\bm{k}) =
-\frac{1}{3} \sqrt{\frac{2}{3}} m^2t_{a}^{3} \mathbb{T}_{3a}(\bm{k}) \left[ (\mathbb{T}_{3a}(\bm{k}))^2-3  (\mathbb{Q}_0(\bm{k}))^2\right].   
\end{align}
As $\mathbb{T}_{3a}(\bm{k}) \propto k_x(k_x^2-3k_y^2)$ and $\mathbb{Q}_0(\bm{k}) \propto 1$ in the $\bm{k}\to \bm{0}$ limit, the essential anisotropy is given by 
\begin{align}
\label{eq:spinsplit_asym_TL2}
g^{z}_5(\bm{k}) &\simeq
 \sqrt{\frac{2}{3}} m^2t_{a}^{3} [\mathbb{Q}_0(\bm{k})]^2 \mathbb{T}_{3a}(\bm{k})  \nonumber \\
 &=24 m^2 t_a^3 \sin \tilde{k}_xa  \left(\cos \tilde{k}_ya -\cos \tilde{k}_xa \right) \nonumber \\
 &  \ \ \  \times \left(2 \cos \tilde{k}_xa \cos \tilde{k}_ya+\cos k_xa\right)^2 \nonumber \\
 &\simeq  \frac{27}{2} m^2 t_a^3  k_x \left(k_x^2-3 k_y^2\right)a^3. 
\end{align} 
Thus, the result in terms of the $\bm{k}$ dependence of the spin splitting is consistent with the numerical result in Fig.~\ref{Fig: spin_split_asym}(b). 
Moreover, the essential model parameters for the spin splitting are obtained; the expression in Eq.~(\ref{eq:spinsplit_asym_TL2}) contains the product of the even number of order parameters as $m^2$ and the form factor of the bond MT multipole $\mathbb{T}_{3a}(\bm{k})$.

From a general aspect, the conditions for the antisymmetric spin splitting with nonzero $g_s^{\mu}(\bm{k})$ are summarized as follows~\cite{Hayami_PhysRevB.102.144441}: 
\begin{enumerate}
\item Odd number of bond MT multipoles are involved. 
\item At least, two spin components leading to noncollinear coplanar spin textures are involved. 
\item Trace of the sublattice degree of freedom (product of cluster multipoles) remains finite. 
\end{enumerate}
According to the rank of the relevant MT multipoles, the momentum dependence of antisymmetric spin-split band structures is different; the MT dipole induces the $p$-wave spin splitting, the MT octupole induces the $f$-wave spin splitting, the MT dotriacontapole induces the $h$-wave spin splitting, and so on.

\begin{figure}[t!]
\begin{center}
\includegraphics[width=1.0 \hsize]{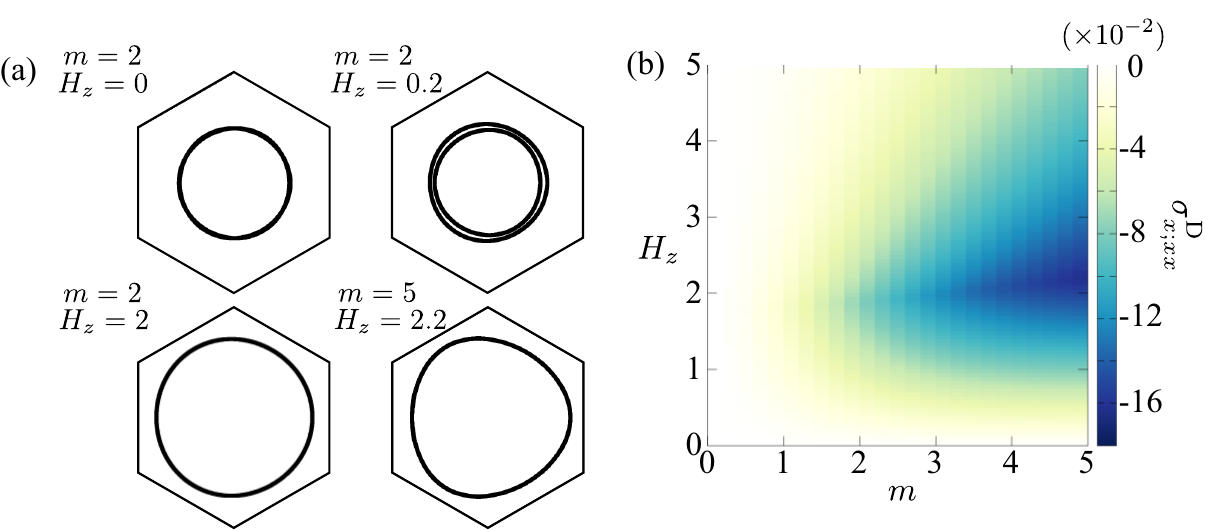}
\caption{
\label{Fig: nonreciprocal}
(a) Fermi surfaces for both spin components at 1/10 filling for several $m$ and $H_z$~\cite{Hayami_PhysRevB.106.014420}. 
(b) Contour plots of nonlinear conductivity $\sigma^{\rm D}_{x;xx}$ in the plane of $m$ and $H_z$ at 1/10 filling. 
Reprinted figure with permission from Ref.~\citen{Hayami_PhysRevB.106.014420}, Copyright (2022) by the American Physical Society.
}
\end{center}
\end{figure}

The above antisymmetric spin splitting in the form of $k_x\left(k_x^2-3 k_y^2\right) \sigma_z$ indicates that antisymmetric band deformation is expected in the presence of the $z$-spin component, e.g., by applying external magnetic field along $z$ direction~\cite{Hayami_PhysRevB.101.220403, Hayami_PhysRevB.102.144441}. 
Figure~\ref{Fig: nonreciprocal}(a) shows the band structure in the 120$^{\rm \circ}$ AFM ordering in the presence of the Zeeman coupling through an external magnetic field along the out-of-plane direction, where the spin polarization at each $\bm{k}$ is not indicated explicitly; 
the band is asymmetrically deformed in the form of $k_x\left(k_x^2-3 k_y^2\right)$. 
A similar situation also occurs in the noncoplanar AFM without the uniform magnetization~\cite{Hayami_doi:10.7566/JPSJ.91.094704}.
In terms of $g_s^{\mu}(\bm{k})$, the condition for the asymmetric band modulation without the spin-orbit coupling is summarized as~\cite{Hayami_PhysRevB.102.144441} 
\begin{enumerate}
\item Odd number of bond MT multipoles are involved. 
\item Three spin components, which are necessary to represent noncoplanar spin structures, are involved.  
\item Trace of the sublattice and spin degrees of freedom remains finite. 
\end{enumerate}
According to the rank of the relevant MT multipoles, the momentum dependence of antisymmetric band deformation is different; the MT dipole induces the $p$-wave band deformation, the MT octupole induces the $f$-wave band deformation, the MT dotriacontapole induces the $h$-wave band deformation, and so on.

This antisymmetric band deformation is the origin of the nonlinear nonreciprocal conductivity $\sigma^{\rm D}_{x;xx}$ in $J_{x}=\sigma^{\rm D}_{x;xx} E^2_x$, where $J_x$ and $E_x$ are the electric current and external electric field, respectively, as shown in Fig.~\ref{Fig: nonreciprocal}(b)~\cite{Hayami_PhysRevB.106.014420}. 
Thus, the AFM ordering with a noncoplanar spin texture can give rise to the antisymmetric band deformation even without the relativistic spin-orbit coupling~\cite{Hayami_doi:10.7566/JPSJ.91.094704, hayami2021phase, Hayami_PhysRevResearch.3.043158}.

\section{Cross correlations}
\label{sec: Cross-correlation phenomena}

\begin{table}[htb!]
\centering
\caption{
Correspondence of (upper panel) the external fields and (lower panel) the responses to the multipoles~\cite{Yatsushiro_PhysRevB.104.054412}. 
The spatial inversion and time-reversal parities of the external field or the response are shown in the columns of $\mathcal{P}$ and $\mathcal{T}$, respectively.
In the column of multipole, $X_{l,m}$ ($l=0$--$3$) means the rank-$l$ multipole ($X=Q,M,T,G$).
 \label{table:field}}
 \scalebox{0.75}{
\begin{tabular}{ccccc}
\hline \hline
max. rank & $\mathcal{P}$ & $\mathcal{T}$ & external field &  multipole \\ \hline
 0  & $+$  & $+$  & $\bm{\nabla}\cdot \bm{E}$ & $Q_0$\\
 & $-$ & $-$ & $\bm{E}\cdot \bm{B}$  & $M_0$\\
   & $-$ & $+$  & $\bm{E}\cdot (\bm{\nabla}\times \bm{E})$, $\bm{B}\cdot (\bm{\nabla}\times \bm{B})$ & $G_{0}$\\
     & $+$ & $-$ & $\bm{B}\cdot (\bm{\nabla}\times \bm{E})$, $\bm{E}\cdot (\bm{\nabla}\times \bm{B})$ & $T_{0}$\\
 \hline
1 & $-$ & $+$ & electric field ${\bm E}$ & $Q_{1,m}$\\
&  & & thermal gradient $-\bm{\nabla} T$ & $Q_{1,m}$\\
 & $+$ & $-$ & magnetic field ${\bm H}$ &  $M_{1,m}$\\
 & $-$ & $-$  & $\bm{\nabla}\times \bm{B}$, $\displaystyle \frac{\partial{\bm{E}}}{\partial t}$ & $T_{1,m}$\\
 & $+$ & $+$  & $\bm{\nabla}\times \bm{E}$, $\displaystyle \frac{\partial{\bm{B}}}{\partial t}$ & $G_{1,m}$\\
 \hline
  2
& $+$ & $+$ & stress $\tau_{ij}$ & $Q_{0}$, $Q_{2,m}$\\
& $+$ & $+$ & nonlinear field $E_i E_j$ ($H_i H_j$) & $Q_{0}$, $Q_{2,m}$\\
& $-$ & $-$ &composite field $E_i H_j$ & $M_{0}$, $T_{1,m}$, $M_{2,m}$\\
\hline
  3
& $-$ & $+$ &nonlinear field $E_i E_j E_k$ & $Q_{1,m}$, $Q_{3,m}$\\
& $+$ & $-$ &nonlinear field $H_i H_j H_k$ & $M_{1,m}$, $M_{3,m}$\\
\hline\hline
max. rank & $\mathcal{P}$ & $\mathcal{T}$ & response &  multipole \\ \hline
 0 & $+$ & $+$ & temperature change $\Delta T$ & $Q_0$\\
   & $-$ & $+$ & chirality & $G_{0}$\\
 \hline
1  & $-$  & $+$  & electric polarization ${\bm P}$ & $Q_{1,m}$\\
  & $+$ & $-$ & magnetization ${\bm M}$ &  $M_{1,m}$\\
 & $-$ & $-$ & electric (thermal) current  ${\bm J}$(${\bm J}^{\rm Q}$) & $T_{1,m}$\\
 &$+$ & $+$ & rotational distortion ${\bm \omega}$ & $G_{1,m}$\\
 \hline
  2
& $+$ & $+$ & strain $\bm{\varepsilon}$ & $Q_{0}$, $Q_{2,m}$\\
& $-$ & $+$ & spin current $J_{i}\sigma_{j}$ & $G_0$, $Q_{1,m}$, $G_{2,m}$\\
\hline\hline
\end{tabular}
}
\end{table}

The symmetry-adapted multipole is also useful to predict and understand cross correlations in materials, since it contains all the information about the time-reversal symmetry and crystallographic point group symmetry of the system. 
For example, every external field and response has a correspondence to symmetry-adapted multipoles from a symmetry viewpoint, as shown in Table~\ref{table:field}. 
Moreover, since any physical tensors are defined by a tensor connecting input and output physical quantities, their component is also expressed by using multipoles. 
In other words, there is a one-to-one correspondence between physical tensors and multipoles. 
In this section, we introduce the multipole representation for cross correlations.

\subsection{Coupling between multipoles}
\label{sec: Coupling between multipoles}

\begin{table}[h!]
\centering
\caption{
Couplings between multipoles (right column) in the presence of monopole ($X_0=Q_0, M_0, T_0, G_0$) and dipole ($\bm{X}=\bm{Q}, \bm{M}, \bm{T}, \bm{G}$) (left column). 
}
\label{table: coupling_mp}
\begin{tabular}{ccccccccccccc}
\hline\hline
multipole & coupling   \\ \hline
$Q_0$ & $X_0 X'_0$, $\bm{X} \cdot \bm{X}'$  \\
$M_0$  & $T_0 G_0$, $\bm{G}\cdot \bm{T}$, $\bm{M}\cdot \bm{Q}$ \\
$T_0$  & $G_0 M_0$, $\bm{G}\cdot \bm{M}$, $\bm{T}\cdot \bm{Q}$  \\
$G_0$ &  $M_0 T_0$, $\bm{Q}\cdot \bm{G}$, $\bm{T}\cdot \bm{M}$ \\
$\bm{Q}$ & $M_0\bm{M}$, $T_0\bm{T}$, $G_0\bm{G}$, $\bm{G}\times \bm{Q}$, $\bm{M}\times \bm{T}$\\
$\bm{M}$ & $M_0\bm{Q}$, $T_0\bm{G}$, $G_0\bm{T}$, $\bm{G}\times \bm{M}$, $\bm{T}\times \bm{Q}$ \\
$\bm{T}$  & $M_0\bm{G}$, $T_0\bm{Q}$, $G_0\bm{M}$, $\bm{G}\times \bm{T}$, $\bm{M}\times \bm{Q}$\\
$\bm{G}$  & $M_0\bm{T}$, $T_0\bm{M}$, $G_0\bm{Q}$, $\bm{X} \times \bm{X'}$ \\
\hline\hline
 \end{tabular}
 \end{table}

\begin{figure}[htb!]
\centering
\includegraphics[width=1.0 \hsize]{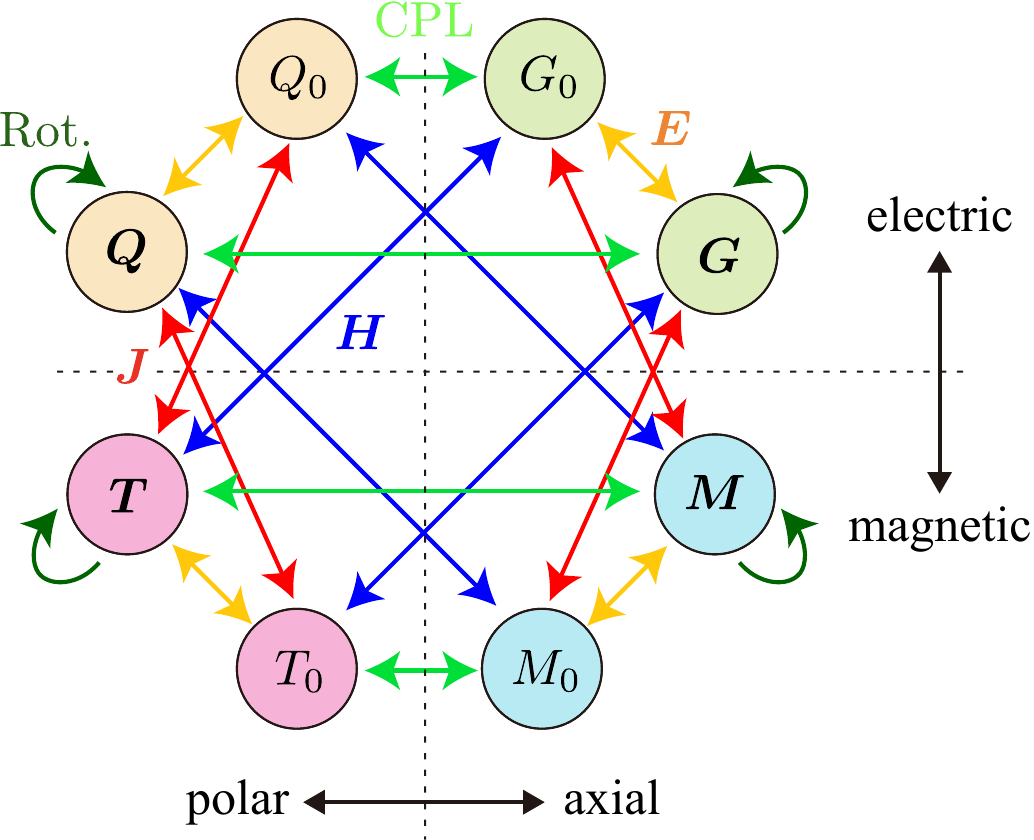} 
\caption{
\label{fig: cross_mp}
Mutual relations between four types of monopole $(Q_0, M_0, T_0, G_0)$ and dipole $(\bm{Q}, \bm{M}, \bm{T}, \bm{G})$, which are connected through external fields. 
$\bm{E}$, $\bm{H}$, $\bm{J}$, CPL, and Rot. stand for the electric field, magnetic field, electric current, circularly-polarized light, and rotational distortion, respectively. 
}
\end{figure}

The cross correlations are intuitively understood from the coupling between multipoles. 
As an example, let us consider the cross correlations triggered by the presence of an ET monopole $G_0$. 
Since $G_0$ corresponds to a rank-0 axial tensor (pseudoscalar) with time-reversal even, it can couple with the inner product of the rank-1 axial tensor with time-reversal even $\bm{G}\equiv (G_{x},G_{y},G_{z})$ and rank-1 polar tensor with time-reversal even $\bm{Q}\equiv (Q_{x},Q_{y},Q_{z})$ as follows: 
\begin{align}
\label{eq: G0_1}
G_0 \leftrightarrow \bm{G}\cdot \bm{Q}. 
\end{align}
From the expression, the cross correlation between the rotational distortion corresponding to $\bm{G}$ and the electric field corresponding to $\bm{Q}$ is expected once the electronic ordering accompanying $G_0$ occurs~\cite{Oiwa_PhysRevLett.129.116401}. 
It is a short-cut argument by using the following free energy,
\begin{multline}
F=\frac{\chi^{-1}}{2}\bm{Q}^{2}+\frac{\kappa^{-1}}{2}\bm{G}^{2}-\bm{Q}\cdot\bm{E}-\bm{G}\cdot(\bm{\nabla}\times\bm{E})
\\
- \alpha G_{0}(\bm{G}\cdot\bm{Q}),
\end{multline}
where $\chi$, $\kappa$, and $\alpha$ are electric susceptibility, ET dipole susceptibility, and coupling constant between the ET monopole and corresponding multipole, respectively.
In the presence of the ET monopole $G_{0}$, the external electric field $\bm{E}$ induces the E dipole $\bm{Q}$, and the ET dipole $\bm{G}$ through the trilinear coupling.
In fact, the equilibrium values for $\bm{Q}$ and $\bm{G}$ can be obtained by minimizing $F$ with respect to them as
\begin{align}
\begin{pmatrix}
\bm{Q} \\ \bm{G}
\end{pmatrix}
=
\frac{\chi\kappa}{1-\chi\kappa(\alpha G_{0})^{2}}
\begin{pmatrix}
\kappa^{-1} & \alpha G_{0} \\
\alpha G_{0} & \chi^{-1}
\end{pmatrix}
\begin{pmatrix}
\bm{E} \\ \bm{\nabla}\times\bm{E}
\end{pmatrix},
\end{align}
which clearly shows that $\bm{E}$ induces $\bm{G}$ in the presence of $G_{0}$.

Similarly, $G_0$ can couple with the inner product of the rank-1 axial tensor with time-reversal odd $\bm{M}\equiv (M_{x},M_{y},M_{z})$ and rank-1 polar tensor with time-reversal odd $\bm{T}\equiv (T_{x},T_{y},T_{z})$ as 
\begin{align}
G_0 \leftrightarrow \bm{M}\cdot \bm{T}. 
\end{align}
Since $\bm{M}$ and $\bm{T}$ correspond to the magnetization and electric current, respectively, the cross correlation between them, i.e., the current-induced magnetization (Edelstein effect), is expected. 
Indeed, these cross correlations were proposed and observed in a chiral crystal Te with $G_0$, as also discussed in Sect.~\ref{sec: Tellurium}~\cite{yoda2015current,furukawa2017observation,yoda2018orbital, Oiwa_PhysRevLett.129.116401}.

\begin{figure}[t!]
\centering
\includegraphics[width=1.0 \hsize]{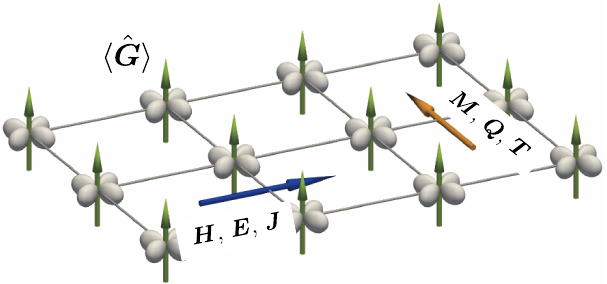} 
\caption{
\label{fig: ETD}
Transverse response in the presence of the ET dipole $\bm{G}$; the input external fields $(\bm{H}, \bm{E}, \bm{J})$ induce conjugate physical quantities $(\bm{M}, \bm{Q}, \bm{T})$ along the perpendicular direction~\cite{Hayami_doi:10.7566/JPSJ.91.113702}.
Reprinted figure with permission from Ref.~\citen{Hayami_doi:10.7566/JPSJ.91.113702}, Copyright (2022) by the Physical Society of Japan.
}
\end{figure}

In a similar manner, one can expect various cross correlations for other multipoles including higher-rank ones, such as a linear magnetoelectric effect in the form of $\bm{M}\times \bm{Q}$ under the MT dipole $\bm{T}$~\cite{popov1999magnetic,schmid2001ferrotoroidics, EdererPhysRevB.76.214404, Spaldin_0953-8984-20-43-434203}, the rotational distortion by an external magnetic field in the form of $\bm{G} \cdot \bm{M}$ in the presence of the MT monopole $T_0$~\cite{Hayami_PhysRevB.108.L140409}, and the current-induced M monopole in the form of $M_0 \bm{T}$ in the presence of the ET dipole $\bm{G}$~\cite{Hayami_PhysRevB.106.144402}. 
We summarize the effective couplings between multipoles up to rank 1 in Table~\ref{table: coupling_mp}. 
We also show a mutual relation between multipoles connected by various external fields, such as electric current, electric field, and magnetic field in Fig.~\ref{fig: cross_mp}. 
It is noteworthy that $\bm{X} \times \bm{X}'$ in the presence of $\bm{G}$ in Table~\ref{table: coupling_mp} means that the transverse responses of the conjugate physical quantities are expected, as shown in Fig.~\ref{fig: ETD}. 
The electric field, magnetic field, and electric current induce $\bm{Q}$, $\bm{M}$, and $\bm{T}$, respectively. 
For example, the electric field along the $x$ direction, $E_x$, induces the spin current along the $y$ direction, $(\bm{J}\times \bm{\sigma})_y$, which corresponds to the longitudinal spin current generation~\cite{Hayami_doi:10.7566/JPSJ.91.113702}.

\subsection{Response tensors}
\label{sec: Response tensors}

Thanks to Neumann's principle, macroscopic physical responses are determined not by the space-group symmetry but by the crystallographic (magnetic) point-group symmetry~\cite{neumann1885vorlesungen,curie1894symetrie}. 
So far, macroscopic responses in materials have been systematically organized by using group theory~\cite{birss1964symmetry,kleiner1966space,kleiner1967space,kleiner1969space,grimmer1993general, Seemann_PhysRevB.92.155138,gallego2019automatic}, although some cautions are required for cross correlations when external fields cause dissipation, such as the case of electric current~\cite{SHTRIKMAN1965147, kleiner1966space, BUTZAL1982518, grimmer1993general, Seemann_PhysRevB.92.155138, Zelezny_PhysRevLett.119.187204, Mook_PhysRevResearch.2.023065}. 
We discuss the correspondence between the response tensor components and multipoles based on the group theory and response theory~\cite{kubo_doi:10.1143/JPSJ.12.570, Watanabe_PhysRevB.96.064432, Hayami_PhysRevB.98.165110, Watanabe_PhysRevX.11.011001, Yatsushiro_PhysRevB.104.054412} in the cases of the linear and second-order responses in Sect.~\ref{sec: Linear response tensor} and \ref{sec: Nonlinear response tensor}, respectively.

\subsubsection{Linear response tensor}
\label{sec: Linear response tensor}

\begin{table*}[htb!]
\centering
\caption{
Correspondence between the linear response functions $\chi^{\rm (J, E)}$ in Eqs.~(\ref{eq:chiJ}) and (\ref{eq:chiE}) or $\chi$ in Eq.~(\ref{eq:susiso}) and multipoles $X_{l,m}$ ($l=1$--$4$, $X=Q,M,T,G$)~\cite{Yatsushiro_PhysRevB.104.054412}. 
In the rightmost column, the parenthesis represents the corresponding multipoles to the response $B$ and the external field $F$. 
\label{table:tensor_parity_MP_linear}}
\begin{tabular}{cccccclccccc}
\hline \hline 
 rank & $\mathcal{P}$ & $\mathcal{T}$ & & multipole &  response tensors \\ \hline 

1 & polar & i & $\chi^{\rm (J)}$ & $T_{1,m}$  & \\
 & & & $\chi^{\rm (E)}$ & $Q_{1,m}$ & pyroelectric $(P_i = \chi_i \Delta T)$ \\ \cline{3-6}
 & & c & $\chi^{\rm (J)}$ & $Q_{1,m}$ &   \\
 & & & $\chi^{\rm (E)}$ & $T_{1,m}$ &  magnetic pyrotoroidic $(T_i = \chi_i \Delta T)$ \\ \cline{2-6}
 
& axial & i & $\chi^{\rm (J)}$ & $M_{1,m}$ & \\
& & & $\chi^{\rm (E)}$ & $G_{1,m}$ & electric pyrotoroidic $(G_i = \chi_i \Delta T)$ \\
\cline{3-6}

& & c & $\chi^{\rm (J)}$ & $G_{1,m}$  & \\
& & & $\chi^{\rm (E)}$ & $M_{1,m}$ & pyromagnetic  $(M_i = \chi_i \Delta T)$\\
\hline

2 & polar & i & $\chi^{\rm (J)}$ & $T_{0}, M_{1,m}, T_{2,m}$ & 
\\
& & & $\chi^{\rm (E)}$ & $Q_{0}, G_{1,m}, Q_{2,m}$ &  magnetic susceptibility $(M_i = \chi_{i;j}H_j)$  
\\
\cline{3-6}

& & c & $\chi^{\rm (J)}$  & $Q_{0}, G_{1,m}, Q_{2,m}$  & electric conductivity $(J_i = \sigma^{\rm (J)}_{i;j}E_j)$ \\ 
& & & $\chi^{\rm (E)}$ & $T_{0}, M_{1,m}, T_{2,m}$&  
electric conductivity $(J_i = \sigma^{\rm (E)}_{i;j}E_j)$ \\ 
\cline{2-6}

 & axial & i & $\chi^{\rm (J)}$ & $M_{0}, T_{1,m}, M_{2,m}$&  \\
 & & & $\chi^{\rm (E)}$ & $G_{0}, Q_{1,m}, G_{2,m}$& electrorotation $(\omega_i = \sigma^{\rm (E)}_{i;j}E_j)$ \\
 \cline{3-6}
 
 & & c & $\chi^{\rm (J)}$ & $G_{0}, Q_{1,m}, G_{2,m}$& Edelstein effect $(M_i = \sigma^{\rm (J)}_{i;j} E_j)$ \\
  & & & $\chi^{\rm (E)}$ & $M_{0}, T_{1,m}, M_{2,m}$&  magnetoelectric $(M_i = \sigma^{\rm (E)}_{i;j}E_j)$\\
 \hline
 
 3 & polar & i & $\chi^{\rm (J)}$ & $T_{1,m}, M_{2,m}, T_{3,m}$ &    \\
 & & & $\chi^{\rm (E)}$ & $Q_{1,m}, G_{2,m}, Q_{3,m}$&  piezoelectric $(P_i = \sigma_{i;jk}\varepsilon_{jk})$  \\
 \cline{3-6}
 
 & & c & $\chi^{\rm (J)}$ & $Q_{1,m}, G_{2,m}, Q_{3,m}$&  \\
 & & & $\chi^{\rm (E)}$ & $T_{1,m}, M_{2,m}, T_{3,m}$ & magnetic piezotoroidal $(T_i = \sigma_{i;jk}\varepsilon_{jk})$ \\ \cline{2-6}

& axial & i & $\chi^{\rm (J)}$ & $M_{1,m}, T_{2,m}, M_{3,m}$&   spin conductivity $(J_i \sigma_j = \sigma^{\rm (J)}_{ij;k}E_{k})$ \\
& & & $\chi^{\rm (E)}$ & $G_{1,m}, Q_{2,m}, G_{3,m}$&  spin conductivity $(J_i \sigma_j = \sigma^{\rm (E)}_{ij;k}E_{k})$ 
\\
\cline{3-6}

& & c & $\chi^{\rm (J)}$ & $G_{1,m}, Q_{2,m}, G_{3,m}$ &  \\
& & & $\chi^{\rm (E)}$ & $M_{1,m}, T_{2,m}, M_{3,m}$&  piezomagnetic $(M_i = \sigma_{i;jk}\varepsilon_{jk})$ \\
\hline

4 & polar & i & $\chi^{\rm (J)}$ & $T_{0}, M_{1,m}, T_{2,m}, M_{3,m}, T_{4,m}$ &   \\

& & & $\chi^{\rm (E)}$  & $Q_{0}, G_{1,m}, Q_{2,m}, G_{3,m}, Q_{4,m}$ & elastic stiffness $(\tau_{ij} = \sigma_{ij;kl}\varepsilon_{kl})$ \\
\cline{3-6}

& & c & $\chi^{\rm (J)}$  &  $Q_{0}, G_{1,m}, Q_{2,m}, G_{3,m}, Q_{4,m}$ &   \\ 

& & & $\chi^{\rm (E)}$& $T_{0}, M_{1,m}, T_{2,m}, M_{3,m}, T_{4,m}$ &  \\ 
\cline{2-6}

& axial & i & $\chi^{\rm (J)}$ & $M_{0}, T_{1,m}, M_{2,m}, T_{3,m}, M_{4,m}$ &   \\

& & & $\chi^{\rm (E)}$ & $G_{0}, Q_{1,m}, G_{2,m}, Q_{3,m}, G_{4,m}$ & piezospincurrent $(J_i \sigma_j = \sigma_{ij;kl}\varepsilon_{kl})$ \\
\cline{3-6}
 
& & c & $\chi^{\rm (J)}$ & $G_{0}, Q_{1,m}, G_{2,m}, Q_{3,m}, G_{4,m}$ &  \\
 
& & & $\chi^{\rm (E)}$ & $M_{0}, T_{1,m}, M_{2,m}, T_{3,m}, M_{4,m}$ & flexomagnetic $(M_{i} = \sigma_{i;jkl}\nabla_j\varepsilon_{kl})$ \\
 
\hline\hline
\end{tabular}
\end{table*}

\begin{table*}[htb!]
\centering
\caption{
Correspondence between the nonlinear response functions $\chi^{\rm (Re), (Im)}$ and multipoles $X_{l,m}$ ($l=1$--$4$, $X=Q,M,T,G$)~\cite{Yatsushiro_PhysRevB.104.054412}. 
In the rightmost column, the parenthesis represents the corresponding multipoles to the response $B$ and the external field $F$. 
\label{table:tensor_parity_MP_nonlinear}}
\begin{tabular}{cccccclccccc}
\hline \hline 
 rank & $\mathcal{P}$ & $\mathcal{T}$ & & multipole &  response tensors \\ \hline 

 3 & polar & i &  $\chi^{\rm (Re)}$ & $Q_{1,m}, G_{2,m}, Q_{3,m}$&  2nd-order magnetoelectric $(P_i = \sigma^{\rm (Re)}_{i;jk}H_j H_k)$   \\
 & & & $\chi^{\rm (Im)}$ & $T_{1,m}, M_{2,m}, T_{3,m}$ & \\
 \cline{3-6}
 
 & & c & $\chi^{\rm (Re)}$ & $T_{1,m}, M_{2,m}, T_{3,m}$ & 2nd-order electric conductivity $(J_i = \sigma^{\rm (Re)}_{i;jk}E_j E_k)$ \\
 & & & $\chi^{\rm (Im)}$ & $Q_{1,m}, G_{2,m}, Q_{3,m}$& 2nd-order electric conductivity $(J_i = \sigma^{\rm (Im)}_{i;jk}E_j E_k)$ \\ \cline{2-6}

& axial & i &  $\chi^{\rm (Re)}$ & $G_{1,m}, Q_{2,m}, G_{3,m}$&   \\
& & & $\chi^{\rm (Im)}$ & $M_{1,m}, T_{2,m}, M_{3,m}$&   \\
\cline{3-6}

& & c & $\chi^{\rm (Re)}$ & $M_{1,m}, T_{2,m}, M_{3,m}$&  2nd-order magnetoelectric $(H_i = \sigma^{\rm (Re)}_{i;jk}E_j E_k)$  \\
& & & $\chi^{\rm (Im)}$ & $G_{1,m}, Q_{2,m}, G_{3,m}$ & \\
\hline

4 & polar & i &  $\chi^{\rm (Re)}$  & $Q_{0}, G_{1,m}, Q_{2,m}, G_{3,m}, Q_{4,m}$ & magnetostriction $(\tau_{ij} = \sigma^{\rm (Re)}_{ij;kl}H_k H_l)$  \\

& & & $\chi^{\rm (Im)}$ & $T_{0}, M_{1,m}, T_{2,m}, M_{3,m}, T_{4,m}$ & \\
\cline{3-6}

& & c &  $\chi^{\rm (Re)}$& $T_{0}, M_{1,m}, T_{2,m}, M_{3,m}, T_{4,m}$ & magnetoelectric gyration $(\Delta g_{ij}= \chi^{\rm (Re)}_{ij;kl} E_k H_l)$  \\ 

& & & $\chi^{\rm (Im)}$  &  $Q_{0}, G_{1,m}, Q_{2,m}, G_{3,m}, Q_{4,m}$ & \\ 
\cline{2-6}

& axial & i &  $\chi^{\rm (Re)}$ & $G_{0}, Q_{1,m}, G_{2,m}, Q_{3,m}, G_{4,m}$ & quadratic gyration $(\Delta g_{ij}= \chi^{\rm (Re)}_{ij;kl} E_k E_l)$  \\

& & & $\chi^{\rm (Im)}$ & $M_{0}, T_{1,m}, M_{2,m}, T_{3,m}, M_{4,m}$ & \\
\cline{3-6}
 
& & c & $\chi^{\rm (Re)}$ & $M_{0}, T_{1,m}, M_{2,m}, T_{3,m}, M_{4,m}$ & piezomagnetoelectric $(P_{i} = \sigma^{\rm (Re)}_{i;jkl}H_j\varepsilon_{kl})$  \\
 
& & & $\chi^{\rm (Im)}$ & $G_{0}, Q_{1,m}, G_{2,m}, Q_{3,m}, G_{4,m}$ & \\
 
\hline\hline
\end{tabular}
\end{table*}

Let us introduce the perturbative Hamiltonian under an external field $F_j(t) = \int_{-\infty}^{\infty} \frac{d\omega}{2\pi} F_{j,\omega} e^{-i\omega t+\delta t}$($\delta>0$) for $j=x,y,z$ as $\mathcal{H}_{\rm ext}=-\sum_j\hat{A}_j F_j(t)$, where $\hat{A}_j$ is a conjugate operator for the external field. 
Then, linear response tensor $\chi_{i; j} (\omega)$ is defined by 
\begin{align}
\braket{\hat{B}_{i,\omega}}
 &=\int_{-\infty}^\infty \frac{d\omega'}{2\pi}\delta(\omega-\omega') \chi_{i;j}(\omega') F _{j,\omega'}, 
 \label{eq:suscep}
\end{align} 
where $\hat{B}_{i,\omega}$ is the operator for the observables. 
Considering the uniform external field with the wave vector $\bm{q} \to \bm{0}$ and then taking the static limit $\omega \to 0$, the linear response function $\chi_{i; j} \equiv \chi_{i; j} (\omega \to 0)$ for the periodic system is represented as
\begin{align}
\label{eq: linear}
\chi_{i; j} = - \frac{i\hbar}{V} \sum_{{\bm k} nm} \frac{f[\varepsilon_n({\bm k})] - f[\varepsilon_m({\bm k})]}{\varepsilon_n({\bm k}) - \varepsilon_m({\bm k})}
\frac{B_{i{\bm k}}^{nm} \dot{A}_{j{\bm k}}^{mn}}{i\hbar\delta + \varepsilon_n({\bm k}) - \varepsilon_m({\bm k})}, 
\end{align}
where $\dot{A}=dA/dt$ and $Z_{{\bm k}}^{nm}\equiv\braket{n{\bm k}|\hat{Z}|m{\bm k}}$ ($Z=A, B$) is the matrix element between the Bloch states $\bra{n{\bm k}}$ and $\ket{m{\bm k}}$ with the band indices $n$ and $m$. 
 $\varepsilon_n({\bm k})$ is the eigenenergy of the Hamiltonian and $f[\varepsilon_n({\bm k})]$ is the Fermi distribution function. 
 $V$, $\hbar$, and $\delta$ are the system volume, the reduced Planck constant, and the broadening factor, respectively. 

Assuming the relaxation-time approximation and mimic the constant $\tau=1/\delta$ as the relaxation time (and we omit higher correction terms of $\delta$), $\chi_{i;j}$ is decomposed as 
\begin{align}
\chi_{i;j} &= \chi_{i;j}^{\rm (J)} + \chi_{i;j}^{\rm (E)},
\\
\label{eq:chiJ}
\chi_{i;j}^{\rm (J)}&= 
-\frac{\tau}{V} \sum_{\bm{k} nm}^{=}
\frac{f[\varepsilon_n(\bm{k})] - f[\varepsilon_m(\bm{k})]}{\varepsilon_n(\bm{k}) - \varepsilon_m(\bm{k})}
B_{i\bm{k}}^{nm} \dot{A}_{j\bm{k}}^{mn},
\\
\label{eq:chiE}
\chi_{i;j}^{\rm (E)}&
=-\frac{i\hbar}{V}\sum_{\bm{k} nm}^{\neq} \frac{f[\varepsilon_n(\bm{k})] - f[\varepsilon_m(\bm{k})]}{[\varepsilon_n(\bm{k}) - \varepsilon_m(\bm{k})]^2} B_{i\bm{k}}^{nm}\dot{A}_{j\bm{k}}^{mn},
\end{align}
where the symbol ``$=$'' in the summation means $\varepsilon_{n}(\bm{k})\to\varepsilon_{m}(\bm{k})$ and $\chi_{i;j}^{\rm (J)}$ includes the intraband (dissipative) contribution proportional to $\tau$, while $\chi_{i;j}^{\rm (E)}$ is the interband (nondissipative) one, which remains finite in the clean limit of $\delta \to 0$.
Thus, $\chi_{i;j}^{\rm (J)}$ is finite only in metals, while $\chi_{i;j}^{\rm (E)}$ has finite contributions both in metals and insulators. 

$\chi^{\rm (J)}$ and $\chi^{\rm (E)}$ have the opposite time-reversal property owing to the presence of $\tau$~\cite{Watanabe_PhysRevB.96.064432, Hayami_PhysRevB.98.165110}. 
They are transformed as 
\begin{align}
\label{eq:TR_linear}
\chi_{i;j}^{\rm (J)} = -t_{B_i}t_{A_j}\chi_{i;j}^{\rm (J)}, \ \chi_{i;j}^{\rm (E)} = t_{B_i}t_{A_j}\chi_{i;j}^{\rm (E)},
\end{align}
where $t_{Z}=\pm 1$ stands for the time-reversal parity for $\hat{Z}$, which satisfies $Z_{\bm k}^{nm}=t_Z Z_{-{\bm k}}^{\bar{m}\bar{n}}$. 
The $\bar{n}$-th band stands for the time-reversal partner of the $n$-th band. 
Equation~\eqref{eq:TR_linear} means that $\chi_{i;j}^{\rm (J)}$ [$\chi_{i;j}^{\rm (E)}$] can be finite when $t_{B_i}t_{A_j}=-1(+1)$. 
In other words, $\chi_{i;j}^{\rm (J)}$ [$\chi_{i;j}^{\rm (E)}$] becomes nonzero when the M and MT (E and ET) multipoles are active for $t_{B_i}t_{A_j}=+1$, while $\chi_{i;j}^{\rm (J)}$ [$\chi_{i;j}^{\rm (E)}$] becomes nonzero when the E and ET (M and MT) multipoles are active for $t_{B_i}t_{A_j}=-1$. 
The multipoles contributing to $\chi_{i;j}^{\rm (J)}$ and $\chi_{i;j}^{\rm (E)}$ are summarized in
Table~\ref{table:tensor_parity_MP_linear}.
Similarly, the linear isothermal susceptibility is given by
\begin{align}
\label{eq:susiso}
\chi_{i;j}&= 
-\frac{1}{V} \sum_{\bm{k} nm}
\frac{f[\varepsilon_n(\bm{k})] - f[\varepsilon_m(\bm{k})]}{\varepsilon_n(\bm{k}) - \varepsilon_m(\bm{k})+i0}
B_{i\bm{k}}^{nm} A_{j\bm{k}}^{mn},
\end{align}
and the second relation in Eq.~(\ref{eq:TR_linear}) holds.

From the symmetry, arbitrary tensor components are related to multipoles. 
For example, we show the correspondence between the 2nd-rank response tensor $\chi_{i;j}^{\rm (2)}$ and multipoles, where $\chi_{i;j}^{\rm (2)}$ is defined by $B_i = \sum_j \chi_{i;j}^{\rm (2)} F_j$ ($B_i$ and $F_j$ are rank-1 tensors). 
$\chi_{i;j}^{\rm (2)}$ is spanned by the multipoles in the following form, 
\begin{align} 
\label{eq:rank11tensor}
\hat{\chi}^{\rm (2)} = 
\begin{pmatrix}
X_{0}-X_{u}+X_{v}& X_{xy}+Y_{z} & X_{zx}-Y_{y} \\
X_{xy}-Y_{z} & X_{0}-X_{u}-X_{v} & X_{yz}+Y_{x} \\
X_{zx}+Y_{y} & X_{yz}-Y_{x} & X_{0}+2X_{u} 
\end{pmatrix},
\end{align}
where $X=Q$ or $T$ ($G$ or $M$) and $Y=G$ or $M$ ($Q$ or $T$) for the polar (axial) 2nd-rank tensor depending on their time-reversal property.
Among 9 multipoles, rank-0 $(X_0)$, rank-1 $(Y_x, Y_y, Y_z)$,
and rank-2 $(X_u, X_v, X_{yz}, X_{zx}, X_{xy})$ multipoles represent the isotropic component, antisymmetric components, and symmetric traceless components, respectively. 
In a similar way, higher-rank response tensors are related to multipoles; see the details in Refs.~\citen{Hayami_PhysRevB.98.165110, Yatsushiro_PhysRevB.104.054412}. 

In the following, we show two fundamental examples of linear-response tensors: one is the electrical conductivity tensor and the other is the magnetoelectric (current) tensor. 

\paragraph{Electric conductivity tensor}
The electric conductivity corresponds to the 2nd-rank polar tensor defined as 
\begin{align}
\bm{J}&=\hat{\sigma}\bm{E}
=\left(\hat{\sigma}^{\rm (J)}+\hat{\sigma}^{\rm (E)}\right) \bm{E},
\end{align}
where $\dot{A}_{i}=B_{i}=J_{i}$ in Eq.~(\ref{eq: linear}). 
From the time-reversal parity of $A_i$ and $B_i$, the breaking of the time-reversal symmetry is necessary for $\hat{\sigma}^{\rm (E)}$, while there is no symmetry restriction for $\hat{\sigma}^{\rm (J)}$.  
In addition, $\hat{\sigma}^{\rm (E)}$ is the antisymmetric tensor $\sigma_{i;j}^{\rm (E)}=-\sigma_{j;i}^{\rm (E)}\equiv\sigma_{i;j}^{\rm (E,A)}$ and $\hat{\sigma}^{\rm (J)}$ is the symmetric tensor $\sigma_{i;j}^{\rm (J)}=\sigma_{j;i}^{\rm (J)}\equiv\sigma_{i;j}^{\rm (J,S)}$ from Eqs.~(\ref{eq:chiJ}) and (\ref{eq:chiE}). 
Accordingly, the corresponding multipole degrees of freedom for $\hat{\sigma}^{\rm (J,S)}$ and $\hat{\sigma}^{\rm (E,A)}$ are
identified as 
\begin{align}
&
\hat{\sigma}^{\rm (J,S)}=
\begin{pmatrix}
Q_{0}-Q_{u}+Q_{v} & Q_{xy} & Q_{zx} \\
Q_{xy} & Q_{0}-Q_{u}-Q_{v} & Q_{yz} \\
Q_{zx} & Q_{yz} & Q_{0}+2Q_{u}
\end{pmatrix}, 
\cr&
\hat{\sigma}^{\rm (E,A)}=
\begin{pmatrix}
0 & M_{z} & -M_{y} \\
-M_{z} & 0 & M_{x} \\
M_{y} & -M_{x} & 0
\end{pmatrix}. 
\end{align}
The result shows that the anomalous Hall conductivity corresponding to the antisymmetric non-dissipative part becomes nonzero in the presence of the M dipole moment, and arises from the interband contribution. 
In other words, the rank-1 quantity should be active when inducing the anomalous Hall effect; the rank-1 anisotropic M dipole in Eqs.~(\ref{eq:AMDx})--(\ref{eq:AMDz}) is an indicator rather than the rank-2 MT quadrupole and the rank-3 M octupole. 
In the metallic case, the inter-band contribution to the anomalous Hall conductivity can be cast into the intra-band contribution by integration by part, as the anomalous Hall conductivity is essentially the Fermi-liquid property~\cite{Haldane_PhysRevLett.93.206602, Wang_PhysRevB.74.195118, Wang_PhysRevB.76.195109}.
On the contrary, in the insulating case, there can exist the intrinsic inter-band contribution that becomes the source of quantum Hall effect.

\paragraph{Magnetoelectric(current) tensor}
The magnetoelectric (current) tensor $\hat{\alpha}$ corresponds to the 2nd-rank axial tensor, which is defined as 
\begin{align}
\bm{M}&=\hat{\alpha}\bm{E}=\left(\hat{\alpha}^{\rm (J)}+\hat{\alpha}^{\rm (E)}\right)\bm{E}.
\end{align}
where $\dot{A}_i = J_i$ and $B_j = M_j$ in Eq.~(\ref{eq: linear}). 
Since $\hat{\alpha}^{\rm (J)}$ becomes nonzero when the spatial inversion symmetry is broken and $\hat{\alpha}^{\rm (E)}$ becomes nonzero when both the time-reversal and spatial inversion symmetries are broken, their tensor components are expressed by multipoles as 
\begin{align}
&
\hat{\alpha}^{\rm (J)}=
\begin{pmatrix}
G_{0}-G_{u}+G_{v} & G_{xy}+Q_{z} & G_{zx}-Q_{y} \\
G_{xy}-Q_{z} & G_{0}-G_{u}-G_{v} & G_{yz}+Q_{x} \\
G_{zx}+Q_{y} & G_{yz}-Q_{x} & G_{0}+2G_{u}
\end{pmatrix},
\cr&
\hat{\alpha}^{\rm (E)}=
\begin{pmatrix}
M_{0}-M_{u}+M_{v} & M_{xy}+T_{z} & M_{zx}-T_{y} \\
M_{xy}-T_{z} & M_{0}-M_{u}-M_{v} & M_{yz}+T_{x} \\
M_{zx}+T_{y} & M_{yz}-T_{x} & M_{0}+2M_{u}
\end{pmatrix}.
\end{align}
The isotropic longitudinal magnetoelectric (current) response is realized in the presence of the M (ET) monopole, the antisymmetric transverse response in the presence of the MT (E) dipole, and the symmetric transverse and traceless longitudinal responses in the presence of the M (ET) quadrupoles.

\subsubsection{Nonlinear response tensor}
\label{sec: Nonlinear response tensor}

By performing a similar approach, the correspondence between the second-order nonlinear response tensor $\chi_{i; jk}$ and multipoles is derived~\cite{kubo_doi:10.1143/JPSJ.12.570, Watanabe_PhysRevX.11.011001,Yatsushiro_PhysRevB.104.054412}. 
Supposing the static limit (${\bm q}\to {\bm 0}$, $\omega \to 0$), $\chi_{i; jk}\equiv \chi_{i; jk} (0,0)$ is given by 
\begin{align}
\label{eq:2nd_c_susceptability_2}
\chi_{i; jk} &= \frac{1}{2} \frac{1}{V} \sum_{\bm k} \sum_{lmn} \frac{B_{i {\bm k}}^{nm}\left(A_{j{\bm k}}^{ml} A_{k {\bm k}}^{ln} +A_{k {\bm k}}^{ml} A_{j {\bm k}}^{ln} \right) }{\varepsilon_n({\bm k}) - \varepsilon_m({\bm k})+2i \hbar\delta }
\nonumber \\ 
&\times \left\{
\frac{f[\varepsilon_n({\bm k})]-f[\varepsilon_l({\bm k})]}{\varepsilon_n({\bm k})-\varepsilon_l({\bm k}) +i\hbar\delta} - \frac{f[\varepsilon_l({\bm k})]-f[\varepsilon_m({\bm k})]}{\varepsilon_l({\bm k})-\varepsilon_m({\bm k}) +i\hbar\delta} 
\right\}.
\end{align}
Similar to the linear response tensor $\chi_{i;j}$, $\chi_{i; jk}$ is decomposed into two parts with different time-reversal properties as 
\begin{align}
\chi_{i; jk} =& \chi_{i;jk}^{\rm (Re)} +  \chi_{i;jk}^{\rm (Im)}, \\
\chi_{i;jk}^{\rm (Re)} \equiv&  
\frac{1}{V} \sum_{{\bm k}lmn}\frac{ {\rm Re}\left(B_{i {\bm k}}^{nm}A_{j{\bm k}}^{ml} A_{k {\bm k}}^{ln}\right)\varepsilon_{nm}({\bm k})}{[\varepsilon_{nm}({\bm k})]^2 + (2\hbar\delta)^2}
\nonumber \\
&\times \left(
\frac{ (f_n-f_l)\varepsilon_{nl}({\bm k})}{[\varepsilon_{nl}({\bm k})]^2 +(\hbar\delta)^2}
 - \frac{(f_l-f_m)\varepsilon_{lm}({\bm k})}{[\varepsilon_{lm}({\bm k})]^2+(\hbar\delta)^2} 
\right) \notag\\
&- \frac{2\hbar^2\delta^2}{V} \sum_{{\bm k} lmn} \frac{{\rm Re}\left(B_{i {\bm k}}^{nm}A_{j{\bm k}}^{ml} A_{k {\bm k}}^{ln}\right) }{[\varepsilon_{nm}({\bm k}) ]^2 + (2 \hbar\delta)^2}
\nonumber \\ 
& \times \left\{
\frac{f_n-f_l}{[\varepsilon_{nl}({\bm k})]^2 +(\hbar\delta)^2} - \frac{f_l-f_m}{[\varepsilon_{lm}({\bm k})]^2 +(\hbar\delta)^2} 
\right\}, \\
\chi_{i;jk}^{\rm (Im)} \equiv& 
\frac{\hbar\delta}{V}
\sum_{{\bm k}lmn}  \frac{{\rm Im}\left(B_{i {\bm k}}^{nm} A_{j{\bm k}}^{ml} A_{k {\bm k}}^{ln} \right)\varepsilon_{nm}({\bm k}) }{[\varepsilon_{nm}({\bm k}) ]^2 + (2 \hbar\delta)^2}
\\ \nonumber 
& \times \left\{
\frac{f_n-f_l}{[\varepsilon_{nl}({\bm k})]^2 +(\hbar\delta)^2} - \frac{f_l-f_m}{[\varepsilon_{lm}({\bm k})]^2 +(\hbar\delta)^2} 
\right\} \notag\\
&+ \frac{2\hbar \delta}{V} \sum_{{\bm k}lmn} \frac{{\rm Im}\left(B_{i {\bm k}}^{nm}A_{j{\bm k}}^{ml} A_{k {\bm k}}^{ln} \right)}{[\varepsilon_{nm}({\bm k}) ]^2 + (2 \hbar\delta)^2}
\nonumber \\
&\times \left(
\frac{(f_n-f_l)\varepsilon_{nl}({\bm k})}{[\varepsilon_{nl}({\bm k})]^2 +(\hbar\delta)^2}
 - \frac{(f_l-f_m)\varepsilon_{lm}({\bm k})}{[\varepsilon_{lm}({\bm k})]^2+(\hbar\delta)^2} 
\right), 
 \end{align}
where $\varepsilon_{nm}({\bm k})= \varepsilon_n({\bm k})- \varepsilon_m({\bm k})$ and $f[\varepsilon_n({\bm k})]= f_n$~\cite{comment_2ndresponse}. 
In contrast to the linear response tensor $\chi_{i;j}$, there are complicated intraband and interband processes in both $\chi_{i;jk}^{\rm (Re)}$ and $\chi_{i;jk}^{\rm (Im)}$.

The following relations 
\begin{align}
\label{eq:nonlinear_t}
\chi_{i;jk}^{\rm (Re)} = t_{B_i}t_{A_j}t_{A_k}\chi_{i;jk}^{\rm (Re)}, \quad \chi_{i;jk}^{\rm (Im)} = -t_{B_i}t_{A_j}t_{A_k}\chi_{i;jk}^{\rm (Im)}, 
\end{align}
are satisfied and they indicate that $\chi_{i;jk}^{\rm (Re)}$ and $\chi_{i;jk}^{\rm (Im)}$ are represented by the E and ET (M and MT) multipoles and the M and MT (E and ET) multipoles, respectively, for $t_{B_i}t_{A_j}t_{A_k}=+1(-1)$. 
In other words, $\chi_{i;jk}^{\rm (Re)}$ is proportional to the even order of $\tau$, while $\chi_{i;jk}^{\rm (Im)}$ is proportional to the odd order of $\tau$. 
By the similar procedure, we obtain the nonlinear response tensor driven by $\dot{A}$ with dissipation, although it is not shown here due to rather complicated expression~\cite{Watanabe_PhysRevX.11.011001, Oiwa_doi:10.7566/JPSJ.91.014701}. Note that the relation Eq.~(\ref{eq:nonlinear_t}) also holds in this case.

Especially for the second-order nonlinear electric conductivity, $\sigma_{i;jk}$ in $J_{i}=\sigma_{i;jk}E_{j}E_{k}$ is decomposed according to the relaxation time $\tau$ dependence in the clean limit by performing a similar procedure to Eqs.~(\ref{eq:2nd_c_susceptability_2})--(\ref{eq:nonlinear_t}) under the consideration of the electric field in the length gauge.
The velocity gauge is also known as conventional gauge choice. 
In the nonlinear conductivity, $\tau^{2}$ and $\tau^{0}$ appearing in $\sigma_{i;jk}^{\rm (Re)}=\sigma^{\mathrm{D}}_{i;jk}+\sigma^{\mathrm{int}}_{i;jk}$ correspond to the Drude ($\sigma^{\mathrm{D}}$) and intrinsic ($\sigma^{\mathrm{int}}$) terms, respectively, and $\tau^{1}$ appearing in $\sigma_{i;jk}^{\rm (Im)}=\sigma^{\mathrm{BCD}}_{i;jk}$ corresponds to the Berry curvature dipole ($\sigma^{\mathrm{BCD}}$) term~\cite{gao2019semiclassical,Watanabe_PhysRevX.11.011001,Oiwa_doi:10.7566/JPSJ.91.014701}.

The other decomposition is often used~\cite{Stepan_SciPostPhysCore.5.3.039, wang2022intrinsic,Das_PhysRevB.108.L201405} as
\begin{align}
\sigma_{i;jk}=\sigma_{i;jk}^{\rm O}+\sigma_{i;jk}^{\rm int,H}+\sigma_{i;jk}^{\rm BCD},
\end{align}
where the relations, $\sigma_{i;jk}^{\rm int}=\sigma_{i;jk}^{\rm int,O}+\sigma_{i;jk}^{\rm int,H}$ and $\sigma_{i;jk}^{\rm O}=\sigma_{i;jk}^{\rm D}+\sigma_{i;jk}^{\rm int,O}$, hold.
In this expression, $\sigma_{i;jk}^{\rm O}$ represents dissipative (ohmic) contribution, while $\sigma_{i;jk}^{\rm int,H}$ is non-dissipative contribution related to the Berry connection polarizability, that is responsible to non-adiabatic Laudau-Zener tunneling process.
$\sigma_{i;jk}^{\rm O}$ is symmetric in all indices, $(i,j,k)$, while $\sigma_{i;jk}^{\rm int,H}$ is anti-symmetric in two indices, $(i,j)$ or $(i,k)$.
Note that $\sigma_{i;jk}^{\rm int,O}$ is impurity-insensitive ohmic conductivity as it remains finite in the clean limit of $1/\tau\to0$.

The relevant multipoles in these conductivity tensors are given by~\cite{Yatsushiro_PhysRevB.105.155157, Hayami_PhysRevB.106.024405, Kirikoshi_PhysRevB.107.155109} 
\begin{subequations}
  \label{NLC}
  \begin{align}
    \sigma^{\mathrm{O}}=&\, \left[\begin{matrix}
    3T_{x}^{\prime}+2T_{x}^{\alpha} & T_{y}^{\prime}-T_{y}^{\alpha}-T_{y}^{\beta} & T_{z}^{\prime}-T_{z}^{\alpha}+T_{z}^{\beta} \\
    T_{x}^{\prime}-T_{x}^{\alpha}+T_{x}^{\beta} & 3T_{y}^{\prime}+2T_{y}^{\alpha} & T_{z}^{\prime}-T_{z}^{\alpha}-T_{z}^{\beta} \\
    T_{x}^{\prime}-T_{x}^{\alpha}-T_{x}^{\beta} & T_{y}^{\prime}-T_{y}^{\alpha}+T_{y}^{\beta} & 3T_{z}^{\prime}+2T_{z}^{\alpha} \\
    T_{xyz} & T_{z}^{\prime}-T_{z}^{\alpha}-T_{z}^{\beta} & T_{y}^{\prime}-T_{y}^{\alpha}+T_{y}^{\beta} \\
    T_{z}^{\prime}-T_{z}^{\alpha}+T_{z}^{\beta} & T_{xyz} & T_{x}^{\prime}-T_{x}^{\alpha}-T_{x}^{\beta} \\
    T_{y}^{\prime}-T_{y}^{\alpha}-T_{y}^{\beta} & T_{x}^{\prime}-T_{x}^{\alpha}+T_{x}^{\beta} & T_{xyz} \\
    \end{matrix}\right]^{\mathrm{T}},
    \label{NLC-O}
    \\
    \sigma^{\mathrm{int, H}}=&\, \left[\begin{matrix}
    0 & 2(T_{y}-M_{zx}) & 2(T_{z}+M_{xy}) \\
    2(T_{x}+M_{yz}) & 0 & 2(T_{z}-M_{xy}) \\
    2(T_{x}-M_{yz}) &  2(T_{y}+M_{zx}) & 0 \\
    M_{u}+M_{v} & -(T_{z}-M_{xy}) & -(T_{y}+M_{zx}) \\
    -(T_{z}+M_{xy}) & -M_{u}+M_{v} & -(T_{x}-M_{yz}) \\
    -(T_{y}-M_{zx}) & -(T_{x}+M_{yz}) & -2M_{v} \\
    \end{matrix}\right]^{\mathrm{T}},
    \label{NLC-H}
   \\
    \sigma^{\mathrm{BCD}}=&\, \left[\begin{matrix}
    0 & 2(Q_{y}-G_{zx}) & 2(Q_{z}+G_{xy}) \\
    2(Q_{x}+G_{yz}) & 0 & 2(Q_{z}-G_{xy}) \\
    2(Q_{x}-G_{yz}) &  2(Q_{y}+G_{zx}) & 0 \\
    G_{u}+G_{v} & -(Q_{z}-G_{xy}) & -(Q_{y}+G_{zx}) \\
    -(Q_{z}+G_{xy}) & -G_{u}+G_{v} & -(Q_{x}-G_{yz}) \\
    -(Q_{y}-G_{zx}) & -(Q_{x}+G_{yz}) & -2G_{v} \\
    \end{matrix}\right]^{\mathrm{T}},
    \label{NLC-BCD}
  \end{align}
\end{subequations}
where the matrix representation of the conductivity tensor $\sigma$ has been expressed as 
\begin{equation*}
  \sigma=
  \left[\begin{matrix}
    \sigma_{x;xx} & \sigma_{y;xx} & \sigma_{z;xx} 
    \\
    \sigma_{x;yy} & \sigma_{y;yy} & \sigma_{z;yy} 
    \\
    \sigma_{x;zz} & \sigma_{y;zz} & \sigma_{z;zz} 
    \\
    \sigma_{x;yz} & \sigma_{y;yz} & \sigma_{z;yz} 
    \\
    \sigma_{x;zx} & \sigma_{y;zx} & \sigma_{z;zx} 
    \\
    \sigma_{x;xy} & \sigma_{y;xy} & \sigma_{z;xy} 
    \\
  \end{matrix}\right]^{\mathrm{T}},
\end{equation*}
where T means the transpose of a matrix. 
We summarize the correspondence between the multipoles and the nonlinear response tensors in Table~\ref{table:tensor_parity_MP_nonlinear}.

\subsection{Essential model parameters for responses}
\label{sec: Essential model parameters for responses}

As discussed above, multipoles and cross correlations are closely linked from a symmetry point of view. 
On the other hand, since multipoles are also related to the microscopic electronic degrees of freedom in the system, it provides useful information to understand which model parameters are essential to cause cross correlations from a microscopic point of view. 
In this section, we introduce a systematic and analytic method to extract the essential model parameters once the model Hamiltonian is given~\cite{Oiwa_doi:10.7566/JPSJ.91.014701}. 
The heart of this method is that the linear response tensor $\chi_{\mu;\alpha}(\omega)$ and nonlinear response tensor $\tilde{\chi}_{\mu;\alpha\beta}(\omega_1,\omega_2)$ [$\tilde{\chi}_{\mu;\alpha\beta}(\omega_1,\omega_2)$ represents a non-symmetrized tensor in terms of $(\alpha,\omega_1)$ and $(\beta,\omega_2)$.] for $\mu,\alpha,\beta=x,y,z$ is transformed to decouple the model-independent $\Lambda$ and model-dependent $\Gamma$ parts by using the Keldysh formalism~\cite{Keldysh1964} and the Chebyshev polynomial expansion method~\cite{Jo2019Chebyshev} as follows: 
\begin{align}
\label{eq:chi_linear}
\chi_{\mu;\alpha}(\omega)&=\sum_{ij} \Lambda_{ij}(\omega;\gamma)\Gamma^{ij}_{\mu;\alpha}, \\
\label{eq:chi_2ndlinear}
\tilde{\chi}_{\mu;\alpha\beta}(\omega_1,\omega_2) &= \sum_{ijk}\Lambda_{ijk}(\omega_1,\omega_2;\gamma)\Gamma_{\mu;\alpha\beta}^{ijk},
\end{align}
where $\Lambda$ and $\Gamma$ satisfy the following relations: 
\begin{align}
\Lambda^*_{ij}(\omega;\gamma)&=\Lambda_{ji}(-\omega;\gamma), \\
\Lambda^*_{ijk}(\omega_1,\omega_2;\gamma)&=\Lambda_{kji}(-\omega_2,-\omega_1;\gamma), \\
\Gamma^{ij*}_{\mu;\alpha}&=\Gamma^{ji}_{\mu;\alpha}=\Gamma^{ij}_{\alpha;\mu}, \\
\Gamma^{ijk*}_{\mu;\alpha\beta}&=\Gamma^{kji}_{\mu;\beta\alpha}, \\
\Gamma^{ijk}_{\mu;\alpha\beta}&=\Gamma^{jki}_{\alpha;\beta\mu}=\Gamma^{kij}_{\beta;\mu\alpha}. 
\end{align}
Moreover, the following relations are obtained from the above ones: 
\begin{align}
&{\rm Re}[\Gamma^{ij}_{\mu;\alpha}]={\rm Re}[\Gamma^{ji}_{\mu;\alpha}]={\rm Re}[\Gamma^{ij}_{\alpha;\mu}],  \\
\label{eq:Gamma_Im_linear}
&{\rm Im}[\Gamma^{ij}_{\mu;\alpha}]=-{\rm Im}[\Gamma^{ji}_{\mu;\alpha}]=-{\rm Im}[\Gamma^{ij}_{\alpha;\mu}], \\
&\mathrm{Re}\left[\Gamma_{\mu ; \alpha\beta}^{i j k}\right] = \mathrm{Re}\left[\Gamma_{\mu ; \beta\alpha}^{k j i}\right], \\ 
&\mathrm{Im}\left[\Gamma_{\mu ; \alpha\beta}^{i j k}\right] = -\mathrm{Im}\left[\Gamma_{\mu ; \beta\alpha}^{k j i}\right]. 
\end{align}
These relations indicate that the real part of $\Gamma^{ij}_{\mu;\alpha}$, ${\rm Re}[\Gamma^{ij}_{\mu;\alpha}]$, corresponds to the symmetric tensor and the imaginary part,  ${\rm Im}[\Gamma^{ij}_{\mu;\alpha}]$, corresponds to the antisymmetric tensor in the linear response tensor. 

Among the expressions in Eqs.~(\ref{eq:chi_linear}) and (\ref{eq:chi_2ndlinear}), $\Lambda$ includes the information independent of the model, such as the chemical potential $\mu$, temperature $T$, frequency $\omega$, and the inverse relaxation time $\delta$. 
Meanwhile, $\Gamma$ includes the information about the model Hamiltonian, input operator $\hat{A}$, and output operator $\hat{B}$, where the model Hamiltonian covers all the terms expressed by a quadratic form of the creation and annihilation operators, such as the (spin-dependent) hopping, relativistic spin-orbit coupling, crystalline electric field, and molecular field. 
Specifically, $\Gamma$ in Eqs.~(\ref{eq:chi_linear}) and (\ref{eq:chi_2ndlinear}) is given by
\begin{align}
\label{eq:Gamma_linear}
\Gamma_{\mu;\alpha}^{ij} &= \sum_{\bm{k}}\Omega_{\mu;\alpha}^{ij}(\bm{k}),  \\
\label{eq:Gamma_2ndlinear}
\Gamma_{\mu;\alpha\beta}^{ijk} &= \sum_{\bm{k}}\Omega_{\mu;\alpha\beta}^{ijk}(\bm{k}), 
\end{align}
with 
\begin{align}
\label{eq: Omega1}
\Omega_{\mu;\alpha}^{ij}(\bm{k}) &= {\rm Tr}\left[ \hat{B}_\mu (\bm{k}) \hat{H}^i(\bm{k}) \hat{A}_\alpha (\bm{k})  \hat{H}^j(\bm{k})\right], \\
\label{eq: Omega2}
\Omega_{\mu;\alpha\beta}^{ijk}(\bm{k}) &= {\rm Tr}\left[ \hat{B}_\mu (\bm{k}) \hat{H}^i(\bm{k}) \hat{A}_\alpha (\bm{k})  \hat{H}^j(\bm{k})
\hat{A}_\beta (\bm{k})  \hat{H}^k(\bm{k})\right], 
\end{align}
where $\hat{O}(\bm{k})=\hat{A}(\bm{k}),\hat{B}(\bm{k}),\hat{H}(\bm{k})$ represent the matrix element in terms of the internal degrees of freedom like spin and orbital at wave vector $\bm{k}$, whose expression of the second quantization is given by 
\begin{align}
\hat{O}=\sum_{\bm{k}}\left[\hat{O}(\bm{k})\right]_{ab}c^\dagger_a(\bm{k})c^{}_b (\bm{k}). 
\end{align}
The superscript $i,j,k$ represents the exponent of the operator originating from the Chebyshev polynomial expansion. 
By evaluating $\Gamma$, one finds the model parameter dependence of arbitrary response tensors, which results in the understanding of the essential microscopic ingredients inducing the responses. 
Since the essential parameters appear as common proportional coefficients in the overall contributions, their analytical expressions could be obtained by evaluating only a few lower-order contributions.  
Once the essential parameters are identified, they provide a microscopic picture of the response, i.e., minimal couplings between the model parameters, such as electrons hopping, spin-orbit coupling, etc., and electric and/or magnetic order parameters.
This method has been used to analyze the essential model parameters of various response tensors, such as the linear spin conductivity tensor~\cite{hayami2022spinconductivity}, linear Hall conductivity tensor~\cite{Hayami_PhysRevB.108.085124}, nonlinear nonreciprocal conductivity tensor~\cite{Yatsushiro_PhysRevB.105.155157, Hayami_PhysRevB.106.024405, Hayami_PhysRevB.106.014420}, nonlinear (spin) Hall conductivity tensor~\cite{Oiwa_doi:10.7566/JPSJ.91.014701, Kirikoshi_PhysRevB.107.155109}, linear magnetoelectric tensor~\cite{Hayami_doi:10.7566/JPSJ.91.123701}, and nonreciprocal magnon dispersion~\cite{Hayami_PhysRevB.105.014404}.

\section{Classification of multipoles}
\label{sec: Classification of multipoles}

\begin{table*}
\caption{
Multipoles and their irreducible representations classified for cubic, tetragonal, orthorhombic, and monoclinic crystals~\cite{Hayami_PhysRevB.98.165110}. 
The upper and lower columns represent even-parity and odd-parity multipoles, respectively. 
The $x$([110]) axis is taken as the $C'_2$ rotation for $D_{2d}$($D'_{2d}$). 
This table is taken and modified from Ref~\citen{Hayami_PhysRevB.98.165110}. 
}
\label{tab_multipoles_table1}
\centering
\scalebox{0.72}{
\begin{tabular}{llll|ccccc|cccccccc|ccc|ccc} \hline\hline
E & ET & M & MT &
$O_{\rm h}$ & $O$ & $T_{\rm d}$ & $T_{\rm h}$ & $T$ &
$D_{\rm 4h}$ & $D_{4}$ & $C_{\rm 4h}$ & $D_{\rm 2d}$ & $D'_{\rm 2d}$ & $C_{\rm 4v}$ & $C_{4}$ & $S_{4}$ &
$D_{\rm 2h}$ & $D_{2}$ & $C_{\rm 2v}$ & $C_{\rm 2h}$ & $C_{\rm 2}$ & $C_{\rm s}$\\ \hline
$Q_{0}$, $Q_{4}$ & --- & --- & $T_{0}$, $T_{4}$ &
${\rm {\rm A}}_{1g}$ & ${\rm {\rm A}}_{1}$ & ${\rm {\rm A}}_{1}$ & ${\rm {\rm A}}_{g}$ & ${\rm {\rm A}}$ &
${\rm {\rm A}}_{1g}$ & ${\rm {\rm A}}_{1}$ & ${\rm {\rm A}}_{g}$ & ${\rm {\rm A}}_{1}$ & ${\rm {\rm A}}_{1}$ & ${\rm {\rm A}}_{1}$ & ${\rm {\rm A}}$ & ${\rm {\rm A}}$ &
${\rm {\rm A}}_{g}$ & ${\rm {\rm A}}$ & ${\rm {\rm A}}_{1}$ & ${\rm {\rm A}}_{g}$ & ${\rm {\rm A}}$ & ${\rm {\rm A}}'$ \\
--- & $G_{xyz}$ & $M_{xyz}$ & --- &
${\rm {\rm A}}_{2g}$ & ${\rm {\rm A}}_{2}$ & ${\rm {\rm A}}_{2}$ & ${\rm {\rm A}}_{g}$ & ${\rm {\rm A}}$ &
${\rm B}_{1g}$ & ${\rm B}_{1}$ & ${\rm B}_{g}$ & ${\rm B}_{1}$ & ${\rm B}_{2}$ & ${\rm B}_{1}$ & ${\rm B}$ & ${\rm B}$ &
${\rm {\rm A}}_{g}$ & ${\rm {\rm A}}$ & ${\rm {\rm A}}_{1}$ & ${\rm {\rm A}}_{g}$ & ${\rm {\rm A}}$ & ${\rm {\rm A}}'$ \\
$Q_{u}$, $Q_{4u}$ & --- & --- & $T_{u}$, $T_{4u}$ &
${\rm E}_{g}$ & ${\rm E}$ & ${\rm E}$ & ${\rm E}_{g}$ & ${\rm E}$ &
${\rm {\rm A}}_{1g}$ & ${\rm {\rm A}}_{1}$ & ${\rm {\rm A}}_{g}$ & ${\rm {\rm A}}_{1}$ & ${\rm {\rm A}}_{1}$ & ${\rm {\rm A}}_{1}$ & ${\rm {\rm A}}$ & ${\rm {\rm A}}$ &
${\rm {\rm A}}_{g}$ & ${\rm {\rm A}}$ & ${\rm {\rm A}}_{1}$ & ${\rm {\rm A}}_{g}$ & ${\rm {\rm A}}$ & ${\rm {\rm A}}'$ \\
$Q_{v}$, $Q_{4v}$ & --- & --- & $T_{v}$, $T_{4v}$ &
& & & & &
${\rm B}_{1g}$ & ${\rm B}_{1}$ & ${\rm B}_{g}$ & ${\rm B}_{1}$ & ${\rm B}_{2}$ & ${\rm B}_{1}$ & ${\rm B}$ & ${\rm B}$ &
${\rm {\rm A}}_{g}$ & ${\rm {\rm A}}$ & ${\rm {\rm A}}_{1}$ & ${\rm {\rm A}}_{g}$ & ${\rm {\rm A}}$ & ${\rm {\rm A}}'$ \\
$Q_{4x}^{\alpha}$ & $G_{x}$, $G_{x}^{\alpha}$  & $M_{x}$, $M_{x}^{\alpha}$ & $T_{4x}^{\alpha}$ &
${\rm T}_{1g}$ & ${\rm T}_{1}$ & ${\rm T}_{1}$ & ${\rm T}_{g}$ & ${\rm T}$ &
${\rm E}_{g}$ & ${\rm E}$ & ${\rm E}_{g}$ & ${\rm E}$ & ${\rm E}$ & ${\rm E}$ & ${\rm E}$ & ${\rm E}$ &
${\rm B}_{3g}$ & ${\rm B}_{3}$ & ${\rm B}_{2}$ & ${\rm B}_{g}$ & ${\rm B}$ & ${\rm {\rm A}}''$ \\
$Q_{4y}^{\alpha}$ & $G_{y}$, $G_{y}^{\alpha}$ & $M_{y}$, $M_{y}^{\alpha}$ & $T_{4y}^{\alpha}$ &
& & & & &
& & & & & & & & 
${\rm B}_{2g}$ & ${\rm B}_{2}$ & ${\rm B}_{1}$ & ${\rm B}_{g}$ & ${\rm B}$ & ${\rm {\rm A}}''$ \\
$Q_{4z}^{\alpha}$ & $G_{z}$, $G_{z}^{\alpha}$ & $M_{z}$, $M_{z}^{\alpha}$ & $T_{4z}^{\alpha}$ &
& & & & &
${\rm {\rm A}}_{2g}$ & ${\rm {\rm A}}_{2}$ & ${\rm {\rm A}}_{g}$ & ${\rm {\rm A}}_{2}$ & ${\rm {\rm A}}_{2}$ & ${\rm {\rm A}}_{2}$ & ${\rm {\rm A}}$ & ${\rm A}$ &
${\rm B}_{1g}$ & ${\rm B}_{1}$ & ${\rm A}_{2}$ & ${\rm A}_{g}$ & ${\rm A}$ & ${\rm A}'$ \\
$Q_{yz}$, $Q_{4x}^{\beta}$ & $G_{x}^{\beta}$ &$M_{x}^{\beta}$ &  $T_{yz}$, $T_{4x}^{\beta}$ &
${\rm T}_{2g}$ & ${\rm T}_{2}$ & ${\rm T}_{2}$ & ${\rm T}_{g}$ & ${\rm T}$ &
${\rm E}_{g}$ & ${\rm E}$ & ${\rm E}_{g}$ & ${\rm E}$ & ${\rm E}$ & ${\rm E}$ & ${\rm E}$ & ${\rm E}$ &
${\rm B}_{3g}$ & ${\rm B}_{3}$ & ${\rm B}_{2}$ & ${\rm B}_{g}$ & ${\rm B}$ & ${\rm A}''$ \\
$Q_{zx}$, $Q_{4y}^{\beta}$ & $G_{y}^{\beta}$ & $M_{y}^{\beta}$ & $T_{zx}$, $T_{4y}^{\beta}$ &
& & & & &
& & & & & & & & 
${\rm B}_{2g}$ & ${\rm B}_{2}$ & ${\rm B}_{1}$ & ${\rm B}_{g}$ & ${\rm B}$ & ${\rm A}''$ \\
$Q_{xy}$, $Q_{4z}^{\beta}$ & $G_{z}^{\beta}$ & $M_{z}^{\beta}$ & $T_{xy}$, $T_{4z}^{\beta}$ &
& & & & &
${\rm B}_{2g}$ & ${\rm B}_{2}$ & ${\rm B}_{g}$ & ${\rm B}_{2}$ & ${\rm B}_{1}$ & ${\rm B}_{2}$ & ${\rm B}$ & ${\rm B}$ &
${\rm B}_{1g}$ & ${\rm B}_{1}$ & ${\rm A}_{2}$ & ${\rm A}_{g}$ & ${\rm A}$ & ${\rm A}'$ \\ \hline
--- & $G_{0}$, $G_{4}$ & $M_{0}$, $M_{4}$ & --- &
${\rm A}_{1u}$ & ${\rm A}_{1}$ & ${\rm A}_{2}$ & ${\rm A}_{u}$ & ${\rm A}$ &
${\rm A}_{1u}$ & ${\rm A}_{1}$ & ${\rm A}_{u}$ & ${\rm B}_{1}$ & ${\rm B}_{1}$ & ${\rm A}_{2}$ & ${\rm A}$ & ${\rm B}$ &
${\rm A}_{u}$ & ${\rm A}$ & ${\rm A}_{2}$ & ${\rm A}_{u}$ & ${\rm A}$ & ${\rm A}''$ \\
$Q_{xyz}$ & --- & --- & $T_{xyz}$ &
${\rm A}_{2u}$ & ${\rm A}_{2}$ & ${\rm A}_{1}$ & ${\rm A}_{u}$ & ${\rm A}$ &
${\rm B}_{1u}$ & ${\rm B}_{1}$ & ${\rm B}_{u}$ & ${\rm A}_{1}$ & ${\rm A}_{2}$ & ${\rm B}_{2}$ & ${\rm B}$ & ${\rm A}$ &
${\rm A}_{u}$ & ${\rm A}$ & ${\rm A}_{2}$ & ${\rm A}_{u}$ & ${\rm A}$ & ${\rm A}''$ \\
--- & $G_{u}$, $G_{4u}$ & $M_{u}$, $M_{4u}$ & --- &
${\rm E}_{u}$ & ${\rm E}$ & ${\rm E}$ & ${\rm E}_{u}$ & ${\rm E}$ &
${\rm A}_{1u}$ & ${\rm A}_{1}$ & ${\rm A}_{u}$ & ${\rm B}_{1}$ & ${\rm B}_{1}$ & ${\rm A}_{2}$ & ${\rm A}$ & ${\rm B}$ &
${\rm A}_{u}$ & ${\rm A}$ & ${\rm A}_{2}$ & ${\rm A}_{u}$ & ${\rm A}$ & ${\rm A}''$ \\
--- & $G_{v}$, $G_{4v}$ & $M_{v}$, $M_{4v}$ & --- &
& & & & &
${\rm B}_{1u}$ & ${\rm B}_{1}$ & ${\rm B}_{u}$ & ${\rm A}_{1}$ & ${\rm A}_{2}$ & ${\rm B}_{2}$ & ${\rm B}$ & ${\rm A}$ &
${\rm A}_{u}$ & ${\rm A}$ & ${\rm A}_{2}$ & ${\rm A}_{u}$ & ${\rm A}$ & ${\rm A}''$ \\
$Q_{x}$, $Q_{x}^{\alpha}$ & $G_{4x}^{\alpha}$ & $M_{4x}^{\alpha}$ & $T_{x}$, $T_{x}^{\alpha}$ &
${\rm T}_{1u}$ & ${\rm T}_{1}$ & ${\rm T}_{2}$ & ${\rm T}_{u}$ & ${\rm T}$ &
${\rm E}_{u}$ & ${\rm E}$ & ${\rm E}_{u}$ & ${\rm E}$ & ${\rm E}$ & ${\rm E}$ & ${\rm E}$ & ${\rm E}$ &
${\rm B}_{3u}$ & ${\rm B}_{3}$ & ${\rm B}_{1}$ & ${\rm B}_{u}$ & ${\rm B}$ & ${\rm A}'$ \\
$Q_{y}$, $Q_{y}^{\alpha}$ & $G_{4y}^{\alpha}$ & $M_{4y}^{\alpha}$ & $T_{y}$, $T_{y}^{\alpha}$ &
& & & & &
& & & & & & & & 
${\rm B}_{2u}$ & ${\rm B}_{2}$ & ${\rm B}_{2}$ & ${\rm B}_{u}$ & ${\rm B}$ & ${\rm A}'$ \\
$Q_{z}$, $Q_{z}^{\alpha}$ & $G_{4z}^{\alpha}$ & $M_{4z}^{\alpha}$ & $T_{z}$, $T_{z}^{\alpha}$ &
& & & & &
${\rm A}_{2u}$ & ${\rm A}_{2}$ & ${\rm A}_{u}$ & ${\rm B}_{2}$ & ${\rm B}_{2}$ & ${\rm A}_{1}$ & ${\rm A}$ & ${\rm B}$ &
${\rm B}_{1u}$ & ${\rm B}_{1}$ & ${\rm A}_{1}$ & ${\rm A}_{u}$ & ${\rm A}$ & ${\rm A}''$ \\
$Q_{x}^{\beta}$ & $G_{yz}$, $G_{4x}^{\beta}$ & $M_{yz}$, $M_{4x}^{\beta}$ & $T_{x}^{\beta}$ &
${\rm T}_{2u}$ & ${\rm T}_{2}$ & ${\rm T}_{1}$ & ${\rm T}_{u}$ & ${\rm T}$ &
${\rm E}_{u}$ & ${\rm E}$ & ${\rm E}_{u}$ & ${\rm E}$ & ${\rm E}$ & ${\rm E}$ & ${\rm E}$ & ${\rm E}$ &
${\rm B}_{3u}$ & ${\rm B}_{3}$ & ${\rm B}_{1}$ & ${\rm B}_{u}$ & ${\rm B}$ & ${\rm A}'$ \\
$Q_{y}^{\beta}$ & $G_{zx}$, $G_{4y}^{\beta}$ & $M_{zx}$, $M_{4y}^{\beta}$ & $T_{y}^{\beta}$ &
& & & & &
& & & & & & & & 
${\rm B}_{2u}$ & ${\rm B}_{2}$ & ${\rm B}_{2}$ & ${\rm B}_{u}$ & ${\rm B}$ & ${\rm A}'$ \\
$Q_{z}^{\beta}$ & $G_{xy}$, $G_{4z}^{\beta}$ & $M_{xy}$, $M_{4z}^{\beta}$ & $T_{z}^{\beta}$ &
& & & & &
${\rm B}_{2u}$ & ${\rm B}_{2}$ & ${\rm B}_{u}$ & ${\rm A}_{2}$ & ${\rm A}_{1}$ & ${\rm B}_{1}$ & ${\rm B}$ & ${\rm A}$ &
${\rm B}_{1u}$ & ${\rm B}_{1}$ & ${\rm A}_{1}$ & ${\rm A}_{u}$ & ${\rm A}$ & ${\rm A}''$ \\
\hline\hline
\end{tabular}
}
\end{table*}

\begin{table*}
\caption{
Multipoles and their irreducible representations classified for hexagonal and trigonal crystals~\cite{Hayami_PhysRevB.98.165110}.
The upper and lower columns represent even-parity and odd-parity multipoles, 
We take the $x$($y$) axis as the $C'_2$ rotation axis for $D_{\rm 3h}$($D'_{\rm 3h}$), $D_{\rm 3d}$($D'_{\rm 3d}$), and $D_{\rm 3}$($D'_{\rm 3}$). 
For $C_{\rm 3v}$($C'_{\rm 3v}$), we take the $zx$($yz$) plane as the $\sigma_v$ mirror plane. 
$C_{\rm 3i}=S_6$. 
This table is taken and modified from Ref~\citen{Hayami_PhysRevB.98.165110}.
}
\label{tab_multipoles_table2}
\centering
\scalebox{0.68}{
\begin{tabular}{llll|cccccccc|cccccccc} \hline\hline
E & ET & M & MT &
$D_{\rm 6h}$ & $D_{6}$ & $C_{\rm 6h}$ & $C_{\rm 6v}$ & $C_{6}$ & $D_{\rm 3h}$ & $D'_{\rm 3h}$ & $C_{\rm 3h}$ &
$D_{\rm 3d}$ &
$D'_{\rm 3d}$ & $D_{3}$ & $D'_{3}$ & $C_{\rm 3v}$ & $C'_{\rm 3v}$ & $C_{\rm 3i}$ & $C_{3}$ \\ \hline
$Q_{0}$, $Q_{u}$, $Q_{40}$ & --- & --- & $T_{0}$, $T_{u}$, $T_{40}$ &
${\rm A}_{1g}$ & ${\rm A}_{1}$ & ${\rm A}_{g}$ & ${\rm A}_{1}$ & ${\rm A}$ & ${\rm A}_{1}'$ & ${\rm A}_{1}'$ & ${\rm A}'$ &
${\rm A}_{1g}$ & ${\rm A}_{1g}$ & ${\rm A}_{1}$ & ${\rm A}_{1}$ & ${\rm A}_{1}$ & ${\rm A}_{1}$ & ${\rm A}_{g}$ & ${\rm A}$
\\
--- &$G_{z}$, $G_{z}^{\alpha}$ & $M_{z}$, $M_{z}^{\alpha}$ & --- &
${\rm A}_{2g}$ & ${\rm A}_{2}$ & ${\rm A}_{g}$ & ${\rm A}_{2}$ & ${\rm A}$ & ${\rm A}_{2}'$ & ${\rm A}_{2}'$ & ${\rm A}'$ &
${\rm A}_{2g}$ & ${\rm A}_{2g}$ & ${\rm A}_{2}$ & ${\rm A}_{2}$ & ${\rm A}_{2}$ & ${\rm A}_{2}$ & ${\rm A}_{g}$ & ${\rm A}$
\\
$Q_{4a}$ & $G_{3a}$ & $M_{3a}$ & $T_{4a}$ &
${\rm B}_{1g}$ & ${\rm B}_{1}$ & ${\rm B}_{g}$ & ${\rm B}_{2}$ & ${\rm B}$ & ${\rm A}_{1}''$ & ${\rm A}_{2}''$ & ${\rm A}''$ &
${\rm A}_{1g}$ &${\rm A}_{2g}$ & ${\rm A}_{1}$ & ${\rm A}_{2}$ & ${\rm A}_{2}$ & ${\rm A}_{1}$ & ${\rm A}_{g}$ & ${\rm A}$
\\
$Q_{4b}$ & $G_{3b}$ & $M_{3b}$ & $T_{4b}$ &
${\rm B}_{2g}$ & ${\rm B}_{2}$ & ${\rm B}_{g}$ & ${\rm B}_{1}$ & ${\rm B}$ & ${\rm A}_{2}''$ & ${\rm A}_{1}''$ & ${\rm A}''$ &
${\rm A}_{2g}$ &${\rm A}_{1g}$ & ${\rm A}_{2}$ & ${\rm A}_{1}$ & ${\rm A}_{1}$ & ${\rm A}_{2}$ & ${\rm A}_{g}$ & ${\rm A}$
\\
$Q_{zx}$, $Q_{4u}^\alpha$ & $G_{x}$, $G_{3u}$ & $M_{x}$, $M_{3u}$ & $T_{zx}$, $T_{4u}^\alpha$ &
${\rm E}_{1g}$ & ${\rm E}_{1}$ & ${\rm E}_{1g}$ & ${\rm E}_{1}$ & ${\rm E}_{1}$ & ${\rm E}''$ & ${\rm E}''$ & ${\rm E}''$ &
${\rm E}_{g}$ &${\rm E}_{g}$ & ${\rm E}$ & ${\rm E}$ & ${\rm E}$ & ${\rm E}$ & ${\rm E}_{g}$ & ${\rm E}$
\\
$Q_{yz}$, $Q_{4v}^\alpha$ & $G_{y}$, $G_{3v}$ & $M_{y}$, $M_{3v}$ & $T_{yz}$, $T_{4v}^\alpha$ &
& & & & & & & &
& & & & & & & 
\\
$Q_{v}$, $Q_{4u}^{\beta 1}$, $Q_{4u}^{\beta 2}$ & $G_{xyz}$ & $M_{xyz}$ & $T_{v}$, $T_{4u}^{\beta 1}$, $T_{4u}^{\beta 2}$ &
${\rm E}_{2g}$ & ${\rm E}_{2}$ & ${\rm E}_{2g}$ & ${\rm E}_{2}$ & ${\rm E}_{2}$ & ${\rm E}'$ & ${\rm E}'$ & ${\rm E}'$ &
${\rm E}_{g}$ &${\rm E}_{g}$ & ${\rm E}$ & ${\rm E}$ & ${\rm E}$ &${\rm E}$ & ${\rm E}_{g}$ & ${\rm E}$
\\
$Q_{xy}$, $Q^{\beta 1}_{4v}$, $Q^{\beta 2}_{4v}$ & $G_{z}^{\beta}$ & $M_{z}^{\beta}$ & $T_{xy}$, $T^{\beta 2}_{4v}$, $T^{\beta 2}_{4v}$ &
& & & & & & & & 
& & & & & &  & 
\\ \hline
--- & $G_{0}$, $G_{u}$, $G_{40}$ & $M_{0}$, $M_{u}$, $M_{40}$ & --- &
${\rm A}_{1u}$ & ${\rm A}_{1}$ & ${\rm A}_{u}$ & ${\rm A}_{2}$ & ${\rm A}$ & ${\rm A}_{1}''$ & ${\rm A}_{1}''$ & ${\rm A}''$ &
${\rm A}_{1u}$ & ${\rm A}_{1u}$ & ${\rm A}_{1}$ &  ${\rm A}_{1}$ & ${\rm A}_{2}$ & ${\rm A}_{2}$ & ${\rm A}_{u}$ & ${\rm A}$
\\
$Q_{z}$, $Q_{z}^{\alpha}$ & --- & --- & $T_{z}$, $T_{z}^{\alpha}$ &
${\rm A}_{2u}$ & ${\rm A}_{2}$ & ${\rm A}_{u}$ & ${\rm A}_{1}$ & ${\rm A}$ & ${\rm A}_{2}''$ & ${\rm A}_{2}''$ & ${\rm A}''$ &
${\rm A}_{2u}$ & ${\rm A}_{2u}$ & ${\rm A}_{2}$ &  ${\rm A}_{2}$ & ${\rm A}_{1}$ &  ${\rm A}_{1}$ & ${\rm A}_{u}$ & ${\rm A}$
\\
$Q_{3a}$ & $G_{4a}$ & $M_{4a}$ & $T_{3a}$ &
${\rm B}_{1u}$ & ${\rm B}_{1}$ & ${\rm B}_{u}$ & ${\rm B}_{1}$ & ${\rm B}$ & ${\rm A}_{1}'$ & ${\rm A}_{2}'$ & ${\rm A}'$ &
${\rm A}_{1u}$ & ${\rm A}_{2u}$ & ${\rm A}_{1}$ & ${\rm A}_{2}$ & ${\rm A}_{1}$ & ${\rm A}_{2}$ & ${\rm A}_{u}$ & ${\rm A}$
\\
$Q_{3b}$ & $G_{4b}$ & $M_{4b}$ & $T_{3b}$ &
${\rm B}_{2u}$ & ${\rm B}_{2}$ & ${\rm B}_{u}$ & ${\rm B}_{2}$ & ${\rm B}$ & ${\rm A}_{2}'$ & ${\rm A}_{1}'$ & ${\rm A}'$ &
${\rm A}_{2u}$ & ${\rm A}_{1u}$ & ${\rm A}_{2}$ & ${\rm A}_{1}$ & ${\rm A}_{2}$ & ${\rm A}_{1}$ & ${\rm A}_{u}$ & ${\rm A}$
\\
$Q_{x}$, $Q_{3u}$ & $G_{zx}$, $G_{4u}^\alpha$ & $M_{zx}$, $M_{4u}^\alpha$ & $T_{x}$, $T_{3u}$ &
${\rm E}_{1u}$ & ${\rm E}_{1}$ & ${\rm E}_{1u}$ & ${\rm E}_{1}$ & ${\rm E}_{1}$ & ${\rm E}'$ & ${\rm E}'$ & ${\rm E}'$ &
${\rm E}_{u}$ & ${\rm E}_{u}$ & ${\rm E}$ & ${\rm E}$ & ${\rm E}$  & ${\rm E}$ & ${\rm E}_{u}$ & ${\rm E}$
\\
$Q_{y}$, $Q_{3v}$ & $G_{yz}$, $G_{4v}^\alpha$ & $M_{yz}$, $M_{4v}^\alpha$ & $T_{y}$, $T_{3v}$ &
& & & & & & & & 
& & & & & & &
\\
$Q_{xyz}$  & $G_{v}$, $G^{\beta1}_{4u}$, $G^{\beta2}_{4u}$ & $M_{v}$, $M_{4u}^{\beta 1}$, $M_{4u}^{\beta 2}$ & $T_{xyz}$ &
${\rm E}_{2u}$ & ${\rm E}_{2}$ & ${\rm E}_{2u}$ & ${\rm E}_{2}$ & ${\rm E}_{2}$ & ${\rm E}''$ & ${\rm E}''$ & ${\rm E}''$ &
${\rm E}_{u}$ & ${\rm E}_{u}$ & ${\rm E}$ & ${\rm E}$ & ${\rm E}$ &  ${\rm E}$ & ${\rm E}_{u}$ & ${\rm E}$
\\
$Q_{z}^{\beta}$ & $G_{xy}$, $G^{\beta1}_{4v}$, $G^{\beta2}_{4v}$ & $M_{xy}$, $M^{\beta 1}_{4v}$, $M^{\beta 2}_{4v}$ & $T_{z}^{\beta}$ &
& & & & & & & &
& & & & & & & 
\\
\hline\hline
\end{tabular}
}
\end{table*}
 
Since electronic multipoles are symmetry-adapted, they are systematically classified into the irreducible representation under 32 point groups~\cite{Hayami_PhysRevB.98.165110, Watanabe_PhysRevB.98.245129} and 122 magnetic point groups~\cite{Yatsushiro_PhysRevB.104.054412}. 
Tables~\ref{tab_multipoles_table1} and \ref{tab_multipoles_table2} show the irreducible representation of multipoles up to rank 4 under 32 point groups, where the compatible relations between parent groups and subgroups are also shown~\cite{Hayami_PhysRevB.98.165110}; similar tables for the classification of multipoles for 122 magnetic point groups are shown in Ref.~\citen{Yatsushiro_PhysRevB.104.054412}. 
These tables are useful for understanding electronic order parameters and cross correlations in a unified manner. 
We discuss four advantages of using these tables.

\begin{table*}[t!]
\centering
\caption{Irreducible representations (irreps) of four types of multipoles in $4/mmm1'$~\cite{Yatsushiro_PhysRevB.104.054412}.
The superscript ``$\pm$'' of irrep stands for the parity with respect to the antiunitary time-reversal operation $\theta$.
MPG represents the magnetic point group when the corresponding irrep belongs to the totally symmetric representation. 
The primary axis is given in column, P. axis. 
The rightmost column represents the examples of materials listed in MAGNDATA~\cite{gallego2016magndata}. 
This table is taken and modified from Ref~\citen{Yatsushiro_PhysRevB.104.054412}.
\label{table: D4h}
}
\vspace{2mm}
\scalebox{0.95}{
\begin{tabular}{cccccccccccccccccccc}
\hline\hline
& irrep 
&  E & ET & MT & M & MPG & P. axis & materials\\ 
\hline

${\rm A}_{\rm 1g}$ & 
${\rm A}_{\rm 1g}^+$ & $Q_0, Q_u,$ & & & & $4/mmm1'$ & $[001]$ & CeRhAl$_4$Si$_2$~\cite{ghimire2015magnetic} \\

 & ${\rm A}_{\rm 1g}^-$ & & & $T_0, T_u,$ &  & $4/mmm$ & $[001]$ & KMnF$_3$~\cite{knight2020nuclear} \\

${\rm A}_{\rm 2g}$ & 
${\rm A}_{\rm 2g}^+$ &  & $G_z, G_z^\alpha$ & & &$4/m1'$ & $[001]$ & FeTa$_2$O$_6$~\cite{eicher1986magnetic}\\

& ${\rm A}_{\rm 2g}^-$ & & &   & $M_z, M_z^\alpha$  & $4/mm'm'$ & $[001]$ &LaBaMn$_2$O$_6$~\cite{millange1998order} \\

${\rm B}_{\rm 1g}$ & 
${\rm B}_{\rm 1g}^+$ & $Q_v$ & $G_{xyz}$ & & & $mmm1'$ & $[100]$ & SrFeO$_2$~\cite{tsujimoto2007infinite} \\

& ${\rm B}_{\rm 1g}^-$ & & & $T_v$ & $M_{xyz}$ & $4'/mmm'$ & $[001]$ & RuO$_2$~\cite{Berlijn_PhysRevLett.118.077201} \\

${\rm B}_{\rm 2g}$ &
${\rm B}_{\rm 2g}^+$ &  $Q_{xy}$ & $G_z^\beta$ & & &$mmm1'$ & $[110]$\\

& ${\rm B}_{\rm 2g}^-$ & & & $T_{xy}$ & $M_z^\beta$& $4'/mm'm$ & $[001]$ \\

${\rm E}_{\rm g}$ &
${\rm E}_{\rm g}^+$ & $Q_{yz},$ & $G_x,G_x^\alpha, G_x^\beta$ & & & $2/m1'$ & $[100]$ & Cr$_2$ReO$_6$~\cite{Jiao_PhysRevB.97.014426} \\
& & $Q_{zx},$ & $G_y,G_y^\alpha, G_y^\beta$ & & & $2/m1'$ & $[010]$ \\

& ${\rm E}_{\rm g}^-$ & & & $T_{yz},$ & $M_x, M_x^\alpha, M_x^\beta$ & $mm'm'$ & $[100]$ & NiF$_2$~\cite{brown1981neutron} \\
& & & & $T_{zx},$ & $M_y, M_y^\alpha, M_y^\beta$ & $m'mm'$ & $[100]$ \\
\hline
${\rm A}_{\rm 1u}$ &
${\rm A}_{\rm 1u}^+$ &  &$G_0, G_u,$ & & & $4221'$ & $[001]$ & ZnV$_2$O$_4$~\cite{reehuis2003crystallographic} \\

& ${\rm A}_{\rm 1u}^-$ & & & & $M_0, M_u,$ & $4/m'm'm'$ & $[001]$ & GdB$_4$~\cite{Blanco_PhysRevB.73.212411} \\

${\rm A}_{\rm 2u}$ &
${\rm A}_{\rm 2u}^+$ &  $Q_z, Q_z^\alpha$ &  & & &$4mm1'$ & $[001]$ & \\

& ${\rm A}_{\rm 2u}^-$ & & & $T_z, T_z^\alpha$ & & $4/m'mm$ & $[001]$ \\

${\rm B}_{\rm 1u}$ &
${\rm B}_{\rm 1u}^+$ & $Q_{xyz}$  & $G_v$ & & & $\bar{4}2m1'$ & $[001]$ & GeCu$_2$O$_4$~\cite{Zou_PhysRevB.94.214406} \\

& ${\rm B}_{\rm 1u}^-$ & & & $T_{xyz}$  & $M_v$  & $4'/m'm'm$ & $[001]$ & BaMn$_2$As$_2$~\cite{Singh_PhysRevB.80.100403} \\

${\rm B}_{\rm 2u}$ &
${\rm B}_{\rm 2u}^+$ & $Q_{z}^\beta$ & $G_{xy}$ & & &$\bar{4}m21'$ & $[001]$\\

& ${\rm B}_{\rm 2u}^-$ & & & $T_{z}^\beta$ & $M_{xy}$& $4'/m'mm'$ & $[001]$ \\

${\rm E}_{\rm u}$ &
${\rm E}_{\rm u}^+$ &  $Q_x, Q_x^\alpha, Q_x^\beta$  & $G_{yz},$  & & & $2mm1'$ & $[100]$ & CeAuSb$_2$~\cite{Waite_PhysRevB.106.224415} \\
& & $Q_y, Q_y^\alpha, Q_y^\beta$  & $G_{zx},$  & & & $m2m1'$ & $[100]$\\

& ${\rm E}_{\rm u}^-$ & & &  $T_x, T_x^\alpha, T_x^\beta$  & $M_{yz}$ & $m'mm$ & $[100]$ & DyB$_4$~\cite{will1979neutron} \\
& &  & & $T_y, T_y^\alpha, T_y^\beta$  & $M_{zx}$ & $mm'm$ & $[100]$ \\

\hline\hline
\end{tabular}
}
\end{table*}

The one is the construction of the Hamiltonian. 
Since the Hamiltonian must be invariant to any symmetry operations in the system, it consists of multipoles belonging to the totally symmetric irreducible representation of the targeting space group. 
For example, the crystalline-electric-field (CEF) Hamiltonian is described by a linear combination of E multipoles belonging to the identity irreducible representation; the number of the independent CEF parameters is determined by the number of E multipoles except for E monopole $Q_0$ (which gives the origin of energy). 
In the case of the $d$-orbital system, $Q_4$ and $Q_{xyz}$ contribute to the CEF Hamiltonian for $T_{\rm d}$, while $Q_u$, $Q_{4}$, $Q_{4u}$, $Q_z$, and $Q_{z}^{\alpha}$ contribute to that for $C_{4v}$. 
In addition, the spin-orbit coupling, hopping, and Coulomb interaction are described by multipoles belonging to the irreducible representation in the form of the product of two or more multipoles. 
Similarly, the free energy in the Landau expansion is also composed of the product of the expectation value of multipoles, which is candidates of the order parameter, belonging to the totally symmetric irreducible representation~\cite{Kusunose_JPSJ.77.064710, kusunose2020complete}. 
Furthermore, an effective coupling between arbitrary multipoles including the hyperfine field of the nuclear spins can be constructed~\cite{Yatsushiro_PhysRevB.102.195147, kusunose2023configuration}.

The second is that the microscopic multipole couplings can be easily found. 
For example, the MT dipole $T_z$ and M quadrupole $M_{u}$ belong to the irreducible representation ${\rm A}_{2u}^{-}$ and ${\rm A}_{1u}^{-}$, respectively, under the point group $D_{{\rm 4h}}$, while both of them belong to ${\rm A}^-_u$ under the point group $C_{\rm 4h}$ once the twofold rotation axis perpendicular to the principal axis is lost. 
The same irreducible representation means that $T_z$ and $M_u$ are no longer distinguishable in terms of symmetry, which indicates the presence of an effective coupling between them. 
When the low-energy physical space includes both the $T_z$ and $M_u$ degrees of freedom under the point group $C_{\rm 4h}$, the coupling term $T_z M_u$ can appear in the effective Hamiltonian as well as $T_z^2$ and $M_u^2$. 
In short, multipoles belonging to the same irreducible representation are correlated with each other. 
In particular, since even-parity and odd-parity multipoles belong to the same irreducible representation in noncentrosymmetric systems, an effective coupling between multipoles with opposite spatial inversion parities occurs, which becomes the origin of the linear magnetoelectric effect and piezo effect.

The third is to identify the candidate order parameters, since Tables~\ref{tab_multipoles_table1} and \ref{tab_multipoles_table2} clarify the relation of irreducible representations between point groups. 
A candidate multipole order parameter can be deduced by comparing which multipole degrees of freedom newly belong to the totally symmetric irreducible representation when the symmetry is lowered. 
For example, when the symmetry is lowered from the point group $D_{\rm 4h}$ to its subgroups $(D_4, C_{\rm 4v}, C_{\rm 4h}, D_{\rm 2h}, D_{\rm 2d}, S_4)$, $(G_0, Q_z, G_z, Q_v, G_v, G_{xy})$ are candidates for the order parameter, respectively, since they newly belong to the totally symmetric irreducible representation. 

As another example, we show the case for the magnetic point group $4/mmm1'$ in Table~\ref{table: D4h}~\cite{Yatsushiro_PhysRevB.104.054412}. 
From the table, the multipole order parameters in magnetic materials shown in the rightmost column are found, where the information about the magnetic materials is referred to as the collection database, MAGNDATA~\cite{gallego2016magndata}. 
For example, KMnF$_3$ is classified into the MT monopole ($T_0$) ordering and GdB$_4$ is classified into the M monopole ($M_0$) ordering. 
Thus, the former KMnF$_3$ exhibits the cross correlations expected in the presence of $T_0$, such as the rotational distortion by an external magnetic field~\cite{Hayami_PhysRevB.108.L140409}, while the latter GdB$_4$ exhibits those expected in the presence of $M_0$, like the longitudinal linear magnetoelectric effect. 
Since such information for the other 121 magnetic point groups except for $4/mmm1'$ is obtained in Ref.~\citen{Yatsushiro_PhysRevB.104.054412}, the multipole order parameters in magnetic materials are uniquely identified in any case. 
In this context, similar analyses were performed for materials hosting unconventional electronic orderings, such as Cd$_2$Re$_2$O$_7$ with the ET quadrupole orderings~\cite{Hayami_PhysRevLett.122.147602} and URu$_2$Si$_2$ with the staggered E dotriacontapolar~\cite{Kambe_PhysRevB.97.235142, kambe2020symmetry} or ET monopole ordering~\cite{hayami2023chiral}.

\begin{figure*}[htb!]
\centering
\includegraphics[width=1.0 \hsize]{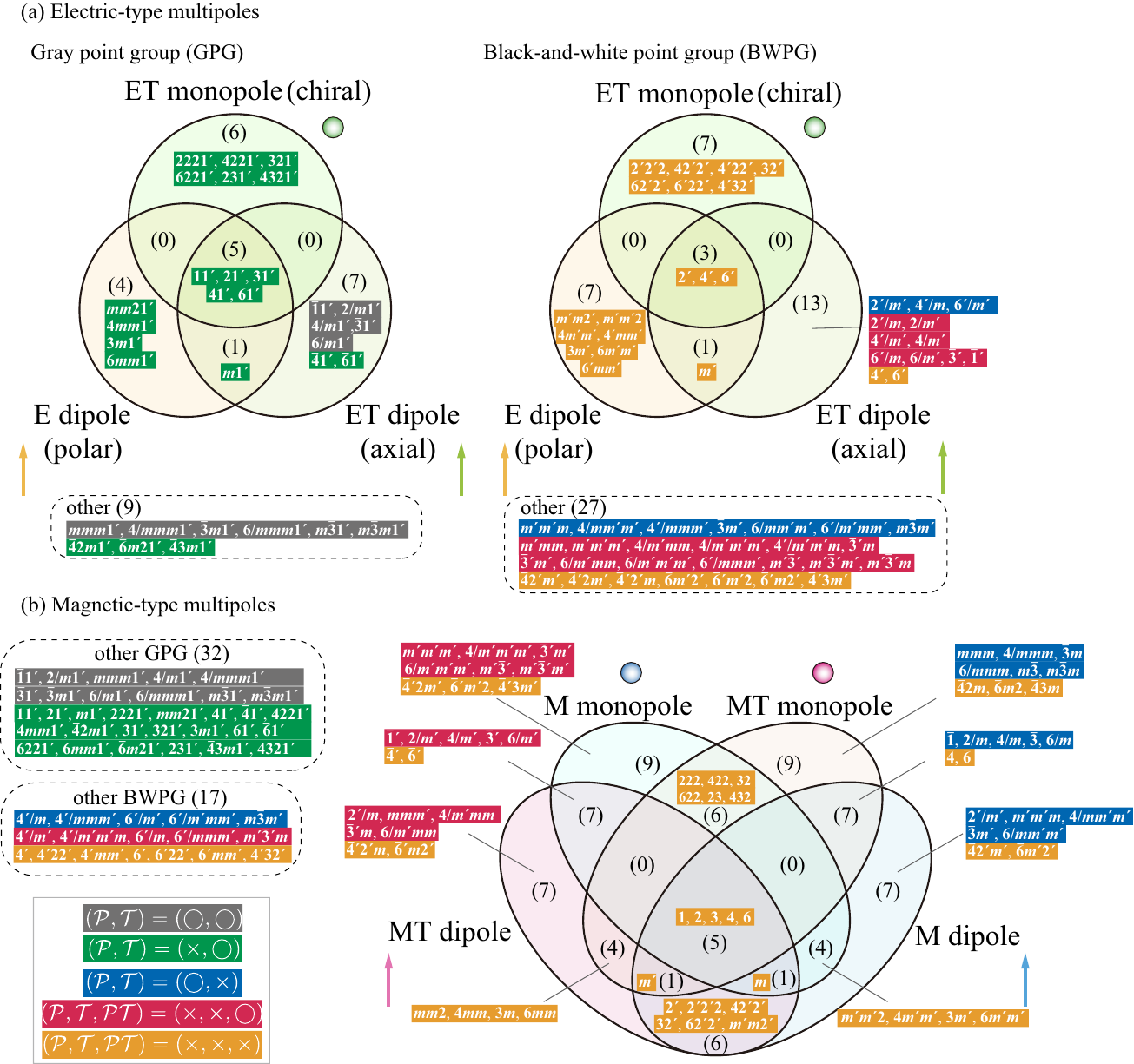}
\caption{
 \label{fig: venn_diagram}
 The Venn diagrams classifying (a) the electric-type multipoles (E dipole, ET monopole, and ET dipole) and (b) the magnetic-type multipoles (M monopole, M dipole, MT monopole, and MT dipole) belonging to the totally symmetric irreducible representation under the 122 magnetic point groups. 
In (a), the multipoles in 32 ordinary crystallographic point groups without the time-reversal operation are categorized as those in the gray point groups in the top-left panel. 
The letters of each magnetic point group are color-coded according to their spatial inversion ($\mathcal{P}$) and time-reversal ($\mathcal{T}$) symmetries, as shown in the left-bottom panel; $\bigcirc$ ($\times$) represents the presence (absence) of the symmetry.
 }
\end{figure*}

\begin{table}[tb!]
\begin{center}
\caption{
Number of rank-0--3 multipoles belonging to the totally symmetric irreducible representation under the 122 magnetic (32 crystallographic) point groups. 
}
\label{tab_mp_number}
\begingroup
\renewcommand{\arraystretch}{1.1}
\scalebox{0.88}{
 \begin{tabular}{lcccc}
 \hline \hline
Type  & $l=0$ & 1 & 2 & 3 \\ \hline
Electric &  122 (32) & 31 (10)   & 106 (27)  & 58 (18)  \\ 
   \hline
Electric toroidal & 32 (11)  & 43 (13)  & 42 (13)  & 71 (21)  \\
Magnetic & 32 (--)  & 31 (--)  &  42 (--) & 58 (--) \\ 
   \hline
Magnetic toroidal & 32 (--)  & 31 (--)  & 42 (--)  & 58 (--)  \\
\hline\hline
\end{tabular}
}
\endgroup
\end{center}
\end{table}

The fourth is the systematic classification of ferroic multipole orderings~\cite{Yatsushiro_PhysRevB.104.054412}. 
Since the ferroic state is characterized by the multipoles belonging to the totally symmetric irreducible representation, any ferroic states are classified into the multipoles according to the (magnetic) point group~\cite{wadhawan2000introduction, cheong2018broken}. 
For example, the ferroelectric, ferromagnetic, ferrotoroidal, and ferroaxial states are characterized by a ferroic alignment of E dipole $\bm{Q}$, M dipole $\bm{M}$~\cite{aizu1966possible, aizu1969possible, aizu1970possible}, MT dipole $\bm{T}$~\cite{litvin2008ferroic}, and ET dipole $\bm{G}$~\cite{Hlinka_PhysRevLett.116.177602}, respectively.

It is noteworthy that the ``toroidal multipole" and ``higher-rank multipole" are not rare compared to the conventional E dipole and M dipole in view of the magnetic point group. 
In order to demonstrate that, we show the Venn diagrams in Fig.~\ref{fig: venn_diagram}, where we classify the electric-type multipoles (E dipole, ET monopole, and ET dipole) [Fig.~\ref{fig: venn_diagram}(a)] and the magnetic-type multipoles (M monopole, M dipole, MT monopole, and MT dipole) [Fig.~\ref{fig: venn_diagram}(b)] belonging to the totally symmetric irreducible representation under the 122 magnetic point groups. 
In the electric-type multipoles in Fig.~\ref{fig: venn_diagram}(a), the E dipole $\bm{Q}$, the ET monopole $G_0$, and the ET dipole $\bm{G}$ belong to the totally symmetric irreducible representation for 31, 32, and 43 out of 122 magnetic point groups, respectively.
In the magnetic-type multipoles in Fig.~\ref{fig: venn_diagram}(b), the M monopole, the M dipole $\bm{M}$, MT monopole $T_0$, and the MT dipole $\bm{T}$ belong to the totally symmetric irreducible representation for 32, 31, 32, and 31 out of 122 magnetic point groups, respectively. 
For higher-rank multipoles with $l \geq 2$, a similar argument holds; the E, ET, M, and MT quadrupoles belong to the totally symmetric irreducible representation for 106, 42, 42, and 42 magnetic point groups, while the E, ET, M, and MT octupoles belong to the totally symmetric irreducible representation for 58, 71, 58, and 58 magnetic point groups. 
We summarize the active multipoles for 32 crystallographic and 122 magnetic point groups in Table~\ref{tab_mp_number}. 
We also show the Venn diagrams representing the different classification of multipoles in terms of $(\mathcal{P}, \mathcal{T}, \mathcal{PT})$ symmetries in Appendix~\ref{app}.

\section{Prototype and candidate materials}
\label{sec: Prototype materials}

We present prototype and candidate materials hosting unconventional multipole orderings. 
We discuss materials with ET monopole/quadrupole in Sect.~\ref{sec: Materials with electric toroidal multipole}, those with ET dipole in Sect.~\ref{sec: Materials with electric toroidal dipole}, those with M octupole and/or MT quadrupole in Sect.~\ref{sec: Materials with magnetic octupole/magnetic toroidal quadrupole}, and those with MT dipole in Sect.~\ref{sec: Materials with magnetic toroidal dipole}. 

\subsection{Materials with electric toroidal monopole and quadrupole}
\label{sec: Materials with electric toroidal multipole}

As discussed in Fig.~\ref{fig: four_mp} and Table~\ref{table:field}, chirality, which is defined by a three-dimensional object without both spatial inversion and mirror symmetries~\cite{barron1986symmetry, barron1986true}, corresponds to the ET monopole $G_0$. 
In other words, chiral materials are microscopically characterized by $G_0$. 
We introduce two prototype chiral materials hosting $G_0$ and higher-rank ET quadrupole. 

\subsubsection{Tellurium}
\label{sec: Tellurium}

\begin{figure}[htb!]
\centering
\includegraphics[width=1.0 \hsize]{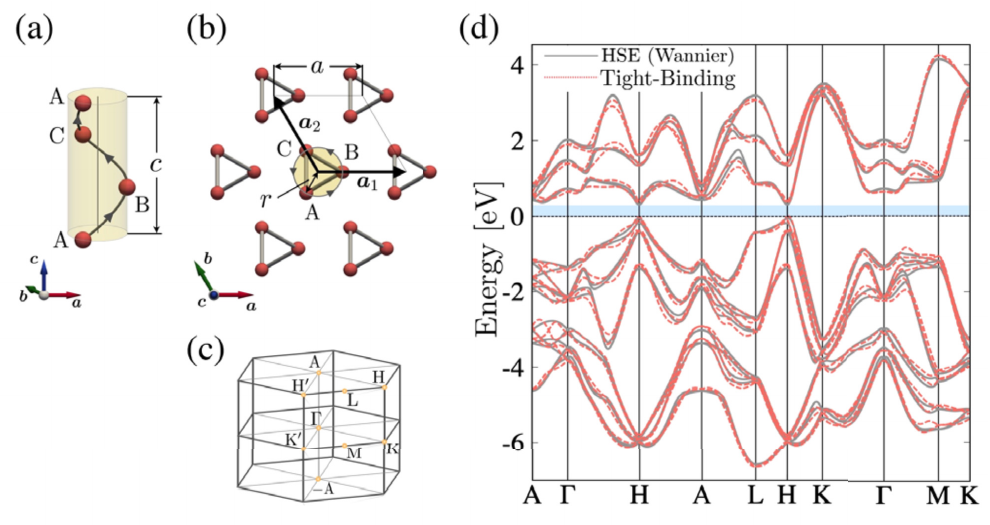} 
\caption{
\label{fig: Te1}
(a, b) Right-handed chiral crystal structure of tellurium (Te) for bulk with sublattices A, B, and C~\cite{Oiwa_PhysRevLett.129.116401}. 
(c) First Brillouin zone of the crystal structure in (b). 
(d) The band dispersion of Te. The gray solid lines represent the bands obtained by the density-functional calculations and the red dashed lines represent the bands obtained by using the symmetry-adapted multipole basis~\cite{Oiwa_PhysRevLett.129.116401}. 
Reprinted figure with permission from Ref.~\citen{Oiwa_PhysRevLett.129.116401}, Copyright (2022) by the American Physical Society.
}
\end{figure}

Elemental tellurium (Te) is a typical chiral material in a bulk, whose space group is characterized by a trigonal space group $P3_121$ $(\#152, D^4_{3})$ or $P 3_2 21$ $(\#154, D_3^6)$ depending on the right- and left-handedness. 
The bulk Te crystal consists of the threefold helical chains with A, B, and C sublattices in a hexagonal arrangement, as shown in Figs.~\ref{fig: Te1}(a) and \ref{fig: Te1}(b). 
The electronic band structure along the high-symmetric lines in the Brillouin zone [Fig.~\ref{fig: Te1}(c)], which was calculated by using the density-functional calculations, is also shown by the gray dashed line in Fig.~\ref{fig: Te1}(d). 
Since the associated point group is $D_3$, the ET monopole $G_0$ as well as the ET quadrupole $G_u$ belongs to the totally symmetric irreducible representation, in which $G_{0}$ is the essential ``order parameter'' of the chirality and $G_{u}$ gives the monoaxial anisotropy along $c$ ($z$) axis.
Thus, physical phenomena induced by $G_0$, such as the Edelstein effect and electrorotation effect, are expected, as shown in Tables~\ref{table:tensor_parity_MP_linear} and \ref{table:tensor_parity_MP_nonlinear}.

\begin{figure}[htb!]
\centering
\includegraphics[width=1.0 \hsize]{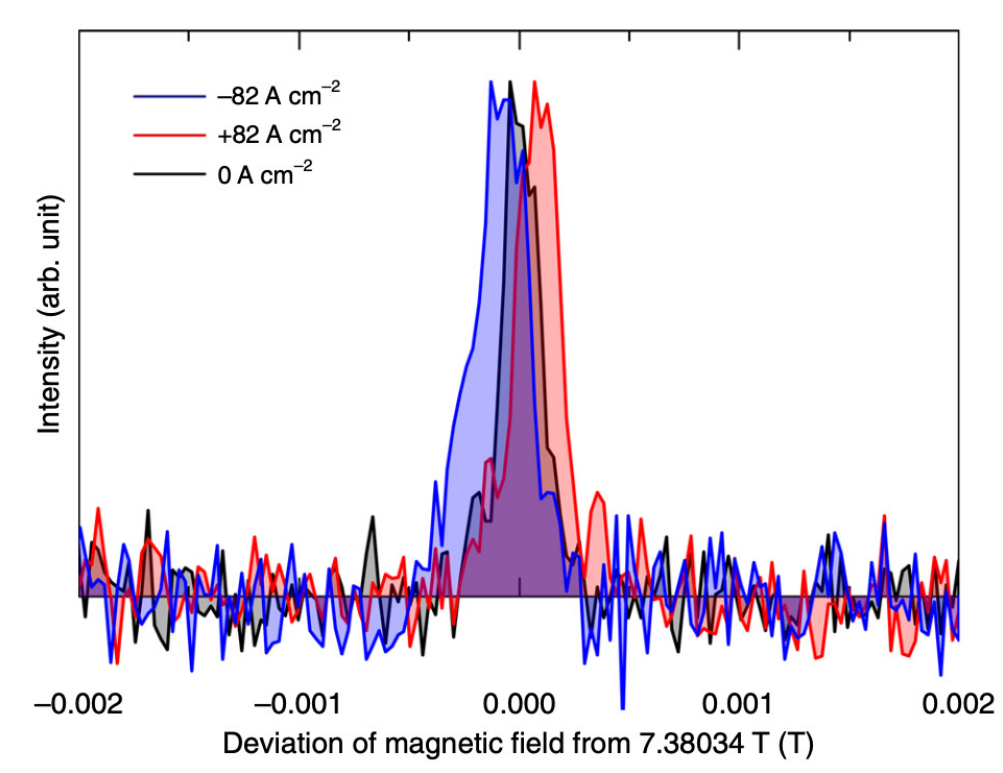} 
\caption{
\label{fig: Te2}
${}^{125}$Te-NMR spectrum on line H for pulsed currents of 0 and $\pm 82$~A cm$^{-2}$~\cite{furukawa2017observation}. 
The horizontal axis represents a deviation of the magnetic field felt by the ${}^{125}$Te nuclei from 7.38034~T, and the vertical axis represents the intensity of the NMR spectrum. 
The figure is reprinted from Ref.~\citen{furukawa2017observation}.
}
\end{figure}

Indeed, the antisymmetric spin-split band structure in the form $c_{\perp}(k_{x}\sigma_{x}+k_{y}\sigma_{y})+c_{\parallel}k_{z}\sigma_{z}$~\cite{Hirayama_PhysRevLett.114.206401, Sakano_PhysRevLett.124.136404}, the current-induced optical activity~\cite{vorob1979optical}, Edelstein effect~\cite{yoda2015current, yoda2018orbital}, and electrorotation effect~\cite{Oiwa_PhysRevLett.129.116401} were proposed and observed in the Te crystal~\cite{shalygin2012current, furukawa2017observation, Furukawa_PhysRevResearch.3.023111}. 
We show one of the typical experiments based on the ${}^{125}$Te-NMR measurement~\cite{furukawa2017observation, Furukawa_PhysRevResearch.3.023111}. 
Figure~\ref{fig: Te2} shows the NMR spectrum by applying pulsed currents of 0 and $\pm 82$~A cm$^{-2}$~\cite{furukawa2017observation}. 
Measured from the result for $0$~A cm$^{-2}$, the spectrum is shifted to a right (left) direction by 10$^{-1}$~mT while keeping its spectral shape for the positive (negative) pulse currents with $82$ ($-82$)~A cm$^{-2}$, which indicates an emergent hyperfine field generated by the current-induced magnetization. 
Such a current-induced NMR shift is reversed when the handedness of the crystal structure is opposite~\cite{Furukawa_PhysRevResearch.3.023111}.

\begin{figure}[htb!]
\centering
\includegraphics[width=1.0 \hsize]{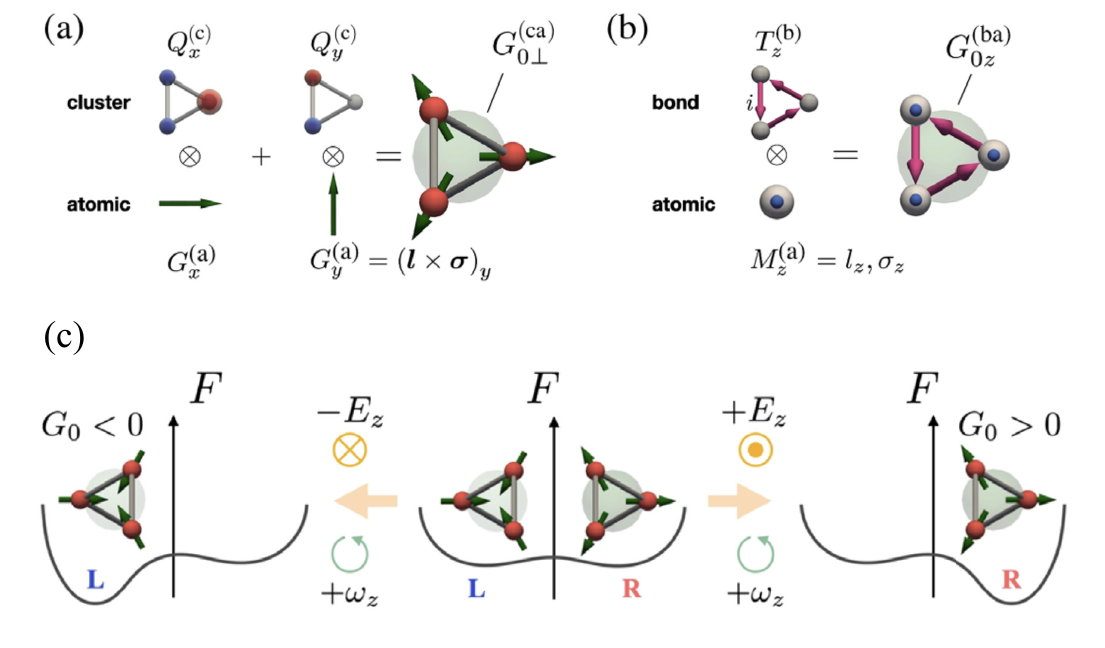} 
\caption{
\label{fig: Te3}
(a, b) The symmetry-adapted multipole bases for the ET monopole in the low-energy model for Te: (a) local ET monopole $G^{\rm (ca)}_{0\perp}$ and (b) itinerant ET monopole $G^{\rm (ba)}_{0z}$~\cite{Oiwa_PhysRevLett.129.116401}. 
(c) Absolute enantioselection by applying both rotation ($\omega_z$) and electric ($E_z$) fields~\cite{Oiwa_PhysRevLett.129.116401}. 
The positive (negative) sign of $\omega_z E_z$ favors the right (left) handedness in terms of the free energy. 
Reprinted figure with permission from Ref.~\citen{Oiwa_PhysRevLett.129.116401}, Copyright (2022) by the American Physical Society.
}
\end{figure}

The microscopic origin of the ET monopole $G_0$ was studied based on the symmetry-adapted multipole basis~\cite{Oiwa_PhysRevLett.129.116401}. 
The red dashed lines in Fig.~\ref{fig: Te1}(d) stand for the electronic band dispersions for an effective tight-binding model constructed by using the symmetry-adapted multipole basis. 
It is found that three types of $G_0$ give dominant contributions to the tight-binding model, whose expressions are represented by 
\begin{align}
G^{\rm (ca)}_{0\perp} &= \frac{1}{\sqrt{2}} (Q^{\rm (c)}_x \otimes G^{\rm (a)}_x +Q^{\rm (c)}_y \otimes G^{\rm (a)}_y), \\ 
G^{\rm (ba)}_{0z} &=  T^{\rm (b)}_z \otimes \sigma^{\rm (a)}_{z},  \\
G^{\rm (ba)}_{0\perp} &= \frac{1}{\sqrt{2}} (T^{\rm (b)}_x \otimes \sigma^{\rm (a)}_x +T^{\rm (b)}_y \otimes \sigma^{\rm (a)}_y),
\end{align}
where the superscripts, c, b, and a, denote cluster, bond, and atomic basis, respectively. 
The matrix for the site-cluster E dipole $Q^{\rm (c)}_\mu$ $(\mu=x,y)$ and bond-cluster MT dipole $T^{\rm (b)}_\mu$ ($\mu=x,y,z$) in the ABC sublattice space is defined as 
\begin{align}
Q_x^{\rm (c)}&= \frac{1}{\sqrt{6}}\left(
\begin{array}{ccc}
-1 & 0 & 0 \\
0 & 2 & 0 \\
0 & 0 & -1
\end{array}
\right),
\cr 
Q_y^{\rm (c)}&= \frac{1}{\sqrt{2}}\left(
\begin{array}{ccc}
-1 & 0 & 0 \\
0 & 0 & 0 \\
0 & 0 & 1
\end{array}
\right), 
\cr  
T_x^{\rm (b)}&= \frac{1}{2}\left(
\begin{array}{ccc}
0 & -i & 0 \\
i & 0 & i \\
0 & -i & 0
\end{array}
\right),
\cr
T_y^{\rm (b)}&= \frac{1}{2\sqrt{3}}\left(
\begin{array}{ccc}
0 & -i & -2i \\
i & 0 & -i \\
2i & i & 0
\end{array}
\right), 
\cr  
T_z^{\rm (b)}&= \frac{1}{\sqrt{6}}\left(
\begin{array}{ccc}
0 & -i & i \\
i & 0 & -i \\
-i & i & 0
\end{array}
\right). 
\end{align}
The schematic pictures of $G^{\rm (ca)}_{0\perp}$ and $G^{\rm (ba)}_{0z}$ are shown in Figs.~\ref{fig: Te3}(a) and \ref{fig: Te3}(b), respectively: The former is described by local degrees of freedom consisting of the flux structure in terms of the ET dipole $\bm{G}^{\rm (a)}$, while the latter is described by itinerant degrees of freedom consisting of the spin-dependent imaginary hopping like the spin-orbit coupling in the form of $c_{\perp}(k_{x}\sigma_{x}+k_{y}\sigma_{y})+c_{\parallel}k_{z}\sigma_{z}$ as shown in Table~\ref{table: asoc}. 
By noting that the electric current is the same symmetry as the MT dipole, one finds that itinerant $G^{\rm (ba)}_{0z}, G^{\rm (ba)}_{0\perp} \propto c_{1}(T_{x}\sigma_{x}+T_{y}\sigma_{y})+c_{2}T_{z}\sigma_{z} $ become the microscopic origin of the Edelstein effect as discussed above~\cite{furukawa2017observation, Furukawa_PhysRevResearch.3.023111}. 

Meanwhile, local $G^{\rm (ca)}_{0\perp}$ can contribute to another cross correlation, i.e., the electrorotation effect, since its expression includes both the E dipole and ET dipole. 
By using such a cross correlation, one may achieve absolute enantioselection in chiral materials by applying a rotation field $\bm{\omega}$ and an electric field $\bm{E}$ simultaneously in parallel direction, as shown in Fig.~\ref{fig: Te3}(c).

\subsubsection{Cd$_2$Re$_2$O$_7$}

\begin{figure}[htb!]
\centering
\includegraphics[width=0.8 \hsize]{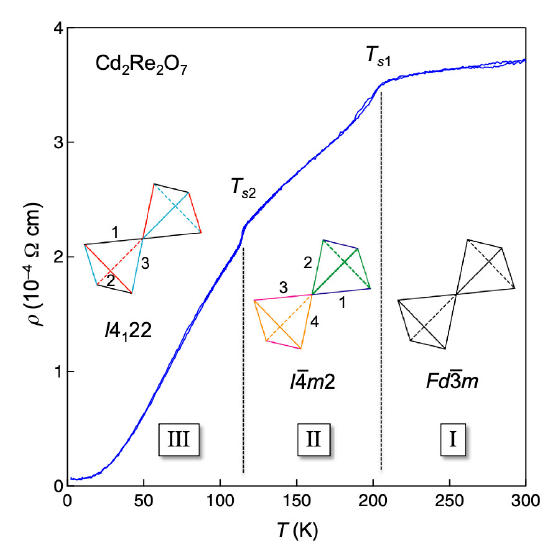} 
\caption{
\label{fig: Cd2Re2O7}
Temperature ($T$) dependence of resistivity $(\rho)$ in Cd$_2$Re$_2$O$_7$, where three phases, I, II, and III, appear~\cite{hiroi2018pyrochlore}. 
The deformation of the Re tetrahedral network is shown in each phase; the identical bonds are distinguished by the colors and numbers. 
Reprinted figure with permission from Ref.~\citen{hiroi2018pyrochlore}, Copyright (2018) by the Physical Society of Japan. 
}
\end{figure}

Another example is the pyrochlore oxide Cd$_2$Re$_2$O$_7$ under the space group $Fd\bar{3}m$, where Re$^{5+}$ ions form a regular corner-sharing tetrahedral network. 
Although this compound has drawn attention as the only superconductor in the family of $\alpha$-pyrochlore oxides~\cite{Hanawa_PhysRevLett.87.187001,sakai2001superconductivity, Jin_PhysRevB.64.180503,hiroi2002high,hiroi2002superconducting}, it also exhibits peculiar phase transitions accompanying the spontaneous inversion symmetry breaking in the strong spin-orbit coupling, as shown in Fig.~\ref{fig: Cd2Re2O7}~\cite{yamaura2002low, hiroi2018pyrochlore}. 
From various measurements, such as the single-crystal x-ray diffraction~\cite{yamaura2002low, Castellan_PhysRevB.66.134528}, powder neutron diffraction~\cite{weller2004pyrochlore}, Raman spectroscopy~\cite{Kendziora_PhysRevLett.95.125503}, nonlinear optics~\cite{petersen2006nonlinear}, polarizing microscope image~\cite{matsubayashi2018formation}, and high-resolution synchrotron x-ray diffraction~\cite{hirai2022successive}, two noncentrosymmetric tetragonal phases have been identified: The one with $I\bar{4}m2$ (phase II) symmetry appears in the temperature range $T_{s2} \leq T \leq T_{s1}$, while the other with $I4_1 22$ (phase III) symmetry appears in $T \leq T_{s2}$~\cite{comment_Cd2Re2O7_inter}, both of which belong to the $E_u$ representation under $Fd\bar{3}m$ symmetry in the high-temperature phase (phase I).  

\begin{figure}[t!]
\centering
\includegraphics[width=1.0 \hsize]{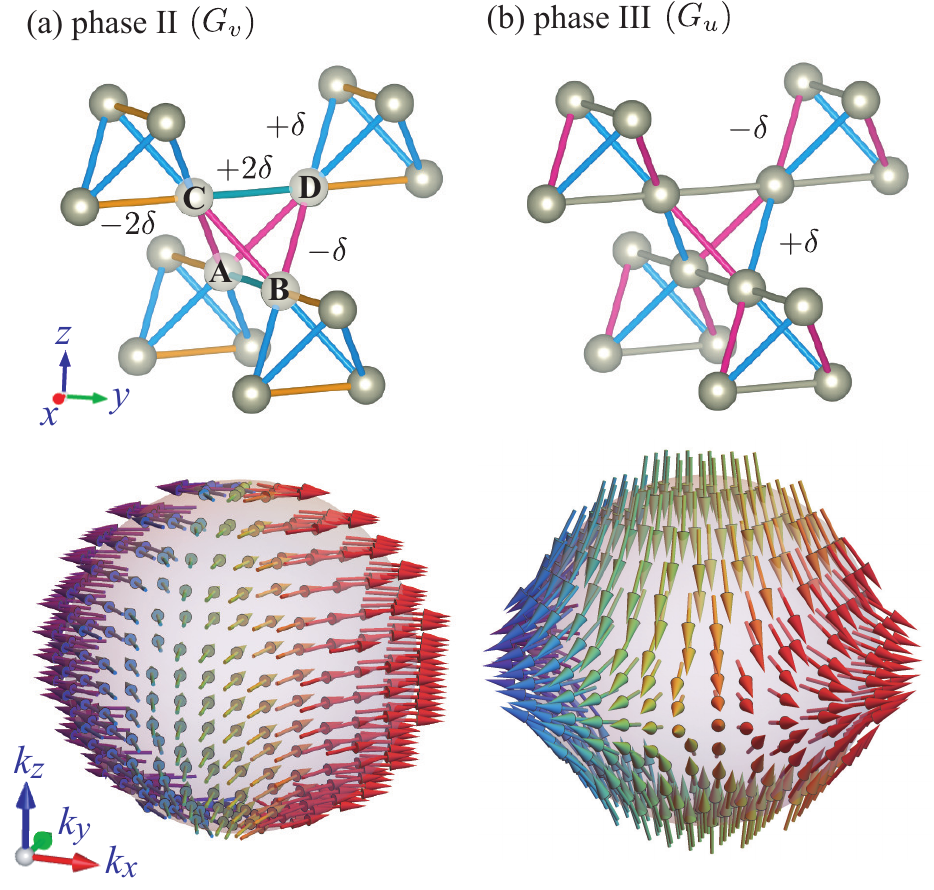} 
\caption{
\label{fig: Cd2Re2O7_2}
Bond modulations under (a) the $x^2-y^2$-type ET quadrupole ($G_v$) and (b) the $3z^2-r^2$-type ET quadrupole ($G_u$)~\cite{Hayami_PhysRevLett.122.147602}. 
The bottom panel shows the schematic spin polarizations on one of the Fermi surfaces split by the above ET quadrupoles. 
Reprinted figure with permission from Ref.~\citen{Hayami_PhysRevLett.122.147602}, Copyright (2019) by the American Physical Society.
}
\end{figure}

From the compatible relation in Table~\ref{tab_multipoles_table1}, the multipole order parameters in phases II and III correspond to the ET quadrupoles with different components: $G_v$ for phase II and $G_u$ for phase III~\cite{Matteo_PhysRevB.96.115156, Hayami_PhysRevLett.122.147602}; it is noted that ET monopole $G_0$ is also activated in phase III since this phase belongs to Sohncke group. 
In this sense, phase III in Cd$_2$Re$_2$O$_7$ is another chiral material hosting $G_0$. 
Although the transition is considered to be of electronic origin owing to the very small tetragonal lattice distortion~\cite{Castellan_PhysRevB.66.134528}, the microscopic electronic degrees of freedom to activate odd-parity ET multipoles have not been directly clarified yet; any four-sublattice orderings by using the charge/spin/orbital degrees of freedom in the pyrochlore structure do not violate spatial inversion symmetry.
Besides, the possibility of the spontaneous parity mixing between orbitals with different parity, which is another route to break the spatial inversion symmetry, can be also ruled out from the density-functional calculations, where the contribution from the $p$-$d$ hybridization is small. 

The remaining possibility is to use the bond degree of freedom. 
In Ref.~\citen{Hayami_PhysRevLett.122.147602}, it was clarified that the four-sublattice bond ordering breaks the spatial inversion symmetry while accompanying the ET quadrupoles. 
Figures~\ref{fig: Cd2Re2O7_2}(a) and (b) show the modulations under the bond orderings with $G_v$ and $G_u$, respectively, whose patterns are consistent with those observed by the single-crystal x-ray diffraction~\cite{yamaura2002low}. 
By combining the spin-orbit coupling, these bond-cluster ET quadrupole orderings exhibit antisymmetric spin splitting in the form of $k_x \sigma_x - k_y \sigma_y$ in phase II [Fig.~\ref{fig: Cd2Re2O7_2}(a)] and $c_1 (k_x \sigma_x + k_y \sigma_y ) + c_2 k_z \sigma_z$ in phase III [Fig.~\ref{fig: Cd2Re2O7_2}(b)] as shown in Table~\ref{table: asoc}.

\subsection{Materials with electric toroidal dipole}
\label{sec: Materials with electric toroidal dipole}

The ET dipole $\bm{G}$, which has been referred to as ferro-rotational order or ferro-axial order, is one of the fundamental dipoles invariant under both spatial inversion and time-reversal operations~\cite{Hlinka_PhysRevLett.113.165502, Hlinka_PhysRevLett.116.177602, jin2020observation, cheong2021permutable, cheong2022linking}. 
Although its physical properties have been obscure owing to the absence of its conjugate electromagnetic fields, recent theoretical studies have clarified the emergence of transverse responses of the conjugate physical quantities, such as antisymmetric thermopolarization~\cite{Nasu_PhysRevB.105.245125}, longitudinal spin current generation~\cite{Roy_PhysRevMaterials.6.045004, Hayami_doi:10.7566/JPSJ.91.113702}, nonlinear transverse magnetization~\cite{inda2023nonlinear, Hayami_PhysRevB.108.094106}, unconventional Hall effect~\cite{Hayami_PhysRevB.108.085124}, and second-order nonlinear magnetostriction~\cite{kirikoshi2023rotational}. 
We introduce two candidate materials exhibiting such cross correlations driven by $\bm{G}$. 

\subsubsection{RbFe(MoO$_4$)$_2$}

\begin{figure}[htb!]
\centering
\includegraphics[width=0.6 \hsize]{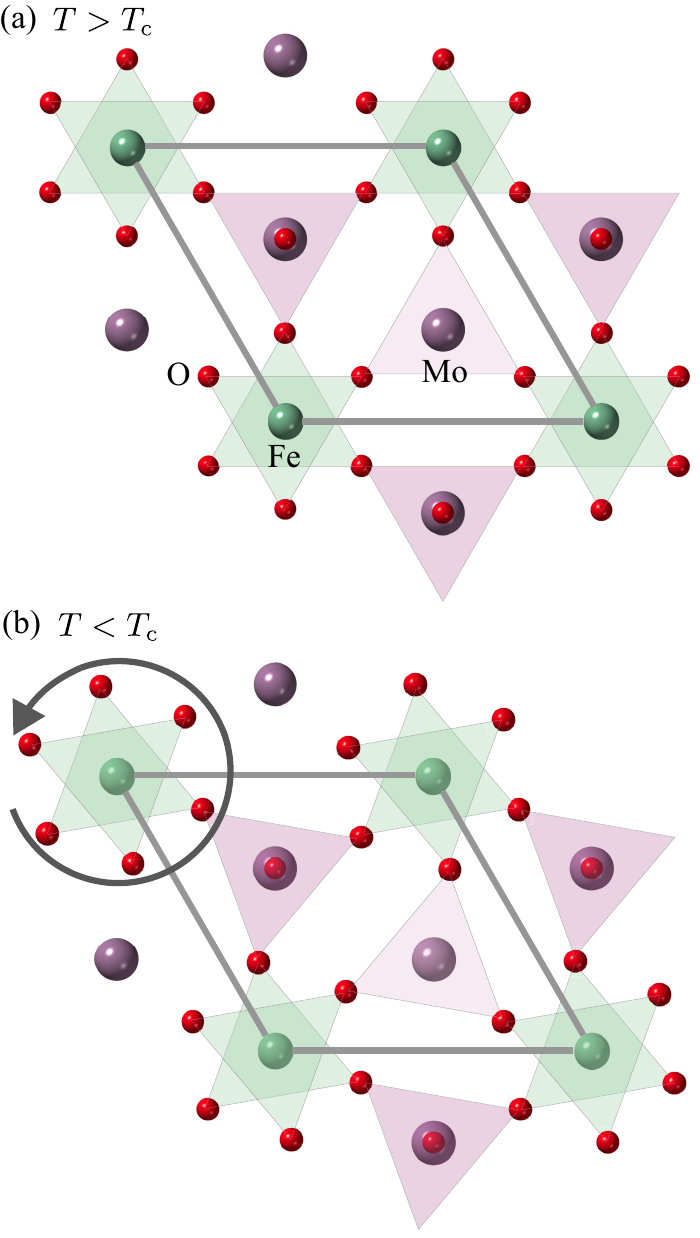} 
\caption{
\label{fig: ReFe(MoO4)2}
The crystal structure of RbFe(MoO$_4$)$_2$ in (a) the high-temperature phase above the critical temperature $(T_{\rm c})$ and (b) the low-temperature phase below $T_{\rm c}$~\cite{jin2020observation}. 
}
\end{figure}

\begin{figure}[htb!]
\centering
\includegraphics[width=1.0 \hsize]{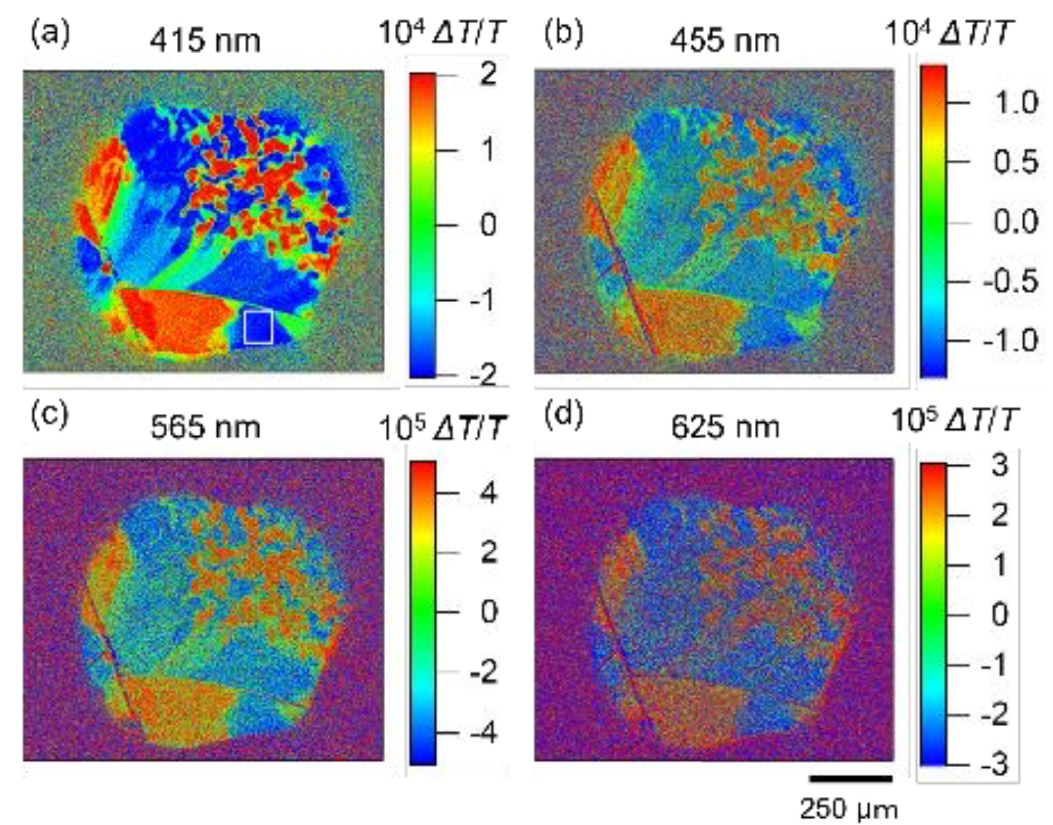} 
\caption{
\label{fig: ReFe(MoO4)2_2}
Electrogyration for RbFe(MoO$_4$)$_2$~\cite{Hayashida_PhysRevMaterials.5.124409}. 
The contour plot shows the relative rotation angle of light polarization $\Delta T/ T$ for different wavelengths: (a) 415~nm, (b) 455~nm, (c) 565~nm, and (d) 625~nm. 
Reprinted figure with permission from Ref.~\citen{Hayashida_PhysRevMaterials.5.124409}, Copyright (2021) by the American Physical Society.
}
\end{figure}

RbFe(MoO$_4$)$_2$ is one of the typical materials showing the phase transition in terms of the ET dipole. 
The crystal structure of RbFe(MoO$_4$)$_2$ is shown in Fig.~\ref{fig: ReFe(MoO4)2}(a), which consists of stacks of FeO$_6$ octahedra sharing vertices with MoO$_4$ tetrahedra. 
This material undergoes a structural phase transition at $T_{\rm c} \sim 195$~K from $\bar{3}m$ ($D_{\rm 3d}$) to $\bar{3}$ ($C_{\rm 3i}$); octahedra (tetrahedra) rotates counterclockwise (clockwise) about the $c$ axis, as shown in Fig.~\ref{fig: ReFe(MoO4)2}(b)~\cite{jin2020observation}. 
From Table~\ref{tab_multipoles_table2}, the site-cluster ET dipole corresponds to the order parameter in this transition. 

The above structural transition was identified by performing the high-sensitivity rotational-anisotropy second-harmonic generation, where the E quadrupole contribution as $P^{\rm eff}_i = \chi^{\rm EQ}_{ijkl} E_j (\omega) \partial_k E_l (\omega)$ is analyzed; $\chi^{\rm EQ}_{ijkl}$ is the rank-4 tensor, which is related to rank 0--4 even-parity electric-type multipoles including the ET dipole degree of freedom. 
The data is explained by the coexistence of two domains with the different signs of the ET dipole. 

Subsequently, domain formation of the ET dipole was detected by electric-field-induced optical rotation, i.e., electrogyration~\cite{Hayashida_PhysRevMaterials.5.124409}. 
This effect is understood from the effective coupling under the ET dipole $\bm{G}$, which has a correspondence of $\bm{G}\leftrightarrow G_0 \bm{Q}$ as similar to Eq.~(\ref{eq: G0_1}),  in Table~\ref{table: coupling_mp} and Fig.~\ref{fig: cross_mp}; $\bm{Q}$ corresponds to the electric field and $G_0$ corresponds to the chirality yielding the optical rotation.

Figures~\ref{fig: ReFe(MoO4)2_2}(a)--\ref{fig: ReFe(MoO4)2_2}(d) represent two-dimensional maps of $\Delta T /T$ obtained by the electrogyration on RbFe(MoO$_4$)$_2$ for different wavelengths, where $\Delta T / T$ is proportional to the rotation angle of the light polarization and its sign depends on the direction of $\bm{G}$. 
In other words, positive and negative $\Delta T / T$ represent different domain states.
As shown in Figs.~\ref{fig: ReFe(MoO4)2_2}(a)--\ref{fig: ReFe(MoO4)2_2}(d), the multi domains are formed in RbFe(MoO$_4$)$_2$ through the phase transition in terms of the ET dipole. 
The behavior of the order parameter was well fitted by considering the first-order displacive-type structural transition.

A similar technique was also used for another candidate material NiTiO$_3$, where the domain size depends on the cooling rate around the transition temperature~\cite{hayashida2020visualization, Hayashida_PhysRevMaterials.5.124409, fang2023ferrorotational}. 
In addition, further experimental techniques, such as three-dimensional imaging using circularly polarized second harmonic generation microscopy, have been developed in order to identify the ET dipole in real space~\cite{yokota2022three}. 
These techniques are powerful tools for detecting the ET dipole in other candidate materials, such as Co$_3$Nb$_2$O$_8$~\cite{Johnson_PhysRevLett.107.137205}, CaMn$_7$O$_{12}$~\cite{Johnson_PhysRevLett.108.067201}, BaCoSiO$_4$~\cite{Xu_PhysRevB.105.184407}, K$_2$Zr(PO$_4$)$_2$~\cite{yamagishi2023ferroaxial}, Na$_2$Hf(BO$_3$)$_2$~\cite{nagai2023chemicalSwitching}, and Na-superionic conductors~\cite{nagai2023chemical}.

\subsubsection{Ca$_5$Ir$_3$O$_{12}$}

\begin{figure}[htb!]
\centering
\includegraphics[width=1.0 \hsize]{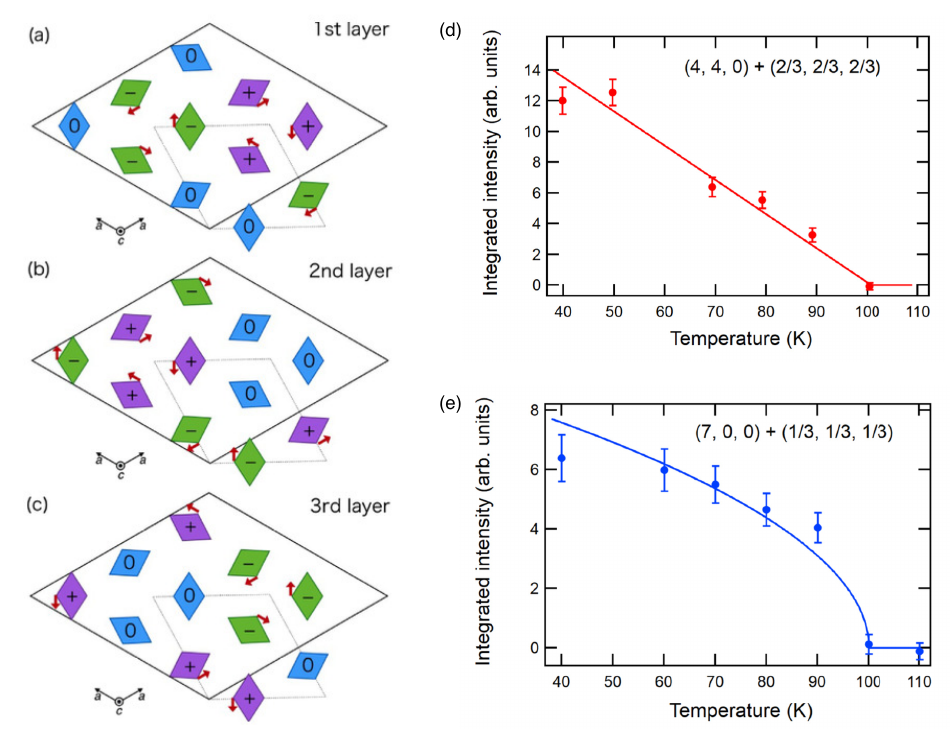} 
\caption{
\label{fig: Ca5Ir3O12}
(a,b,c) Superlattice structure below $T_{\rm s}=105$~K for (a) the first layer $(0 \leq z \leq 1/3)$, (b) the second layer $(1/3 \leq z \leq 2/3)$, and (c) the third layer $(2/3 \leq z \leq 1)$~\cite{hanate2021first}. 
Each rhombus represents an IrO$_6$ octahedron with displacements of O(1) atoms; the arrows represent the displacement direction and the sign corresponds to that of the $z$ component of the ET dipole.  
(d,e) Temperature dependence of the integrated intensity of the superlattice reflection at  (d) $\bm{Q}_{nn0}=(4,4,0)+(2/3,2/3,2/3)$ and (e) $\bm{Q}_{n00}=(7,0,0)+(1/3,1/3,1/3)$, which were obtained by synchrotron x-ray diffraction measurements~\cite{hanate2023space}. 
The solid lines are the fitting ones. 
Reprinted figures (a)--(c) with permission from Ref.~\citen{hanate2021first}, Copyright (2021) by the Physical Society of Japan.
Reprinted figures (d) and (e) with permission from Ref.~\citen{hanate2023space}, Copyright (2023) by the Physical Society of Japan.
}
\end{figure}

Ca$_5$Ir$_3$O$_{12}$ is another candidate material exhibiting the phase transition in terms of the ET dipole. 
The crystal structure of Ca$_5$Ir$_3$O$_{12}$ is a hexagonal structure with a space group of $P\bar{6}2m$~\cite{wakeshima2003electrical}. 
This material undergoes two phase transitions at $T_{\rm s}=105$~K and $T_{\rm m}=7.8$~K~\cite{wakeshima2003electrical, Matsuhira_doi:10.7566/JPSJ.87.013703}: The former transition corresponds to a nonmagnetic structural phase transition, while the latter transition corresponds to a magnetic phase transition. 

The former transition has been identified through several experiments, such as Raman scattering experiments~\cite{Hasegawa_doi:10.7566/JPSJ.89.054602}, inelastic x-ray scattering experiments~\cite{Hanate_doi:10.7566/JPSJ.89.053601}, and synchrotron x-ray diffraction experiments~\cite{hanate2023space}. 
Figures~\ref{fig: Ca5Ir3O12}(a)--\ref{fig: Ca5Ir3O12}(c) show the $\sqrt{3}a\times \sqrt{3}a\times 3c$ superlattice structure in the hexagonal coordinate, which was obtained by inelastic x-ray scattering experiments~\cite{hanate2021first}. 
The displacement of O(1) atoms in each IrO$_6$ chain has been found below $T_{\rm s}$ while keeping the threefold rotational symmetry but breaking the mirror symmetry, which means that the order parameter corresponds to the ${\rm A}_2$ representation at $\bm{q}=(1/3,1/3,1/3)$.
This is consistent with the ordering of the three-site cluster ET dipole $G_{z}$ with the modulation vector $\bm{q}$.
In contrast to RbFe(MoO$_4$)$_2$, the site-cluster ET dipole shows spatial dependence characterized by $\bm{q}=(1/3,1/3,1/3)$. 
In other words, the order parameter is regarded as the density wave of the site-cluster ET dipole. 
Such a superlattice structure was also confirmed by synchrotron x-ray diffraction experiments~\cite{hanate2023space}; the characteristic temperature dependence of the integrated intensity has been found as shown in Figs.~\ref{fig: Ca5Ir3O12}(d) and \ref{fig: Ca5Ir3O12}(e). 
When the reciprocal lattice vectors are $\bm{Q}_{nn0}=(4,4,0)+(2/3,2/3,2/3)$ ($\bm{Q}_{n00}=(7,0,0)+(1/3,1/3,1/3)$), the integrated intensity is proportional to $T_{\rm s}-T$ ($\sqrt{T_{\rm s}-T}$). 
The recent symmetry and microscopic analyses indicate that such an ET dipole ordering is constructed from the trimer-type E quadrupole ordering~\cite{hayami2023cluster}. 

Although a detailed structure for the latter magnetic transition has not been revealed so far, a possibility of multipole magnetic ordering was implied~\cite{HANATE2022170072} from powder neutron diffraction experiments and ${}^{193}$Ir synchrotron-radiation-based M\"{o}ssbauer spectroscopy experiments~\cite{hayami2023cluster}. 
A possible multipole structure consists of the uniform and $(1/3,1/3,1/3)$ modulations of three-site cluster MT quadrupole $T_u$~\cite{hayami2023cluster}, which consists of three M octupoles.

\subsection{Materials with magnetic octupole and magnetic toroidal quadrupole}
\label{sec: Materials with magnetic octupole/magnetic toroidal quadrupole}

As discussed in Sect.~\ref{sec: Cluster multipole}, the unconventional physical phenomena in AFMs can be expected and understood by introducing the concept of cluster multipoles with a ferroic alignment of higher-rank multipoles. 
We introduce two prototypes as examples.

\subsubsection{Mn$_3$Sn}

\begin{figure}[htb!]
\centering
\includegraphics[width=1.0 \hsize]{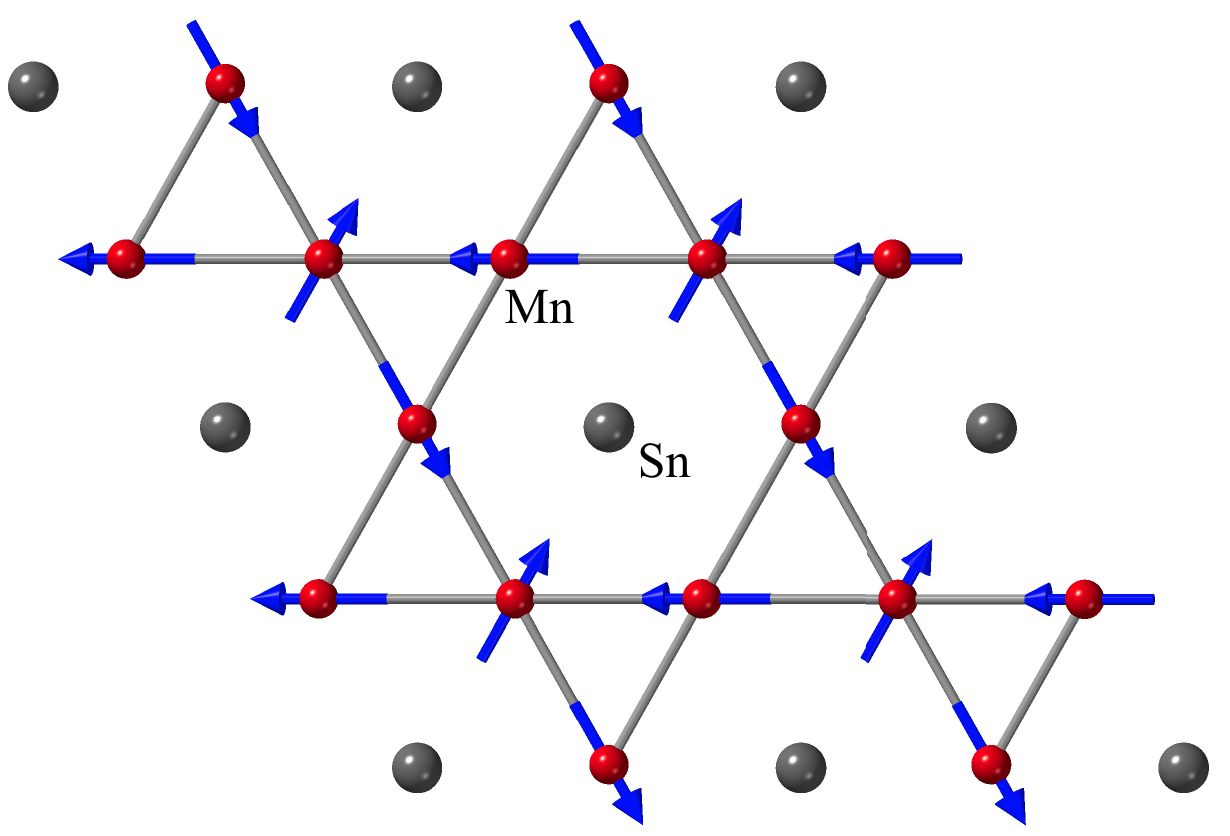} 
\caption{
\label{fig: Mn3Sn}
Crystal and antiferromagnetic structures of Mn$_3$Sn in the $ab$ plane~\cite{nakatsuji2015large}. 
}
\end{figure}

Mn$_3$Sn belongs to a hexagonal Ni$_3$Sn-type structure with space group $P6_3/mmc$. 
The Mn ions form the slightly distorted kagome network in the $ab$ plane, as shown in Fig.~\ref{fig: Mn3Sn}. 
The AFM phase transition occurs at $T_{\rm N}\simeq 420$~K, where Mn ions show noncollinear $120^{\rm \circ}$ magnetic ordering with the vector chirality, as shown in Fig.~\ref{fig: Mn3Sn}~\cite{nakatsuji2015large}; the Mn ion per site has magnetic moments of 3$\mu_{\rm B}$, whereas it has a small net ferromagnetic moment of $\sim 0.002 $ $\mu_{\rm B}$

In spite of the small net magnetization, this compound exhibits the large anomalous Hall effect at room temperature~\cite{nakatsuji2015large}. 
The magnitude of the Hall conductivity is larger than that found in conventional ferromagnets, such as Fe, Co, and Ni. 
Subsequently, it was revealed that Mn$_3$Sn shows various physical phenomena expected in ferromagnets~\cite{nakatsuji2022topological, smejkal2022anomalous}, such as the anomalous Nernst effect~\cite{ikhlas2017large}, magneto-optical Kerr effect~\cite{higo2018large}, magnetic spin Hall effect~\cite{kimata2019magnetic}, and switching by spin-orbit torque~\cite{tsai2020electrical}.

\begin{figure}[htb!]
\centering
\includegraphics[width=0.7 \hsize]{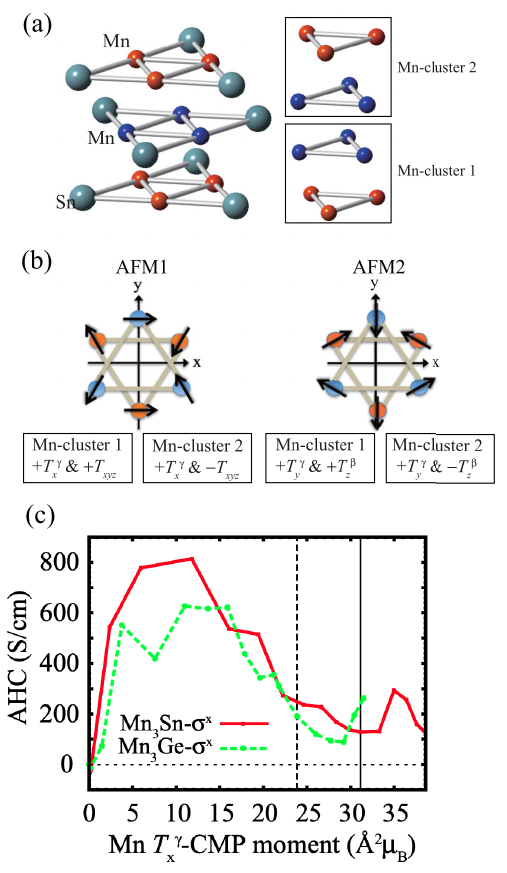} 
\caption{
\label{fig: Mn3Sn_3}
(a) Crystal structure and the Mn clusters in Mn$_3$Sn~\cite{Suzuki_PhysRevB.95.094406}. 
(b) Antiferromagnetic structure corresponding to Fig.~\ref{fig: Mn3Sn}(c). 
The AFM1 and AFM2 represent the AFM spin patterns realized under the magnetic fields along the $x$ and $y$ directions, respectively. 
$T^{\gamma}_{x, y}$, $T_{xyz}$, and $T_z^\beta$ represent the site-cluster M octupoles. 
(c) The anomalous Hall conductivity (AHC) against the site-cluster M octupole $T^\gamma_x$, which was obtained by the first-principles calculations for Mn$_3$Sn and Mn$_3$Ge~\cite{Suzuki_PhysRevB.95.094406}. 
Reprinted figure with permission from Ref.~\citen{Suzuki_PhysRevB.95.094406}, Copyright (2017) by the American Physical Society.
}
\end{figure}

The emergence of the above ferromagnetic-related physical phenomena was understood from the cluster multipole theory~\cite{Suzuki_PhysRevB.95.094406}. 
Suzuki and his coworkers demonstrated that the six-sublattice AFM structure in Figs.~\ref{fig: Mn3Sn_3}(a) and \ref{fig: Mn3Sn_3}(b) corresponds to the site-cluster M octupole, which belongs to the same irreducible representation of M dipole~\cite{comment_Mn3Sn}, i.e., ferromagnetic moment. 
They have also shown the correlation between the magnitude of the cluster multipole moment and the anomalous Hall conductivity by analyzing the realistic tight-binding model obtained from the first-principles band structures and the WANNIER90, as shown in Fig.~\ref{fig: Mn3Sn_3}(c).  

The above ferromagnetic-like physical phenomena under the AFM ordering have drawn attention in recent years~\cite{nakatsuji2022topological, smejkal2022anomalous}. 
Various materials have been proposed and identified so far: Mn$_3$Ir~\cite{Chen_PhysRevLett.112.017205, Chen_PhysRevB.101.104418}, antiperovskite AFM Mn$_3$$A$N ($A=$ Ga, Sn, and Ni)~\cite{Gurung_PhysRevMaterials.3.044409, Zhou_PhysRevB.99.104428, Boldrin_PhysRevMaterials.3.094409, Huyen_PhysRevB.100.094426, you2021cluster}, the pyrochlore oxides~\cite{Tomizawa_PhysRevB.80.100401, kim2020strain}, the bilayer MnPSe$_3$~\cite{Sivadas_PhysRevLett.117.267203}, $\kappa$-type organic conductors~\cite{Naka_PhysRevB.102.075112}, and other materials/situations~\cite{vsmejkal2020crystal,yamasaki2020augmented, Chen_PhysRevB.106.024421}. 
It is noteworthy that the conditions of emergent anomalous Hall effect in AFMs are well explained by the anisotropic M dipole~\cite{Hayami_PhysRevB.103.L180407}, as discussed in Sect.~\ref{sec: Site-cluster multipole}.

\subsubsection{Cd$_2$Os$_2$O$_7$}

\begin{figure}[htb!]
\centering
\includegraphics[width=0.75 \hsize]{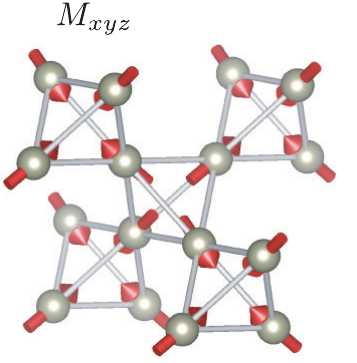} 
\caption{
\label{fig: pyrochlore_Mxyz}
All-in/all-out order in the pyrochlore structure. 
}
\end{figure}

As another example hosting a higher-rank cluster multipole, we introduce the pyrochlore oxide Cd$_2$Os$_2$O$_7$, which exhibits the AFM transition at 225~K; the all-in/all-out spin arrangement in Fig.~\ref{fig: pyrochlore_Mxyz} is realized below the critical temperature~\cite{Yamaura_PhysRevLett.108.247205}. 
A similar magnetic order has also been found in Er$_2$Ti$_2$O$_7$~\cite{poole2007magnetic}. 

The all-in/all-out ordering corresponds to the $xyz$ component of the site-cluster M octupole, i.e., $M_{xyz}$. 
In contrast to Mn$_3$Sn, $M_{xyz}$ belongs to the different irreducible representation from that of M dipole; no ferromagnetic-like physical phenomena appear. 
Instead, this material exhibits physical phenomena characteristic of the M octupole. 
For example, this material is expected to exhibit the symmetric spin-split band structure in the form of $k_y k_z \sigma_x + k_z k_x \sigma_y + k_x k_y \sigma_z$. 
In addition, magnetic-field-induced striction, i.e., magnetostriction, is expected~\cite{Arima_doi:10.7566/JPSJ.82.013705}; $xy$-, $yz$-, and $zx$-type deformations are induced by the magnetic field along the $x$, $y$, and $z$ directions, respectively, which is easily understood by the above symmetric spin-orbit coupling.

\subsection{Materials with magnetic toroidal dipole}
\label{sec: Materials with magnetic toroidal dipole}

The MT dipole has long been studied in the context of multiferroics in magnetic insulators~\cite{Spaldin_0953-8984-20-43-434203, kopaev2009toroidal}. 
Meanwhile, it can be present even in magnetic metals, which leads to a linear magnetoelectric effect and nonreciprocal transport~\cite{Hayami_PhysRevB.90.024432}. 
We here introduce two materials accompanying the MT dipole with an emphasis on the metallic system.

\subsubsection{UNi$_4$B}

\begin{figure}[htb!]
\centering
\includegraphics[width=1.0 \hsize]{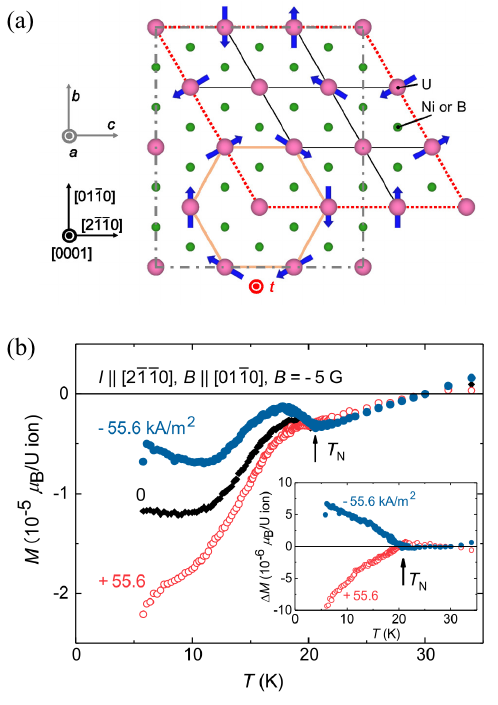} 
\caption{
\label{fig: UNi4B}
(a) Crystal and magnetic structures of UNi$_4$B~\cite{saito2018evidence}. 
Blue arrows indicate magnetic moments of U ions below the critical temperature. 
(b) The temperature $T$ dependence of the magnetization $M$ in the presence of the electric current density $I=\pm 55.6$~kA/m$^2$ and $B=-5$~G~\cite{saito2018evidence}. 
The inset shows $T$ dependence of $\Delta M$. 
Reprinted figure from Ref.~\citen{saito2018evidence}, Copyright (2018) The Authors.
}
\end{figure}

\begin{figure}[htb!]
\centering
\includegraphics[width=0.8 \hsize]{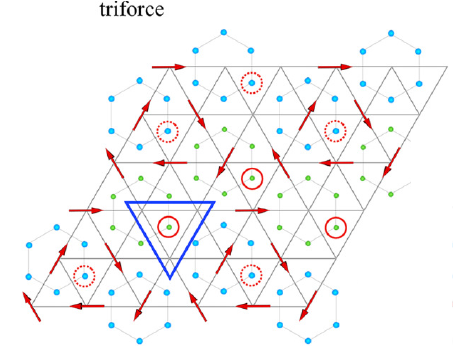} 
\caption{
\label{fig: UNi4B_2}
A different candidate magnetic structure of UNi$_4$B~\cite{Ishitobi_PhysRevB.107.104413}. 
Reprinted figure with permission from Ref.~\citen{Ishitobi_PhysRevB.107.104413}, Copyright (2023) by the American Physical Society.
}
\end{figure}

One of the recent examples hosting the MT dipole in metals is the UNi$_4$B~\cite{Mentink1994, Oyamada2007}. 
This material exhibits a partially-disordered AFM state at $T_{\rm N}= 20.4$~K; 2/3 of U ions exhibit the vortex-type ordering, while the remaining 1/3 of U ions are still disordered, as shown in Fig.~\ref{fig: UNi4B}(a). 
This AFM structure is classified into the site-cluster MT dipole with the $z$ component, which becomes the origin of the linear magnetoelectric effect and nonreciprocal transport~\cite{Hayami_PhysRevB.90.024432}.  
Indeed, careful magnetization measurements under direct electric current were performed by Saito and his collaborators~\cite{saito2018evidence}; they observed that the magnetization is induced perpendicular to the electric current and MT dipole, i.e., $\bm{M} \perp \bm{I} \perp \bm{T}$, which is expected from the MT dipole ordering, as shown in Fig.~\ref{fig: UNi4B}(b). 
Furthermore, recent nonlinear transport experiments indicated the emergence of zero-field current-induced Hall effect, which was also expected from the MT dipole ordering~\cite{ota2022zero}.

Meanwhile, magnetization measurements also indicated the emergence of current-induced magnetization even when the electric current is applied to the direction parallel to $\bm{T}$, i.e., $\bm{M} \perp \bm{I} \parallel \bm{T}$, which was not explained by the magnetic ordering as shown in Fig.~\ref{fig: UNi4B}(a)~\cite{saito2018evidence}. 
In order to reconcile such discrepancy, reinvestigations of both crystal and magnetic structures have been performed~\cite{Haga,tabata2021x, Willwater_PhysRevB.103.184426, Yanagisawa_PhysRevLett.126.157201}, which implied the slight symmetry lowering from the hexagonal symmetry to orthorhombic one. 
From the theoretical side, a new magnetic structure so as to be consistent with the observation of the magnetoelectric effects in experiments has been recently proposed, as shown in Fig.~\ref{fig: UNi4B_2}~\cite{Ishitobi_PhysRevB.107.104413}.  
To settle this issue, further experiments have been suggested, such as the resonant x-ray with the associated ordering wave vectors and the NMR/NQR for ${}^{11}$B sites.

In a similar context, Ce$_3$TiBi$_5$ is another candidate material hosting the MT dipole in metals~\cite{motoyama2018magnetic}, since the linear magnetoelectric effect was observed below the AFM transition temperature~\cite{shinozaki2020magnetoelectric,shinozaki2020study}. 
Although the AFM structure has been theoretically proposed based on the present experimental results~\cite{Hayami_doi:10.7566/JPSJ.91.123701}, the detailed magnetic structure has not been fully elucidated. 
The diffraction measurements are highly desired.

\subsubsection{CuMnAs}

\begin{figure}[htb!]
\centering
\includegraphics[width=0.8 \hsize]{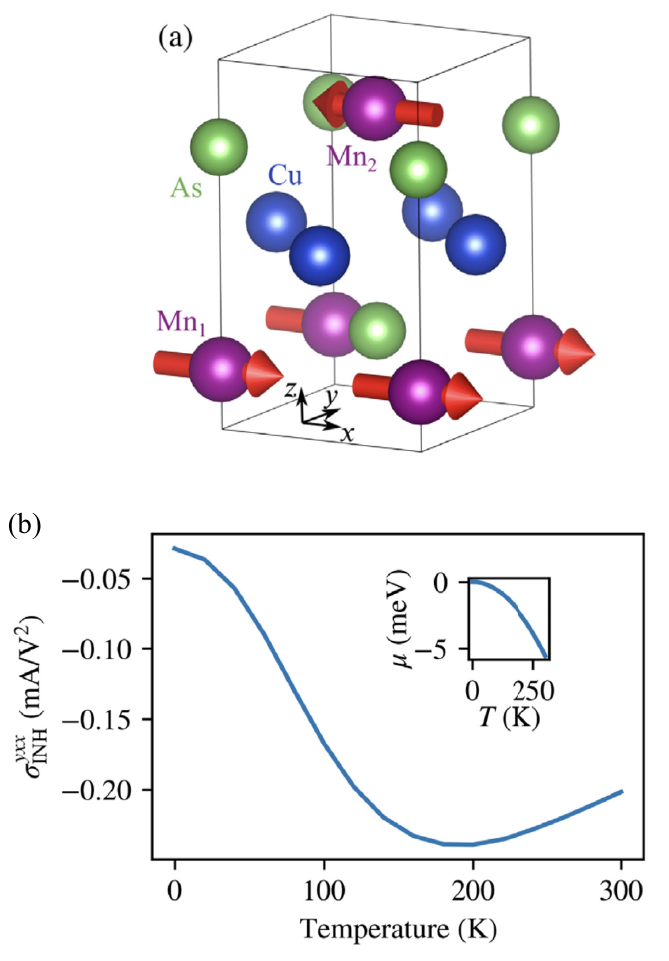} 
\caption{
\label{fig: CuMnAs}
(a) The crystal and magnetic structures of CuMnAs~\cite{Wang_PhysRevLett.127.277201}. 
(b) Temperature dependence of intrinsic nonlinear Hall conductivity in CuMnAs. The inset shows the temperature dependence of the chemical potential~\cite{Wang_PhysRevLett.127.277201}. 
Reprinted figure with permission from Ref.~\citen{Wang_PhysRevLett.127.277201}, Copyright (2021) by the American Physical Society.
}
\end{figure}

Another example is the collinear AFM CuMnAs, which has been extensively studied in the field of spintronics from the viewpoint of its electrical switching of the AFM structure~\cite{wadley2016electrical, Manchon_RevModPhys.91.035004}. 
As shown in Fig.~\ref{fig: CuMnAs}(a), the crystal structure of CuMnAs belongs to the tetragonal point group $D_{\rm 4h}$ ($4/mmm 1'$), and its magnetic structure is described by $2'/m$~\cite{wadley2013tetragonal, Wang_PhysRevLett.127.277201}, where the MT dipole belongs to the totally symmetric irreducible representation~\cite{Yatsushiro_PhysRevB.104.054412}. 
Thus, the linear magnetoelectric effect and nonreciprocal transport are expected similar to UNi$_4$B. 
As one of such physical phenomena, it was shown that the intrinsic nonlinear Hall conductivity in Eq.~(\ref{NLC-H}) can emerge, where its temperature dependence is shown in Fig.~\ref{fig: CuMnAs}(b)~\cite{Wang_PhysRevLett.127.277201}.

\section{Summary and perspectives}
\label{sec: Summary and perspectives}

In this review, we introduced the concept of symmetry-adapted multipole basis set, which gives a unified description of electronic orderings and cross correlations, with an emphasis on the basic principles and advantages. 
The heart of this concept is that four types of multipoles (E, M, ET, and MT) form the symmetry-adapted basis in atoms, molecules, and crystals; any physical quantities, band structures, responses, and transports are expressed by this complete basis set. 
We have shown that a systematic representation based on multipoles is a powerful tool for exploring new electronic orderings and their associated physical phenomena, such as the E dipole inducing the nonlinear Hall effect based on the Berry curvature dipole mechanism, the anisotropic M dipole inducing the anomalous Hall effect without the uniform magnetization, M quadrupole inducing the spin-orbital-momentum locking, the MT monopole inducing the rotational distortion by the magnetic field, the MT dipole inducing the nonlinear nonreciprocal transport, the MT quadrupole inducing the symmetric spin splitting, the ET monopole inducing the chirality, and the ET dipole inducing the transverse response of conjugated physical quantities. 
To stimulate further experimental identification of these orderings and phenomena, we explain how to refer to collected tables for the classification of multipoles under 32 (122 magnetic) point groups. 
We also list several prototype and candidate materials hosting unconventional multipole orderings. 

Since the foundation of multipole representation in electron systems has been theoretically established, one can investigate/understand the electronic band structure, phase transition, and physical properties in materials along this line. 
Thus, the research field is moving on to the next stage. 
We would like to raise several research directions as follows. 
\begin{itemize}
\item {\it Effective Hamiltonian based on density-functional calculations:}

One of the intriguing directions is to evaluate multipole moments in real materials by using density-functional calculations. 
Since an effective Hamiltonian can be constructed by symmetry-adapted multipole basis~\cite{Oiwa_PhysRevLett.129.116401, Kusunose_PhysRevB.107.195118}, one can investigate the correlation between the expectation values of multipole moments and responses. 
By performing high-throughput calculations, one might obtain which model parameters play an essential role in enhancing the physical responses. 
Comparison with experimental results at the quantitative level is an important issue. 

\item {\it Superconducting order parameter:} 

The symmetry-adapted multipole basis can be applied to a superconducting order parameter~\cite{Sigrist_RevModPhys.63.239, Nomoto_PhysRevB.94.174513, Sumita_PhysRevResearch.2.033225, kirikoshi2023classification}. 
Since recent studies have implied the possibility of exotic superconducting states, such as a ferroelectric superconducting state in Sr$_{1-x}$Ca$_x$TiO$_{\delta}$~\cite{nphys4085}, multipole superconducting state~\cite{Kurihara_JPSJ.86.064706}, and superconducting state with the Bogoliubov Fermi surface~\cite{PhysRevLett.118.127001, PhysRevB.98.224509}, it is useful to systematically classify the superconducting order parameters based on the latest multipole representation, as discussed in this review~\cite{kirikoshi2023classification}. 
The exploration of cross correlation and transport phenomena under exotic superconducting states is also attractive~\cite{Daido_PhysRevLett.128.037001, Watanabe_PhysRevB.105.024308, Chazono_PhysRevB.105.024509}.

\item {\it Magnon and phonon systems:} 

The concept of symmetry-adapted multipole basis can be also applied to other elementary excitations, such as magnons and phonons. 
For example, recent theoretical studies clarified that the microscopic origin of nonreciprocal magnon dispersions is described by the odd-rank MT multipoles~\cite{Matsumoto_PhysRevB.101.224419, Matsumoto_PhysRevB.104.134420, Hayami_PhysRevB.105.014404} and that of chiral phonon dispersions in chiral materials~\cite{Pine_PhysRev.188.1489,PhysRevB.2.2049, PhysRevB.4.356, teuchert1974symmetry, PhysRevB.13.1383, PhysRevB.54.826, ghosh2008origin, Zhang_PhysRevLett.115.115502, zhu2018observation, Zhang_PhysRevResearch.4.L012024, chen2022chiral, ishito2023truly, ishito2023chiral, Kato_doi:10.7566/JPSJ.92.075002} is described by the even-rank ET multipoles~\cite{tsunetsugu2023theory}. 
A systematic derivation of the relation between internal degrees of freedom in magnons/phonons and multipoles will be an interesting issue.

\item {\it Mesoscale electronic orderings:} 

Extensions of the formalism to spatially nonuniform electronic orderings characterized by nonzero multiple propagation vectors are another intriguing issue. 
$\alpha$-Mn and Co$M_3$S$_6$ ($M=$ Nb, Ta) are typical examples, where the large anomalous Hall effect in the AFM structure can be explained by the extended cluster multipole theory~\cite{Yanagi_PhysRevB.107.014407}. 
The systematic foundation will be useful to clarify the cross correlation and transport phenomena under complicated multiple-$Q$ spin textures, such as magnetic skyrmion crystals and magnetic hedgehog crystals~\cite{nagaosa2013topological, hayami2021topological, Hayami_PhysRevB.105.104428, Bhowal_PhysRevLett.128.227204}.

\item {\it Coupling to electromagnetic waves}

Exploration of the possibility of controlling the multipoles via electromagnetic waves is also important. 
From the symmetry, $\partial{\bm{E}}/\partial t$ ($\partial{\bm{B}}/\partial t$) is a conjugate field to the MT (ET) dipole, but their mutual coupling has been veiled. 
In this aspect, the Floquet theory might be useful to investigate the role of such fields on multipoles~\cite{Eckardt_RevModPhys.89.011004, oka2019floquet, rudner2020band, Yambe_PhysRevB.108.064420, hayami2024analysis}.  
The dynamics of the multipoles under electromagnetic waves are also intriguing. 

\end{itemize}

\begin{acknowledgments}
We would like to thank R. Arita, H. Hanate, K. Hattori, A. Inda, A. Kirikoshi, J. Kishine, K. Matsuhira, T. Matsumoto, Y. Motome, M. Naka, J. Nasu, T. Nomoto, R. Oiwa, H. Seo, M.-T. Suzuki, S. Tsutsui, R. Yambe, Y. Yanagi, and M. Yatsushiro for fruitful collaborations and constructive discussions. 
We also thank H. Amitsuka, H. Harima, H. Hidaka, D. Hirai, Z. Hiroi, S. Hoshino, H. Ikeda, T. Ishitobi, K. Izawa, T. Kimura, F. Kon, T. Miyamoto, G. Motoyama, S. Ohara, T. Onimaru, A. Oyamada, H. Saito, T. J. Sato, R. Shiina, M. Shimozawa, M. Shinozaki, C. Tabata, H. Tanida, H. Tou, T. Yanagisawa, Y. Yanase, and H. Watanabe for their helpful discussions. 
This research was supported by JSPS KAKENHI Grants Numbers JP21H01031, JP21H01037, JP22H04468, JP22H00101, JP22H01183, JP23K03288, JP23H04869, JP23H00091 and by JST PRESTO (JPMJPR20L8) and JST CREST (JPMJCR23O4), and the grants of Special Project (IMS program 23IMS1101), and OML Project (NINS program No. OML012301) by the National Institutes of Natural Sciences. 
\end{acknowledgments}

\appendix
\section{Venn diagrams for multipoles according to $(\mathcal{P}, \mathcal{T}, \mathcal{PT})$ symmetries}
\label{app}

\begin{figure*}[htb!]
\centering
\includegraphics[width=1.0 \hsize]{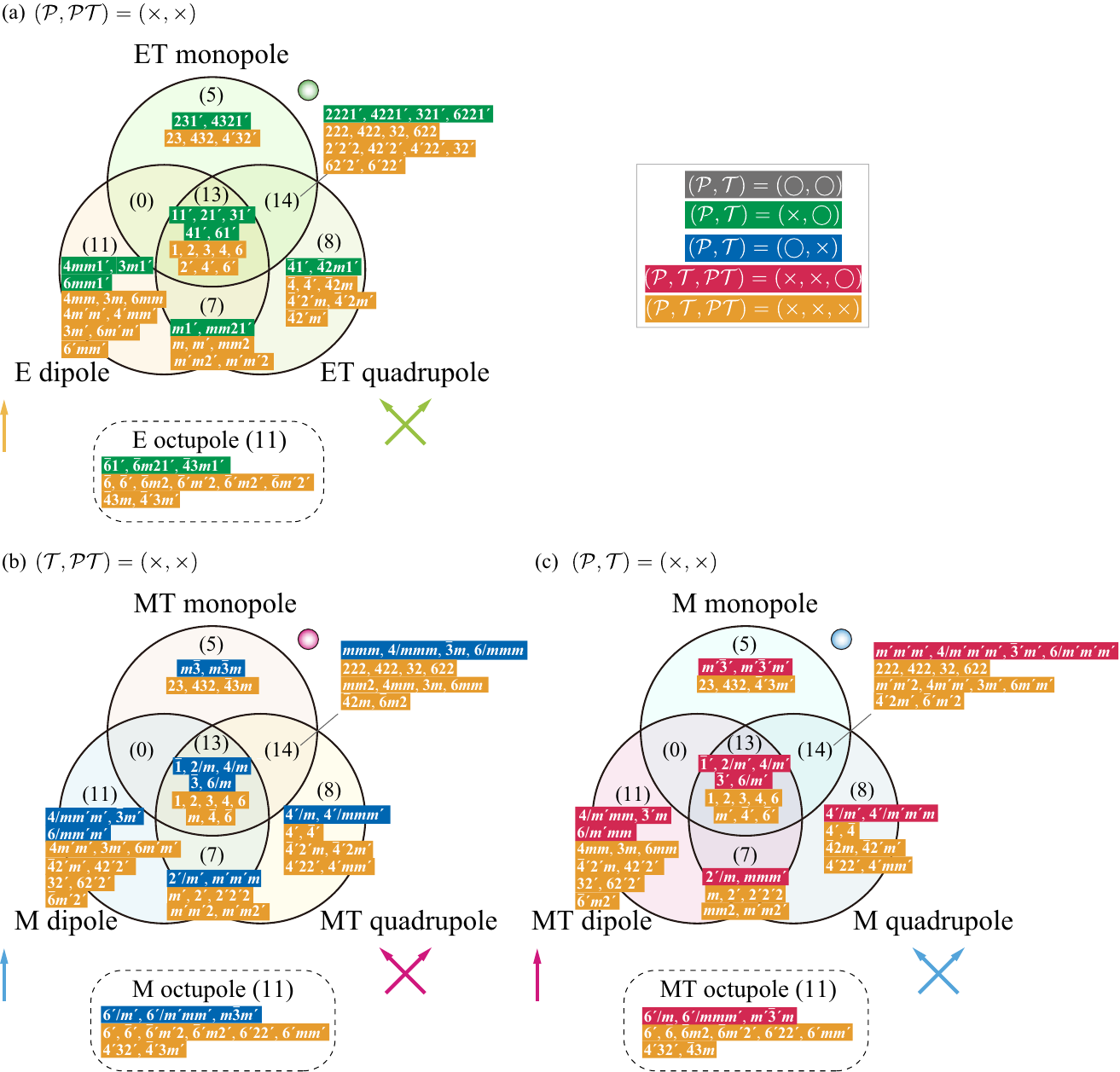}
\caption{
 \label{fig: venn_diagram_2}
The Venn diagrams classifying (a) the odd-parity electric-type multipoles (ET monopole, E dipole, and ET dipole) under $(\mathcal{P}, \mathcal{PT})=(-1, -1)$, (b) the even-parity magnetic-type multipoles (MT monopole, M dipole, and MT quadrupole) under $(\mathcal{T}, \mathcal{PT})=(-1, -1)$, and (c) the odd-parity magnetic-type multipoles (M monopole, MT dipole, and M quadrupole) under $(\mathcal{P}, \mathcal{T})=(-1, -1)$ belonging to the totally symmetric irreducible representation under the 122 magnetic point groups. 
The letters of each magnetic point group are color-coded according to their ($\mathcal{P}, \mathcal{T}, \mathcal{PT}$) parities, as shown in the right-top panel; $\bigcirc$ ($\times$) represents the presence (absence) of the symmetry.
}
\end{figure*}

In this Appendix, we classify the odd-parity electric-type multipoles (ET monopole, E dipole, and ET dipole) under $(\mathcal{P}, \mathcal{PT})=(-1, -1)$ in Fig.~\ref{fig: venn_diagram_2}(a), the even-parity magnetic-type multipoles (MT monopole, M dipole, and MT quadrupole) under $(\mathcal{T}, \mathcal{PT})=(-1, -1)$ in Fig.~\ref{fig: venn_diagram_2}(b), and the odd-parity magnetic-type multipoles (M monopole, MT dipole, and M quadrupole) under $(\mathcal{P}, \mathcal{T})=(-1, -1)$ in Fig.~\ref{fig: venn_diagram_2}(a) belonging to the totally symmetric irreducible representation under the 122 magnetic point groups. 
One finds that there are active 69 multipoles in all the cases, and looks into a beautiful relation that the same numbers of magnetic point groups belong to each category.

\bibliographystyle{JPSJ}
\bibliography{ref}

\begin{thebibliography}{100}
\bibitem{comment_2ndresponse}
It is noted that the nonlinear response function of electric field in the
  length gauge needs rederivation by applying $\hat{A}=-e\hat{r}$ ($\hat{r}$:
  position operator), since the matrix element of $\hat{r}$ in a periodic
  system includes differential operator of ${\bm k}$.
\bibitem{comment_Mn3Sn}
Based on the formulation in Sect.~\ref{sec: Site-cluster multipole}, the
  antiferromagnetic structure in Mn$_3$Sn is classified into the anisotropic
  magnetic dipole.
\end{thebibliography}

\end{document}